\documentclass[fleqn,usenatbib]{mnras}

\usepackage{newtxtext,newtxmath}

\usepackage[T1]{fontenc}

\DeclareRobustCommand{\VAN}[3]{#2}
\let\VANthebibliography\thebibliography
\def\thebibliography{\DeclareRobustCommand{\VAN}[3]{##3}\VANthebibliography}

\usepackage{graphicx}	
\usepackage{amsmath}	
\usepackage{caption}
\usepackage{subcaption}

\usepackage[usenames,dvipsnames,table]{xcolor}     

\title[]{Effect of galaxy-informed merger timescales on massive black hole binaries in LISA}

\author[J. S. Roth et al.]{Jaelyn S. Roth,$^{1}$\thanks{E-mail: jaelyn.roth@vanderbilt.edu }
Kelly Holley-Bockelmann,$^{3,2,1}$
Michael Tremmel,$^{4}$
Jillian Bellovary,$^{5,6,7}$
Riddhi Patel,$^{1,2}$
\and
Andrea Derdzinski,$^{1,2}$
KeShawn Ivory,$^{1}$
Fazeel Mahmood Khan,$^{8,9}$
Andrea Valerio Macciò$^{8,9}$
\\
$^{1}$Department of Physics and Astronomy, Vanderbilt University, Nashville, TN 37235, USA\\
$^{2}$Department of Life and Physical Sciences, Fisk University, Nashville, TN 37208, USA\\
$^{3}$SRON Space Research Organisation Netherlands, Niels Bohrweg 4, 2333 CA Leiden, the Netherlands\\
$^{4}$School of Physics, University College Cork, College Road, Cork, T12 K8AF, Ireland\\
$^{5}$Department of Physics, Queensborough Community College, Bayside, NY, 11364, USA\\
$^{6}$Department of Astrophysics, American Museum of Natural History, New York, NY 10024, USA\\
$^{7}$Astrophysics Program, CUNY Graduate Center, New York, NY, 10016, USA\\
$^{8}$New York University Abu Dhabi, PO Box 129188, Abu Dhabi, UAE\\
$^{9}$Center for Astrophysics and Space Science, NYUAD Research Institute, New York University Abu Dhabi, PO Box 129188, Abu Dhabi, UAE\\
}

\date{Accepted XXX. Received YYY; in original form ZZZ}

\pubyear{\the\year{}}

\begin{document}

\label{firstpage}
\pagerange{\pageref{firstpage}--\pageref{lastpage}}
\maketitle


\begin{abstract}
Over a decade of direct $N$-body simulations tell us that massive black hole binary (MBHB) merger timescales are strongly dependent on the structural and kinematic properties of the host galaxy merger remnant. The physical stages of stellar scattering and gravitational wave (GW) emission of MBH coalescence occur below the resolution limit of most cosmological simulations, requiring post-processing to achieve accurate merging times. We apply a recently developed galaxy-informed merger timescale for MBHBs to the \texttt{Romulus25} simulation and compare with current MBH merger rate and event rate estimates for the upcoming Laser Interferometer Space Antenna (LISA). We find a median stellar scattering time of $\sim10^{8.2}$ yr, roughly an order of magnitude faster than those produced by timescales that rely on extrapolations of the stellar density profile. We predict a merger rate of $\sim11.5 \text{yr}^{-1}$ corresponding to 46 merger events over a 4-year LISA mission. We also investigate the importance of the relative orbital inclination of the MBHB with respect to the host galaxy and of nuclear star cluster (NSC) modeling, finding that the quickest merger events correspond to MBHBs that counterrotate and galaxies assigned to host NSCs, which may be important for host galaxy identification in the multi-messenger era.
\end{abstract}

\begin{keywords}
black hole physics -- galaxies: kinematics and dynamics -- methods: numerical -- gravitational waves
\end{keywords}



\defcitealias{KHB2025}{HB25}
\defcitealias{Sesana2015}{SK15}
\defcitealias{Vasiliev2015}{V15}

\section{Introduction}

Galaxies can grow by merging together, which offers opportunities in turn to form massive black hole binaries (MBHBs)~\citep{Begelman1980}. 
The journey of the two MBHs toward pairing and coalescence is typically categorized into phases depending on the mechanism dominating the energy loss of the binary. In the canonical picture for MBHBs ($10^4 M_{\odot} \lesssim M_{\text{tot}} \lesssim 10^7 M_{\odot}$), dynamical friction dominates on $\sim$ kpc scales, driving the MBHs towards the center of their newly-shared host \citep{Chandrasekhar1943, Binney1987}. Stellar scattering effects become important on $\sim 10-1$ pc scales, hardening the binary further (provided there is a sufficient supply of stars) \citep{Milosavljevic2003}. The presence of an accretion disk on $\sim$ sub-pc scales in gas-rich nuclei may also aid the hardening of the binary \citep{Mayer2007, Cuadra2009, Haiman2009, Siwek2023}. Once $\sim$ mpc separations are reached, the decay is dominated by the emission of gravitational waves (GWs) until coalescence \citep{Peters1963}. 

The timescale governing each of these phases is complicated, however, and is highly dependent on the properties of the host galaxy. Dynamical friction is enhanced when MBHs are encased in a dense nuclear star cluster (NSC), increasing the effective mass and invoking tidal distortion that sinks the MBHs quickly \citep{Ogiya2020, Mukherjee2025}. Spikes from self-interacting dark matter may also aid the efficiency of sinking \citep{Alonso2024}. In contrast, MBHs can fail to sink for a multitude of reasons, including (but not limited to) getting trapped in galactic bar-induced resonances \citep{Bortolas2022, Noahabstract}, encountering star-forming clumps in non-relaxed systems, as would be common at high redshift \citep{Ma2021}, or being stripped of their associated stellar systems in minor mergers, leaving them ``wandering" for longer than the age of the Universe \citep{Bellovary2010, Zhou2025}. 

At scales where the hardening is dominated by stellar scattering, $N$-body simulations and analytic estimates alike show that non-spherical galactic potentials are required to overcome the ``final parsec problem" \citep{Yu2002, Berczik2006, Khan2013, Baile15,Gualandris2017}. Additionally, any degree of rotation of the stellar component \citep{KHB2015, Varisco2021}, inclination of the MBHB with respect to the host \citep{Mirza2017}, and the presence of a nuclear star cluster (NSC) \citep{Khan2021,Khan2026} also encourage efficient hardening. When embedded in a gaseous accretion disk, the orbital evolution of the binary depends on both binary properties, like mass ratio and initial eccentricity \citep{Siwek2023}, as well as the disk structure and thermodynamic properties \citep{Tiede2020, Dorazio2021, Dittmann2026, Clyburn2026}. Simulations in the thin disk regime (which is expected for astrophysical disks surrounding MBHs in gas rich nuclei) do not yet agree on the sign of the imparted torque on the binary within the full range of possible parameter space \citep{Duffell2024,Derdzinski2025}. All of these effects combine in highly nonlinear ways, requiring expensive detailed modeling to achieve self-consistent predictions, albeit only for a few galaxy types at a time.

With their rich variety of host galaxy properties and ancestries, cosmological simulations offer a promising way to investigate the broader possible range of MBH coalescence times. A large challenge with these models, however, is that many of the aforementioned physical processes occur below the resolution limit of the simulation, which can be as large as $\sim$ few kpc. Instead, to prolong the lifetime of the MBHB, simulations either invoke sub-grid models, which operate on-the-fly, or post-processing models, which can be performed on simulation outputs. Both of these methods require analytical estimates for MBH merger timescales since the dynamics are not resolved outright.

\citet{Sesana2015} (hereafter \citetalias{Sesana2015}) propose one such recipe for the merger timescale. They conjecture that the hardening parameter, ($S$), defined as the rate of change of the inverse of the binary semi-major axis ($a$), is equivalent that for a MBHB in an unbound, homogenous stellar system \citep{Quinlan1996}. The relation follows

\begin{equation}\label{eq:sesana2015}
    S_{\text{SK15}} \equiv \frac{d}{dt} \Big( \frac{1}{a} \Big) = H \frac{G \rho_{\text{inf}}}{\sigma_{\text{inf}}},
\end{equation}

\noindent where $H \approx 15-20$ is a dimensionless rate, $G$ is the gravitational constant, and $\rho_{\text{inf}}$ and  $\sigma_{\text{inf}}$ are the stellar density and velocity dispersion measured at the binary influence radius ($r_{\text{inf}}$), which is defined as the radius surrounding the binary that encompasses twice the binary's mass in stars.

This timescale has been well-used in cosmological simulations \citep[e.g.][]{Kelley2017} and semi-analytic models \citep[e.g.][]{Biava2019, Bortolas2021} alike, as one can estimate the coalescence time using properties easily extractable from the stellar density profile of the host. However, the resolution limit in cosmological simulations is oftentimes orders of magnitude larger than $r_{\text{inf}}$ of the binary. This requires us to either use analytic prescriptions or scaling relations for $\rho_{\text{inf}}$ and $\sigma_{\text{inf}}$ or to inwardly extrapolate galaxy properties over very large scales. 

Another recipe by \citet{Vasiliev2015} (hereafter \citetalias{Vasiliev2015}) considers a separation-dependent hardening parameter that follows

\begin{equation}\label{eq:vasiliev2015}
    S_{\text{V15}}(a) = \mu S_{\text{inf}} (a/a_{\text{h}})^{\nu},
\end{equation}
\noindent where $\mu$ is the filling fraction of the loss cone (fiducial value $\mu = 0.3$) and $\nu$ determines the decay of $S$ with $a$ as the loss-cone depletes (fiducial value $\nu = 0.4$). $S_{\text{inf}}$ is the full-loss-cone hardening rate at $r_{\text{inf}}$, obtained by substituting $\sigma_{\text{inf}}^2 \sim GM_{\text{tot}}/r_{\text{inf}}$ and $\rho_{\text{inf}}=2M_{\text{tot}}/r_{\text{inf}}^3$ into Eq. \ref{eq:sesana2015}:

\begin{equation}
    S_{\text{inf}} \equiv A(GM_{\text{tot}})^{1/2}r_{\text{inf}}^{-5/2}
\end{equation}
with a fiducial $A=4$. Lastly, $a_{\text{h}}$ is the hard binary separation, i.e. the separation at which the binding energy of the binary increases rather than decreases with stellar interactions. $a_{\text{h}}$ depends on the binary mass ratio ($q \equiv M_2/M_1 \leq 1$), and is given by $a_{\text{h}} = \frac{q}{4(1+q)^2} r_{\text{inf}}$. 

The \citetalias{Vasiliev2015} timescale has been used in some form for post-processing delays in cosmological simulations with orbital information \citep[e.g.][]{Katz2020, Chen2022_orb, DeCun2023}. Despite the control that this method has over the effect of loss-cone depletion in slowing the scattering process, it cannot reproduce the conditions of a rotating stellar background that are thought to bring the binary closer to the full loss-cone regime again \citep{KHB2015, Varisco2021}. Similarly to \citetalias{Sesana2015}, it also requires either analytic prescriptions or inwards extrapolation to measure $r_{\text{inf}}$ and $\sigma_{\text{inf}}$.

\begin{figure*}
    \centering
    \includegraphics[width=1.0\linewidth]{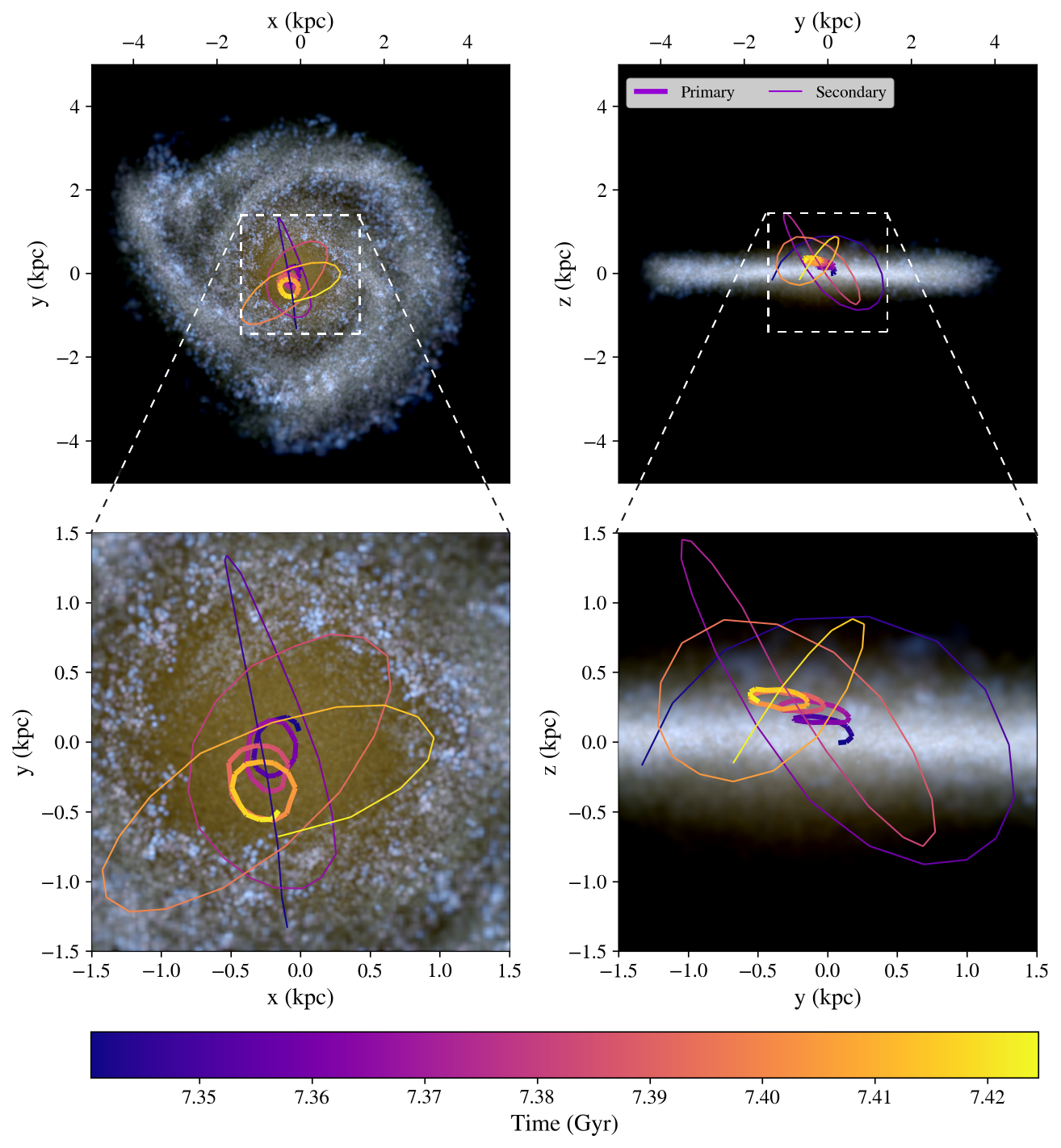}
    \caption{Representative orbit of a MBHB around a Milky Way mass galaxy in \texttt{Romulus25}. The left panel shows the face-on view with respect to the galaxy and the right panel shows the side-on view, oriented such that the stellar angular momentum vector points in the positive $z$ direction. The primary has a mass of $M_1 \sim 3.3 \times 10^8 M_{\odot}$ and its trajectory is shown by the thick line; the secondary has a mass of $M_2 \sim 1.2 \times 10^8 M_{\odot}$ and is shown by the thin line. The line color indicates the time in Gyr in the simulation; the MBH information is stored at a cadence of 1.7 Myr, while the galaxy information is stored at an average cadence of $\sim110$ Myr. The host galaxy has $M_{\star} \sim 7.8 \times 10^{10} M_{\odot}$ and is captured at the closest timestep ($t=7.67$ Gyr) following the numerical merger of the MBHs at $t=7.42$ Gyr. The bottom panel zooms in to the central $\sim$3 kpc of the galaxy. In this instance, the numerical merger occured when the MBHs were separated by $\sim0.47~$kpc and their center-of-mass position was located $\sim0.61~$kpc from the galaxy center. The angle $\Theta$ between the orbital angular momentum and galaxy stellar angular momentum is $\Theta \sim 83^{\circ}$, making this a corotating merger.}
    \label{fig:orbits}
\end{figure*}

In this paper, we adopt a galaxy-informed timescale proposed in \citet{KHB2025} (hereafter \citetalias{KHB2025}) which incorporates kinematic and structural properties of the host galaxy and orbital information about the binary into the estimate for the merger timescale through the late dynamical friction and stellar scattering phases of MBH coalescence. By relying largely on bulk host galaxy properties that are ostensibly simple to obtain from simulations and observations alike, the timescale aims to incorporate the galaxy properties relevant to stellar scattering while maintaining relative simplicity of implementation. It is important then to test the timescale on a cosmologically-realized population of host galaxies with realistic stellar structures and kinematics. Our choice of simulation, \texttt{Romulus25} \citep{Tremmel2017} also succeeds in accurately tracing MBH dynamics down to sub-kpc scales \citep{Tremmel2015}, offering a diverse array of MBHB orbital configurations as well (see Fig.  \ref{fig:orbits}).

The aim of this work is not solely to provide another MBH merger rate estimate to the incredibly vast range of predictions, which span up to $\sim 6$ orders of magnitude across simulations \citep{IzquierdoVillalba2026_MBHCat}, but also to examine the prevalence of certain galaxy features, highlighted in \citetalias{KHB2025} to be conducive to MBH mergers, among a cosmological population of merger hosts. These are key questions surrounding the upcoming LISA mission, set to launch in 2035, which will be able to detect gravitational waves from MBHBs to extremely high redshift \citep[$z \gtrsim 12$,][]{AmaroSeoane2023}. 

The organization of the paper is as follows. We outline our technique in Section \ref{sec:methods}, including an introduction to our post-processing model and base simulation, as well as details on how we determined galaxy properties from the simulation. Section \ref{sec:results_and_discussion} discusses our results, compares them to other standard coalescence timescale models, and calculates the MBH merger rate for LISA. We discuss caveats of our work in Section \ref{sec:caveats} and conclude in Section \ref{sec:conclusions}.

\section{Methods}\label{sec:methods}

In this section, we present a brief description of both the galaxy-informed timescale (\ref{subsec:timescale}) and the \texttt{Romulus25} cosmological simulation (\ref{subsec:romulus}). We also describe how we sourced galaxy information from the galaxies containing MBH mergers within \texttt{Romulus25}, for both the galaxy-informed timescale and the timescales by \citetalias{Sesana2015} and \citetalias{Vasiliev2015} (\ref{subsec:galaxyparams}). 

\subsection{Direct N-body simulations and galaxy-informed coalescence timescale}\label{subsec:timescale}
The galaxy-informed timescale used in this work is introduced in \citetalias{KHB2025}. We briefly summarize the testing suite and timescale here.

\citetalias{KHB2025} collected 28 direct N-body simulations, accumulated over the course of a decade, of realistic galaxy centers hosting MBHBs. The galaxies were constructed in equilibrium and modeled after observations of real galaxies that vary in size, content, and structure. All galaxies exhibit some degree of triaxiality, and many contain NSCs and bulges. These galaxies are described in more detail in the original publications \citep{KHB2015, Khan2015, Khan2021, Khan2024} and summarized in Table 1 of \citetalias{KHB2025}.

The simulations focus on the stellar interactions with the MBHB by simulating only the central few hundred parsecs (depending on the model) of a gas and dark-matter (DM)-free galaxy. The secondary MBH is sent inwards on either planar prograde ($\Theta=0^{\circ}$) or retrograde ($\Theta=180^{\circ}$) orbits, where $\Theta$ is the angle between the orbital angular momentum and the stellar angular momentum of the host. The testing suite in \citetalias{KHB2025} varied $\cos\Theta$ between -1 and 1.

\citetalias{KHB2025} performs a best-subset selection and linear regression over tens of parameters relating to the MBHB (e.g. total mass, initial eccentricity, mass ratio, etc.) and the galaxies (e.g., stellar mass, S\'ersic index, NSC mass, etc.) to determine which few best describe the coalescence time ($T_{\text{coal}}$) of the binary. The resulting four-term equation (Eq. 5 of \citetalias{KHB2025}) with an $R^2 = 0.86$ is

\begin{multline}\label{eq:timescale_4term}
    \log(T_{\text{coal}}) = 8.2 +0.19 \log(M_{\text{tot}}) - 0.16 \log(M_{\text{NSC}}) \\ + 0.71(v/\sigma |_{\text{proj}}) - 0.32(R_{\text{eff}}),
\end{multline}

\noindent where $M_{\text{tot}}$ is the total mass of the MBHB, $M_{\text{NSC}}$ is the mass of the NSC (if the galaxy is nucleated), $v/\sigma$ is the ratio of the circular stellar velocity to the stellar velocity dispersion measured at the binary influence radius, $v/\sigma |_{\text{proj}} = v/\sigma \cos \Theta$ denotes the projection of $v/\sigma$ onto the plane of the binary, and $R_{\text{eff}}$ is the effective radius of the galaxy. \citetalias{KHB2025} finds that Eq. \ref{eq:timescale_4term} reproduces the true coalescence times from the simulations to within $\sim 30 \%$.

\subsection{The \texttt{Romulus25} simulation}\label{subsec:romulus}

We implemented the galaxy-informed timescale onto the MBHB population of \texttt{Romulus25}. \texttt{Romulus25} \citep{Tremmel2017} is a cosmological hydrodynamic simulation with a volume of $(25 \text{cMpc)}^3$ run to $z=0$ with the massively parallel Tree+SPH code, \textsc{ChaNGa} \citep{Menon2015}. \textsc{ChaNGa} builds upon physics modules used in the code \textsc{Gasoline} \citep{Wadsley2004}; those used to run \texttt{Romulus25} include a cosmic UV background \citep{Haardt2012}, star formation and ``blastwave" supernovae feedback \citep{Stinson2006} and low-temperature metal cooling \citep{Guedes2011}. 

\begin{figure*}
    \centering
    \includegraphics[width=1.0\linewidth]{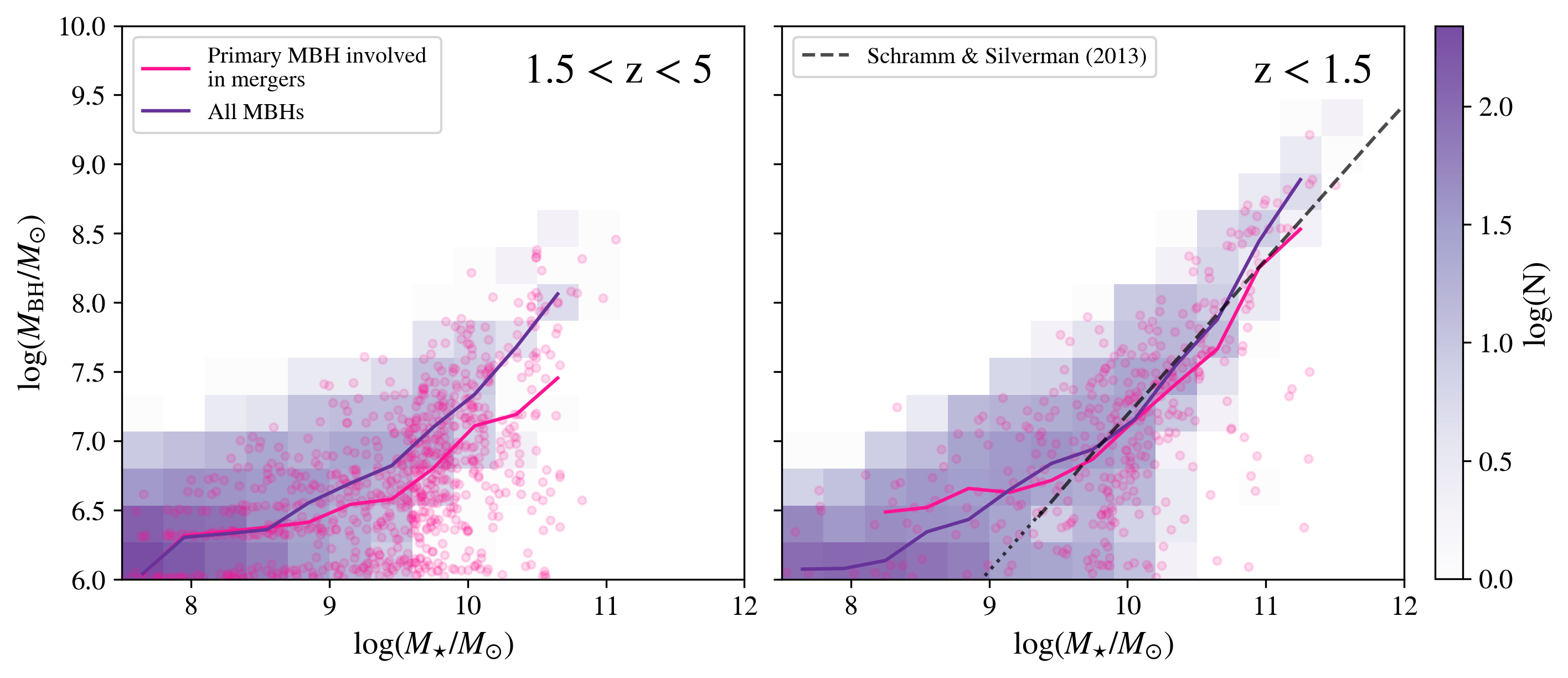}
    \caption{Black hole mass to galaxy stellar mass relation for all galaxies with central MBHs compared to galaxies hosting merging MBHs. We show the relation at high redshift ($1.5 \lesssim z \lesssim 5$, middle panel), and low redshift ($z \lesssim 1.5$, right panel). The colorbar indicates the logged counts of the nonmerging central MBHs in \texttt{Romulus25}, with the associated purple line indicating the median per stellar mass bin. The magenta scatter points show all primary MBH masses involved in MBH mergers and their associated stellar masses, with medians per bin indicated by the magenta line. We correct the galaxy stellar masses by a factor of 0.6 to account for the systematic observational underestimation of stellar mass compared to simulation \citep[see text,][]{Munshi2013}. We add the observed $M_{\text{BH}}-M_{\star}$ relation by \citet{Schramm2013} to the low redshift panel as a dashed line (dotted when extrapolating beyond the sample masses).}
    \label{fig:mbh_mstar}
\end{figure*}

BHs in \texttt{Romulus25} are seeded based upon physical properties of the gas in the simulation. Gas particles are converted into BH particles if they are quite metal poor ($Z<3\times10^{-4}Z_{\odot}$), exceed 15 times the density required to undergo star formation ($n > 3 \text{cm}^{-3}$), and fall within a temperature range which is cooler than the temperature required to form stars ($9500 \text{K} < T < 10,000\text{K}$), implying a collapse timescale that is quicker than the timescale for star formation. These conditions are consistent with seeding scenarios which require rapidly collapsing gas, including the channels for direct-collapse \citep{Begelman2006, Lodato2006}, stellar runaway in dense clusters \citep{Devecchi2009, Davies2011} or even rapid accretion onto the remnant BH of a Population III star \citep{Volonteri2005, Inayoshi2016, Sassano2023}. This prescription provides a more physical interpretation of the MBH occupation fraction compared to models that seed MBHs when a halo mass threshold is met \citep[e.g.][]{Sijacki2015, Schaye2015, Chen2022_DF}. The seed mass for the resulting BHs in \texttt{Romulus25} is $M_{\text{BH,seed}} = 10^6 M_{\odot}$. Fig. \ref{fig:mbh_mstar} shows the $M_{\text{BH}}-M_{\star}$ relation for all galaxies in \texttt{Romulus25} hosting central MBHs (purple) compared to those hosting MBH mergers (pink) across redshift. Many MBHs begin to be seeded and even merge at high redshift ($z \gtrsim 6$), though we don't display these because they occur in low mass galaxies ($M_{\star} < 10^{7.5} M_{\odot}$) that are excluded from our sample. The merging MBHs show a general agreement with the median $M_{\text{BH}}-M_{\star}$ of the nonmerging population, with a slight underrepresentation of high BH masses in the high $M_{\star}$ regime. Generally, though, the merging population is not preferential to a certain subset of the MBH population in \texttt{Romulus25}.

\begin{figure}
    \centering
    \includegraphics[width=1.0\linewidth]{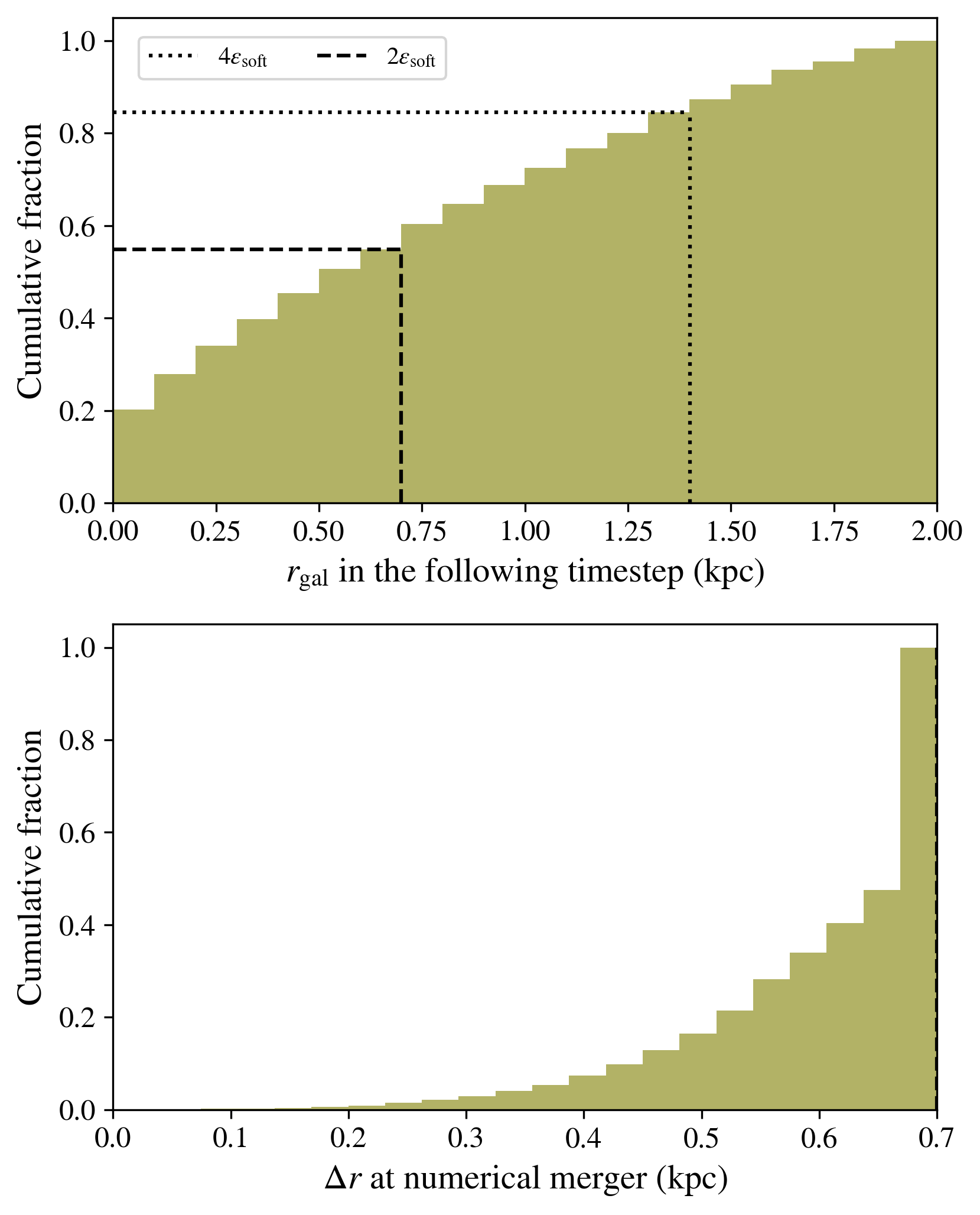}
    \caption{Top: Cumulative fraction of distances between the remnant black hole and the galaxy center ($r_{\text{gal}}$) in the timestep most closely following the numerical merger time. We exclude mergers that have $r_{\text{gal}} > 2$kpc. Bottom: Cumulative fraction of MBHB orbital separations ($\Delta r$) at the time of numerical merger. Separations of two softening lengths ($2\epsilon_{\text{soft}}=0.7$kpc) and four softening lengths ($4\epsilon_{\text{soft}}=1.4$kpc) are marked by the dashed and dotted lines, respectively.}
    \label{fig:sep_and_offsets}
\end{figure}

Once seeded, the BHs are not pinned to the centers of their halos \citep[e.g.][]{Sijacki2015, Schaye2015} but rather their dynamics are modified with a sub-grid force correction term \citep{Tremmel2015}. This term is meant to approximate the dynamical friction that a MBH undergoes when traveling through a sea of stars of much lower mass, which cannot be properly resolved in most cosmological simulations due to the relatively similar masses of BH and star particles. \citet{Tremmel2015} finds that this dynamical friction term appropriately delays the sinking timescale of the MBH towards the halo center, decoupling the MBH merger rate from the galaxy merger rate \citep[see also][for other examples of sub-grid dynamical friction prescriptions]{Chen2022_DF, Ma2023, Genina2024}. This, combined with the physical seeding mechanism, leads to plenty of off-center, or ``wandering" black holes, which are thought to be common in large galaxies \citep{Bellovary2010, Ricarte2021}. 

Despite the high prevalence of these wanderers, we restrict our sample to MBHBs that are roughly central with their hosts, as the $N$-body simulations of \citetalias{KHB2025} explicitly model galactic centers. We limit our sample to mergers where the remnant black hole in the closest following timestep has a galactocentric distance $r_{\text{gal}}<2$kpc; we choose this distance as galaxy merger remnants can have particularly ill-defined centers and may be uncertain up to this distance. In the top panel of Fig. \ref{fig:sep_and_offsets}, we show the distribution of galactocentric distances for the remnant black hole as measured in the timestep most closely following the numerical merging time. Just over 80\% of the mergers in the sample occur within the central four softening lengths ($4\epsilon_{\text{soft}}=1.4$kpc) of the galaxy center.

BHs in \texttt{Romulus25} are allowed to numerically merge if the following criteria \citep{Bellovary2011} are met:
\begin{enumerate}
    \item The relative distance ($\Delta \textbf{r}$) of the two BHs is within two softening lengths: $\Delta \textbf{r} < 2 \epsilon_{\text{soft}}=0.7\text{kpc}$
    \item The two BHs have low enough relative velocities ($\Delta \textbf{v}$) and accelerations ($\Delta \textbf{a}$) that they may be considered bound: $\frac{1}{2}\Delta \textbf{v} < \Delta \textbf{a} \cdot \Delta \textbf{r}$
\end{enumerate}
At $\Delta r\leq0.7$kpc, BHs in \texttt{Romulus25} are still squarely in the dynamical friction phase, and we consider them a bound, close pair \citep{Tremmel2018_pairs}. The cumulative distribution of separations of the numerical mergers from \texttt{Romulus25} is shown in the bottom panel of Fig. \ref{fig:sep_and_offsets}. The peak at $\Delta r=0.7$kpc corresponds to numerical mergers that occurred as soon as the distance threshold was met.

Since the $N$-body simulations of \citetalias{KHB2025} model the central few hundred pc of the galaxies, which is approximately the scale that \texttt{Romulus25} cannot model, we choose not to model additional dynamical friction time to connect the two phases. We exclude mergers of BHs that merged within 10 Myr of their formation time or that were formed within $4 \epsilon_{\text{soft}}$ of another BH. We start with a sample of 1867 mergers that fit these criteria. In the next section, we gather the relevant host galaxy properties, and our final sample of MBH mergers where all required galaxy properties were available amounts to 1457.

\subsection{Sourcing galaxy properties}\label{subsec:galaxyparams}
By design, the galaxy properties of Eq. \ref{eq:timescale_4term} are relatively easy to gather from cosmological simulations. BH information in \texttt{Romulus25} is saved with a finer time resolution than that of the larger simulation volume, thus properties like $M_{\text{tot}}$ are saved at the moment of numerical merger regardless of the time of the simulation snapshots (or "timesteps"). For galaxy properties, we take the state of the galaxy at the timestep most closely following the merger time to approximate the post-merger remnant that the MBHB would inhabit, which ranges anywhere from almost instantaneous to $\sim 400$ Myr after the numerical merger, with $\sim80\%$ occuring within $100$ Myr. To accurately trace the location of the MBHB with respect to the galaxy center, we linearly interpolate along each dimension of the galaxy center's motion to approximate its position between timesteps. Fig. \ref{fig:orbits} shows the orbital motion of a selected MBHB around the galaxy center as projected on the $xy$ and $yz$ planes of the host. We also exclude mergers assigned to galaxies with $\log( M_{\star} / M_{\odot}) < 7.5$, as these are poorly resolved in \texttt{Romulus25} and correspond to galaxies with too few ($\lesssim 150$) star particles. Throughout this section, we make use of the \texttt{TANGOS} \citep{Pontzen2018} and \texttt{pynbody} \citep{Pontzen2013} libraries to source information from our MBH merger hosts.

\begin{figure}
    \centering
    \includegraphics[width=0.9\linewidth]{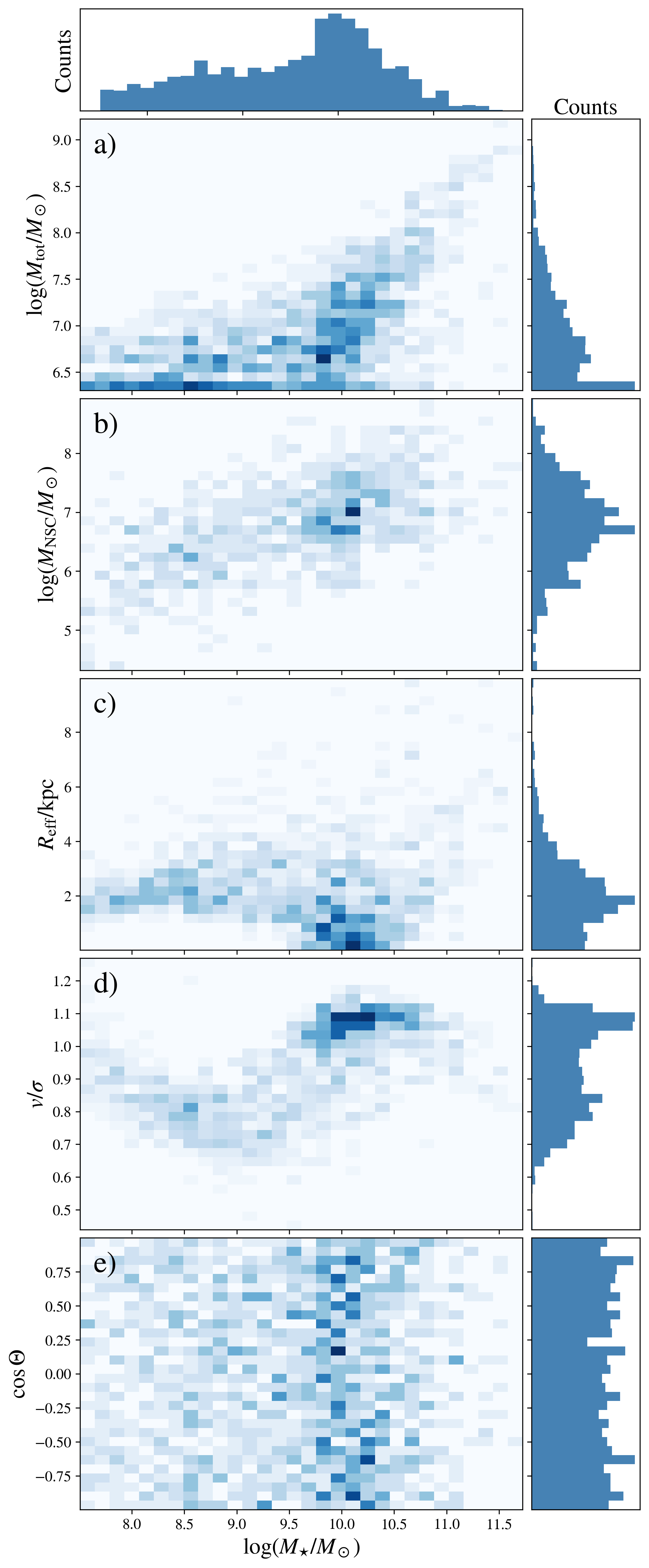}    {\phantomsubcaption\label{fig:ingredients_logMtot}%
     \phantomsubcaption\label{fig:ingredients_logMnsc}%
     \phantomsubcaption\label{fig:ingredients_Reff}%
     \phantomsubcaption\label{fig:ingredients_vsigma}%
     \phantomsubcaption\label{fig:ingredients_cosTheta}%
     }

    \caption{Two-dimensional distributions of each of the ingredients for the galaxy-informed timescale (Eq. \ref{eq:timescale_4term}) vs. the stellar mass of the host galaxy. The topmost panel shows the distribution of $M_{\star}$, and each of the rightmost panels shows the one-dimensional distribution of the ingredient for the timescale. a) Distribution of $M_{\text{tot}}$ of the MBHB at the time of numerical merger by $M_{\star}$. b) Distribution of assumed NSC masses of galaxies assigned NSCs by $M_{\star}$. c) Distribution of effective radii of galaxies hosting MBH mergers by $M_{\star}$. d) Distribution of stellar velocity divided by velocity dispersion by $M_{\star}$, where $v/\sigma$ is measured at four times the gravitational softening length ($4\epsilon_{\text{soft}}$) from the center of the galaxy. e) Distribution of $\cos\Theta$ by $M_{\star}$,  where $\Theta$ is the angle between the galaxy stellar angular momentum and the MBHB orbital angular momentum.}

    \label{fig:ingredients}
\end{figure}

\subsubsection{Stellar kinematics: measured $v/\sigma$}\label{sec:methods_vsigma}
We calculate $v/\sigma$ for each merger using the stellar kinematic properties of the galaxy at the timestep following the numerical merger. The stellar velocity profile is calculated assuming circular orbits around the enclosed stellar and DM mass at each radius, binned every 100 pc. For velocity dispersion, we calculate the spherically-averaged, three-dimensional, enclosed velocity dispersion profile for each galaxy, also binned every 100 pc. For bins with too few stars ($<5$) to calculate velocity dispersion, we fall back to using gas and/or DM particles. We choose to measure $v/\sigma$ at $4 \epsilon_{\text{soft}}=1.4$kpc from the galaxy center, as $\sigma$ is an enclosed quantity and dynamics only just become resolved at $2 \epsilon_{\text{soft}}$ from the galaxy center. While we cannot resolve to scales of $r_{\text{inf}}$ of the MBHB as requested by the galaxy-informed timescale, we expect that the approximation that $(v/\sigma)_{\text{inf}} \approx (v/\sigma)_{4\epsilon_{\text{soft}}}$ only has a minor downstream effect on the coalescence timescale given the small range of measured $v/\sigma$.

In Fig. \ref{fig:ingredients_vsigma}, we show the measured distribution of $v/\sigma$. All of our merger hosts show some degree of ordered rotation, with a minimum value of $v/\sigma \sim0.5$ and a median of $v/\sigma \sim 0.9$. About 40\% of our sample is rotation-dominated ($v/\sigma >1.0$), which differs from the expectation that merger hosts are mostly dispersion-dominated \citep{Bournaud04, Cox06}. The largest values ($v/\sigma \sim1.2$) occur in the Milky-Way mass regime ($M_{\star} \sim 10^{10-11} M_{\odot}$). Note that $v/\sigma$ here is not directly analogous to that observed in real galaxies, as we are not accounting for viewing angle or dust radiative transfer necessary for obtaining accurate line-of-sight velocities from cosmological simulations \citep[e.g.][]{Bardati2024_kinematics, Horlaville2026} (see Sec. \ref{sec:caveats_rotation} for the caveats of this method).

\subsubsection{Orbital configuration}\label{subsubsec:orbital_config}
We calculate the angle $\Theta$ between the MBHB orbital angular momentum ($\vec{l}$) and the galaxy's stellar angular momentum ($\vec{L}$), where $\cos \Theta = \vec{L} \cdot \vec{l} / |\vec{L} \vec{l}|$. We measure $\vec{L}$ by taking the mean of the inner 1.4 kpc ($\lesssim4 \epsilon_{\text{soft}}$) of the spherically-averaged stellar angular momentum profile of the galaxy. We calculate $\vec{l}=M_{1}\vec{r}_{1} \times \vec{v}_{1} + M_{2}\vec{r}_{2} \times \vec{v}_{2}$ where $\vec{r}_{1}$, $\vec{r}_{2}$  and $\vec{v}_{1}$, $\vec{v}_{2}$ are the positions and velocities of the primary and secondary MBHs with respect to the MBHB center-of-mass at the moment of numerical merger.

We define co-rotation (or prograde orbits) such that $0^{\circ} \leq \Theta \leq 90^{\circ}$ and counter-rotation (or retrograde orbits) such that $90^{\circ} < \Theta \leq 180^{\circ}$. We show the measured distribution of $\cos \Theta$ in Fig. \ref{fig:ingredients_cosTheta}, which resembles that of a uniform distribution of inclinations. This implies that $\sim 50 \%$ of the MBHBs in our sample are counterrotating with respect to their hosts. To determine if this was an effect of MBH dynamics becoming uncertain below the resolution limit, we also measured $\cos\Theta$ by calculating $\vec{l}$ at the moment the secondary MBH came within $2 \epsilon_{\text{soft}}$ of the galaxy center. We found $\cos\Theta$ to remain uniformly distributed, and conclude that this measurement is not particularly sensitive to the radius of measurement within the inner few kpc of the galaxy.

\subsubsection{Nuclear star clusters}
As NSCs are unresolved in \texttt{Romulus25}, we implement a very simple procedure to add NSCs onto (some) galaxies hosting MBH mergers. We assign a NSC to a merger host with a probability given by the occupation fraction constraints of \citet[][see their Fig. 3]{Neumayer2020}. While the occupation fraction differs slightly between early and late-type galaxies---high mass late-type galaxies host many more NSCs than early-types of the same mass---we leave a more detailed NSC assignment that distinguishes between galaxy types for a future work. We implement a scatter around the occupation fraction of $0.2$ for $M_{\star} \leq 10^{8} M_{\odot}$, $0.1$ for $10^{8} M_{\odot} \leq M_{\star} < 10^{10} M_{\odot}$, and $0.4$ for $M_{\star} \geq 10^{10} M_{\odot}$, reflecting the large uncertainty in the presence of NSCs in high mass galaxies. It may be true that MBH hosts are preferentially nucleated; \citet{Tremmel2024} finds the occupation fraction for MBHs in \texttt{Romulus25} to match well with the observed occupation fraction of NSCs in the Local Volume \citep{Hoyer2021, Carlsten2022}. Still, we choose the more conservative approach to seed based on stellar mass instead of MBH presence.

For galaxies assigned a NSC, we allocate a mass given the scaling relation also from \citet{Neumayer2020}:

\begin{equation}\label{eq:NSC_scaling_relation}
    \log M_{\text{NSC}} = 0.48 \log \bigg( \frac{M_{\star}}{10^9 M_{\odot}} \bigg) + 6.51,
\end{equation}
with an intrinsic scatter of $0.6$ dex. For galaxies not assigned a NSC, we omit the NSC term of Eq. \ref{eq:timescale_4term} from the coalescence time, necessarily implying shorter coalescence times for nucleated galaxies (see Sec. \ref{subsec:caveats_NSCs} for further discussion about this choice). 

Of our MBH merger host sample, $\sim50\%$ are assigned a NSC. Their masses are displayed in Fig. \ref{fig:ingredients_logMnsc}. The masses span the range $10^{4.3} \lesssim M_{\text{NSC}} / M_{\odot} \lesssim 10^{8.9}$, with a median NSC mass of $M_{\text{NSC}} \sim 10^{6.7} M_{\odot}$. We are able to produce the full observed mass range of NSCs, which approximately spans from $\sim10^{4.5}-10^9 M_{\odot}$ \citep{Georgiev2016, Neumayer2020}. 

\subsubsection{Effective radii}\label{subsec:methods_reff} 
We calculate the effective radii of the \texttt{Romulus25} MBH merger hosts by integrating over their simulated $i-$band surface brightness profiles. We invoke a limiting surface brightness of $\mu_{i} \leq 32$ and only fit over as far as the nearest satellite galaxy, if one exists in the halo. We make a cut for any very large effective radii ($R_{\text{eff}} > 10$kpc) as the galaxy-informed timescale \citepalias[Eq. \ref{eq:timescale_4term},][]{KHB2025} begins to break down at these large scales. The resulting distribution of $R_{\text{eff}}$ is shown in Fig. \ref{fig:ingredients_Reff}. Approximately $\sim78\%$ of our MBH merger hosts have $R_{\text{eff}} < 3$kpc, with a slight trend towards larger $R_{\text{eff}}$ with larger $M_{\star}$. There is an overdensity in the $M_{\star} \sim 10^{10} M_{\odot}$ mass regime, with many cases of $R_{\text{eff}} \sim 1$kpc. 

\subsubsection{Extrapolated stellar properties at the influence radius}

Although most of our terms involve bulk quantities, some information is required near the influence radius of the black hole, both for our case and for commonly-accepted prescriptions.
To compare our timescale to \citetalias{Sesana2015} and \citetalias{Vasiliev2015}, we follow the procedure laid out in Sec. 3.2 of \citetalias{Sesana2015}. First, we make the approximation that the binary spends most of its time at the transition separation, $a_{\text{tr}}$, between stellar hardening, which follows $\propto a^2$, and GW hardening, which follows $\propto a^{-3}$. The combined hardening rate is then
\begin{equation}\label{eq:dadt_star_gw}
    \frac{da}{dt} = \frac{da}{dt} \Bigg|_{\star} + \frac{da}{dt} \Bigg|_{\text{GW}} = -Aa^2 - \frac{B}{a^3},
\end{equation}
\noindent where \citetalias{Sesana2015} posits that $A=S_{\text{SK15}}$ as described by Eq. \ref{eq:sesana2015}, and 
\begin{equation}
    B = \frac{64G^3 M_1 M_2 M_{\text{tot}}F(e)}{5c^2}, 
\end{equation}
where $M_1$ and $M_2$ are the mass of the primary and secondary, $M_{\text{tot}}=M_1 + M_2$ is the total mass of the binary, and $F(e) = (1-e^2)^{-7/2}[1 - (73/24)e^2 - (37/96)e^4]$ describes the evolution of orbital eccentricity ($e$) during GW inspiral \citep{Peters1963}. We obtain the \citetalias{Sesana2015} transition separation $a_{\text{tr,SK15}}$  by setting $(da/dt)_{\star} = (da/dt)_{\text{GW}}$, which yields
\begin{equation}\label{eq:sesana_transition_sep}
    a_{\text{tr,SK15}} = \bigg[ \frac{64 G^2 \sigma_{\text{inf}} M_1 M_2 M_{\text{tot}} F(e)}{5c^2} \bigg]^{1/5}. 
\end{equation}
The timescale can then be approximated as:
\begin{equation}\label{eq:sesana_timescale}
\begin{split}
    t_{\text{SK15}} &= \frac{1}{S_{\text{SK15}}~a_{\text{tr,SK15}}}\\
    &= \frac{\sigma_{\text{inf}}}{GH \rho_{\text{inf}} a_{\text{tr,SK15}}}\\
    &\approx 15.8 \text{ Gyr} \Bigg( \frac{\sigma_{\text{inf}}}{\text{km s}^{-1}} \Bigg) \Bigg( \frac{\rho_{\text{inf}}}{M_{\odot}\text{pc}^{-3}} \Bigg)^{-1} \Bigg( \frac{a_{\text{tr,SK15}}}{10^{-3}\text{pc}} \Bigg)^{-1}.
\end{split}
\end{equation}

For \citetalias{Vasiliev2015}, the transition between stellar and GW hardening $a_{\text{tr,V15}}$ occurs at the separation 

\begin{equation}\label{eq:vasiliev_transition_sep}
    a_{\text{tr,V15}} = \Bigg( \frac{64G^3 M_1 M_2 M_{\text{tot}}}{5c^2F(e)} \frac{a_{\text{h}}^{\nu} \sigma_{\text{inf}}}{\mu S_{\text{V15}}} \Bigg) ^{1/(5+\nu)}
\end{equation}
The timescale, then, is provided by 
\begin{equation}\label{eq:vasiliev_timescale}
    \begin{split}
    t_{\text{V15}} & = \frac{1}{S_{\text{V15}}(a_{\text{tr,V15}})~a_{\text{tr,V15}}}\\
    & \approx 1.7 \times10^8 \text{yr} ~\Bigg( \frac{r_{\text{inf}}}{30\text{pc}} \Bigg)^{\frac{10+4\nu}{5+\nu}} \Bigg( \frac{M_{\text{tot}}}{10^8 M_{\odot}} \Bigg) ^{-\frac{5+3\nu}{5+\nu}}\\
    &\times \mu^{-\frac{4}{5+\nu}} \Bigg( \frac{4q}{(1+q)^2} \Bigg)^{\frac{3\nu-1}{5+\nu}} F(e)^{-\frac{1+\nu}{5+\nu}} ~20^{\nu}
    \end{split}
\end{equation}
where $S_{\text{V15}}$ is provided by Eq. \ref{eq:vasiliev2015}. We take the fiducial values of $\mu=0.3$ and $\nu=0.4$ as provided by \citetalias{Vasiliev2015}. For both timescales, we calculate $F(e)$ by assuming a distribution of eccentricities to match the expectation for high eccentricities that typically result from cosmological simulations \citep{Gualandris2022, Chen2022_orb, Francois2026}. We assume a skewed Gaussian distribution centered roughly at $e=0.8$ with a skew factor $a=-4$ and scale of $0.25$, and save a detailed study of eccentric MBHB orbits in \texttt{Romulus25} for a future work. 

To measure $\rho_{\text{inf}}$ for MBH merger hosts in \texttt{Romulus25}, we fit the slope of the stellar density profile using a power law index $\gamma$ such that $\rho \propto r^{-\gamma}$. We fit $\gamma$ between $0.7$ and $1.4$ kpc from the galaxy center in attempt to describe the inner stellar structure while avoiding the region where the dynamics are not well resolved ($<2\epsilon_{\text{soft}}$). To calculate $r_{\text{inf}}$, we use the relation from \citet{Sesana2010}:
\begin{equation}\label{eq:sesana2010_rinf}
    r_{\text{inf}} = (3-\gamma) \frac{GM_{\text{tot}}}{\sigma_{\text{inf}}^2},
\end{equation}
where we take $\sigma_{\text{inf}} \approx \sigma_{4\epsilon_{\text{soft}}}$. We choose $4\epsilon_{\text{soft}}=1.4$kpc as the radius of measurement as dynamical processes only just become resolved at $2\epsilon_{\text{soft}}=0.7$kpc, and $\sigma$ measures an enclosed quantity up to the radius of measurement. Note that $\sigma_{\text{inf}}$ here is the same as the velocity dispersion used in our $v/\sigma$ measurements (Sec. \ref{sec:methods_vsigma}).

\begin{figure}
    \centering
    \includegraphics[width=1.0\linewidth]{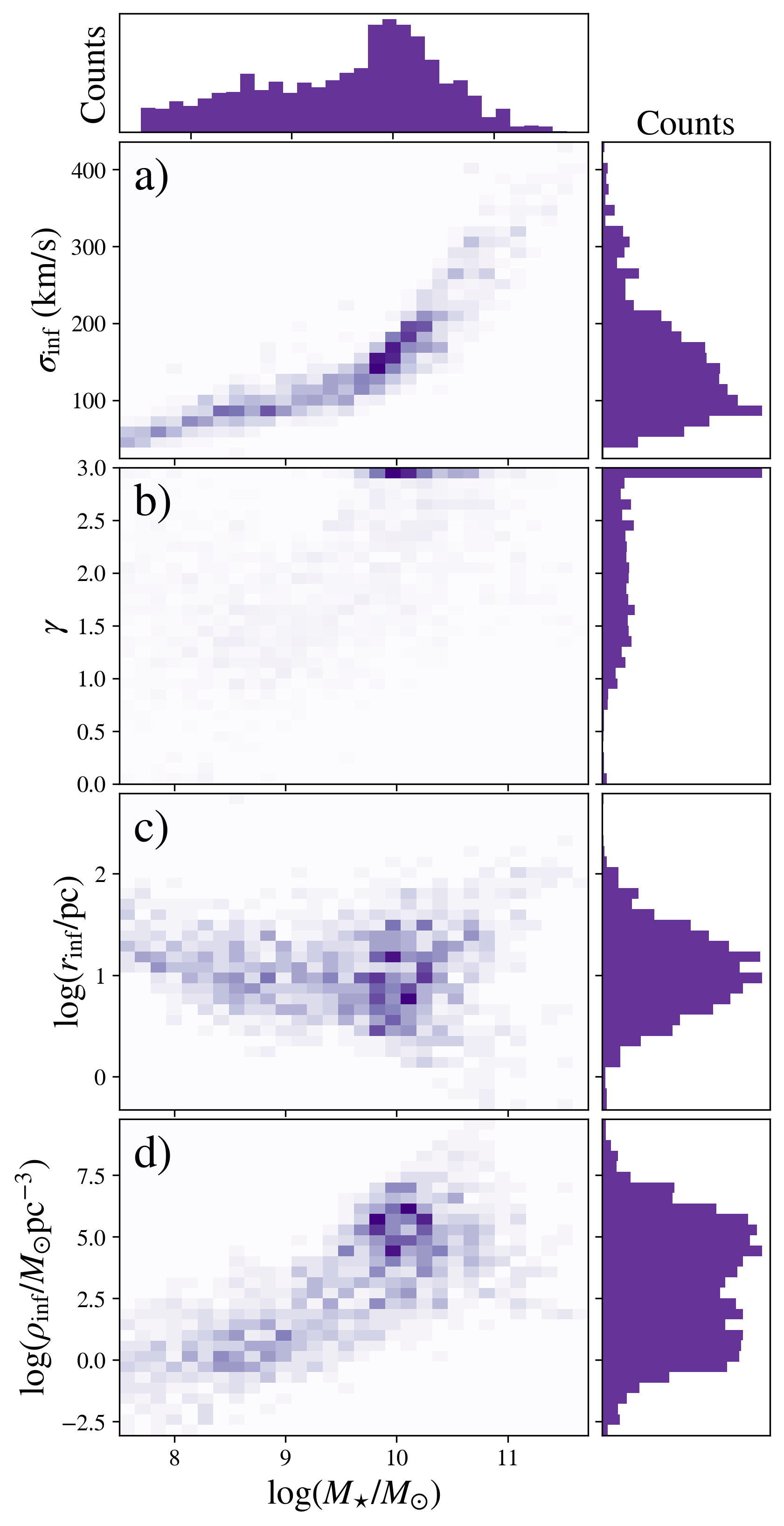}
    {\phantomsubcaption\label{fig:extrapolated_vdisp}%
     \phantomsubcaption\label{fig:extrapolated_gamma}%
     \phantomsubcaption\label{fig:extrapolated_rinf}%
     \phantomsubcaption\label{fig:extrapolated_rho}%
     }
    \caption{Extrapolated galaxy properties for use in the timescales by \citetalias{Sesana2015} and \citetalias{Vasiliev2015}  vs. host galaxy stellar mass. The topmost panel shows the one-dimensional distribution of $M_{\star}$, and each of the rightmost panels shows the one-dimensional distribution of each of the following properties: a) Approximated stellar velocity dispersion at the MBHB influence radius ($\sigma_{\text{inf}}$) where we take $\sigma_{\text{inf}}\approx \sigma_{4\epsilon_{\text{soft}}}$ vs. $M_{\star}$. b) Power-law index ($\gamma$) fit to the inner region (0.7-1.4kpc) of the stellar density profile vs. $M_{\star}$. c) Influence radius ($r_{\text{inf}}$) of the binary defined as the radius enclosing twice the binary's mass in stars, calculated using Eq. \ref{eq:sesana2010_rinf}. d) Extrapolated stellar density at the MBHB influence radius ($\rho_{\text{inf}}$) vs. $M_{\star}$.}
    \label{fig:extrapolated_properties}
\end{figure}

We show the distribution of $\sigma_{\text{inf}}$, $\gamma$, $\rho_{\text{inf}}$, and $r_{\text{inf}}$ against the host galaxy stellar mass in Fig. \ref{fig:extrapolated_properties}, excluding galaxies where the fit resulted in positive slopes ($\gamma<0$) and capping our steepest slopes at $\gamma \leq 3$. Fig. \ref{fig:extrapolated_vdisp} shows the relatively tight relation between stellar mass and stellar velocity dispersion, as expected, with increased dispersion in large galaxies. We find a median dispersion value in our sample of $\sigma_{\text{inf}}\sim130~\text{km s}^{-1}$. Our sample contains a median $\gamma$ of $\sim2.2$, corresponding to a steep cusp. Fig. \ref{fig:extrapolated_gamma} shows that a sizeable portion ($\sim28\%$) of the sample was capped at $\gamma=3$; we choose to retain these cases rather than remove them as to not bias the entire sample toward smaller galaxy masses ($M_{\star} \lesssim 10^{9.5} M_{\odot}$). The rest of the sample shows an increase in cuspiness with higher $M_{\star}$, albeit with a large scatter. Fig. \ref{fig:extrapolated_rinf} shows the influence radii for the merger hosts calculated using Eq. \ref{eq:sesana2010_rinf}, ranging between 1 and 100 pc, with the capped $\gamma=3$ population contributing to the tail toward smaller $r_{\text{inf}}$ values at high $M_{\star}$.

Lastly, Fig. \ref{fig:extrapolated_rho} shows the extrapolated stellar densities at the influence radius, featuring two peaks in the distribution near $\rho_{\text{inf}}\sim100 M_{\odot}\text{pc}^{-3}$ and $\rho_{\text{inf}}\sim10^5 M_{\odot}\text{pc}^{-3}$. Our largest $\rho_{\text{inf}}$ values reach to $\sim10^9 M_{\odot}\text{pc}^{-3}$, a similar maximum as found in other studies \citep[e.g.][]{Volonteri2020}, though these are unphysically dense; NSCs, the densest stellar systems in the universe, can reach densities up to $\sim10^{5-6} M_{\odot}\text{pc}^{-3}$ at their centers \citep{Pechetti2020, Hannah2024}. Since we take no steps to implement NSC cusps into the measured density profiles of our nucleated galaxies, we determine these density overestimates to be caused by poor stellar resolutions in the inner region of the galaxy, a limitation of all uniform-volume cosmological simulations. This implies the downstream calculations of $T_{\text{coal}}$ that rely on $\rho_{\text{inf}}$ and $r_{\text{inf}}$ will be underestimates of the true coalescence time. This $\rho_{\text{inf}}$ measurement is also sensitive to the range over which the slope is fit; in a similar analysis of MBH merger hosts in the \texttt{ASTRID} simulation, \citet{Chen2022_orb} finds that measuring the slope out to $r=1.5\epsilon_{\text{soft}}$ versus $r=2.0\epsilon_{\text{soft}}$ manifests in an order of magnitude difference in $\rho_{\text{inf}}$ (though this effect would be less extreme in \texttt{Romulus25}, as the volume and resolution of \texttt{ASTRID} are much larger). 

\section{Results \& Discussion}\label{sec:results_and_discussion}

In this section, we present our distribution of coalescence times compared to those produced by the \citetalias{Sesana2015} and \citetalias{Vasiliev2015} timescales, analyzed by host galaxy stellar mass and MBHB chirp mass (Sec. \ref{subsec:tcoal_dists}). In Sec. \ref{subsec:tcoal_dists_gal_orb}, we analyze the influence of NSCs and co/counterotation on the galaxy-informed timescale distribution. We also present the impact of the new coalescence times on the MBH merger rate and LISA event rate (Sec. \ref{subsec:merger_rate}), comparing with the merger rates produced by our comparison timescales, and situate these within the large range of existing merger rates predicted by cosmological simulations. Lastly, we discuss implications regarding host galaxy identification in the multi-messenger era under the assumption of short stellar scattering timescales (Sec. \ref{subsec:merger_signatures}).

\subsection{Coalescence time distributions and comparison to commonly used timescales}\label{subsec:tcoal_dists}

\begin{figure}
    \centering
    \includegraphics[width=1.0\linewidth]{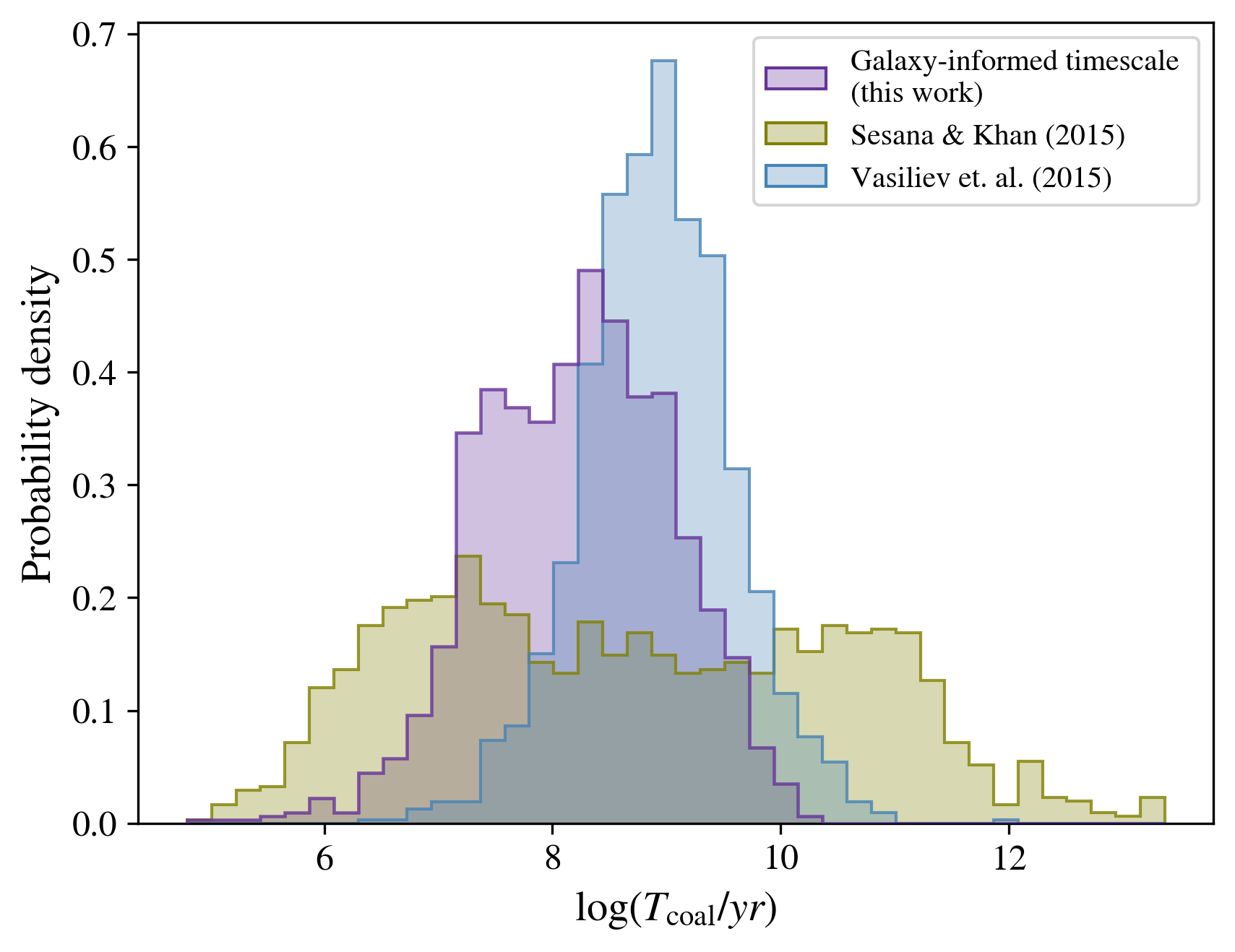}
    \caption{Distribution of coalescence timescales for the galaxy-informed timescale by \citetalias{KHB2025} (purple) compared to that of \citetalias{Sesana2015} (green) and \citetalias{Vasiliev2015} (blue). The median coalescence times for the three timescales, respectively, are $10^{8.2}$ yr, $10^{8.6}$ yr, and $10^{8.9}$ yr, respectively. When excluding hosts where the fit resulted in unphysically dense central stellar densities ($\rho_{\text{inf}}>10^6M_{\odot}$), the \citetalias{Sesana2015} median becomes $\sim10^{8.9}$yr, and the \citetalias{Vasiliev2015} becomes $\sim10^{9.0}$yr.}
    \label{fig:tcoal_dist}
\end{figure}

\begin{figure*}
    \centering
    \includegraphics[width=1.0\linewidth]{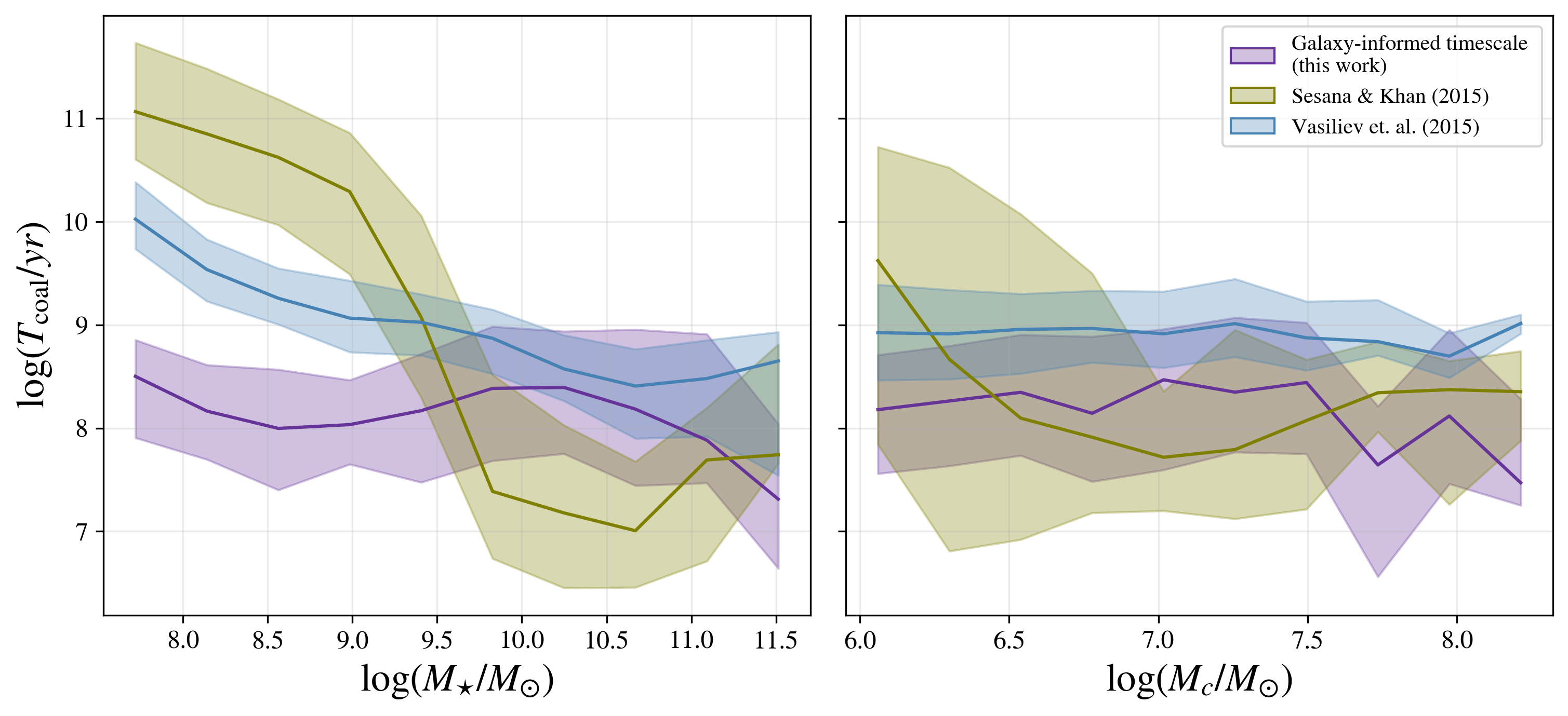}
    \caption{The median (solid line) and 25th and 75th percentile spread (shaded region) in the coalescence time for the galaxy-informed timescale (\citetalias{KHB2025}, purple) and the timescales by \citetalias{Sesana2015} (green) and \citetalias{Vasiliev2015} (blue) binned by remnant galaxy stellar mass (left panel) and MBHB chirp mass (right panel).}
    \label{fig:tcoal_mstar_mchirp}
\end{figure*}

The probability distribution of coalescence timescales for a population of realistic galaxy hosts is shown in Fig. \ref{fig:tcoal_dist}. For the galaxy-informed time delay by \citetalias{KHB2025} that we implemented in this work, the distribution is centered at $\sim 10^{8.2}$ yr and spans $\sim 10^{4.8}-10^{10.2}$ yr, the smallest range of the three timescales ($\sim5.4$ dex). The shape is roughly Gaussian, with a slight indication of a double-peak that we identify in the next section as being related to the NSC assignment.

To compare to other commonly implemented stellar hardening timescales, we also present the distributions of the \citetalias{Sesana2015} and \citetalias{Vasiliev2015} timescales as applied to the \texttt{Romulus25} merger hosts. The median of the \citetalias{Sesana2015} distribution is $\sim 10^{8.6}$yr, with the largest range of the three timescales ($\sim8.4$ dex), including many cases that extend longer than the age of the Universe. The \citetalias{Sesana2015} distribution features a distinct double-peak that directly follows that of the double-peak in $\rho_{\text{inf}}$ seen in Fig. \ref{fig:extrapolated_rho}; the coalescence times by \citetalias{Sesana2015} that fall below $\lesssim10^{8.5}$yr correspond to MBH merger hosts with central densities $\rho_{\text{inf}} \gtrsim10^3 M_{\odot}\text{pc}^{-3}$. Excluding unphysically dense systems ($\rho_{\text{inf}}>10^6M_{\odot}$) results in a \citetalias{Sesana2015} median of $\sim10^{8.9}$yr. The \citetalias{Vasiliev2015} timescale yields a median of $\sim 10^{8.9}$yr and ranges over $\sim5.6$ dex. This distribution does not feature the double-peaked shape, as it relies on $\rho_{\text{inf}}$ only implicitly. When excluding unphysically large densities, the median becomes $\sim10^9$yr. The ranges found here are comparable to those found by other cosmological simulations that have implemented the \citetalias{Sesana2015} \citep[e.g.][]{Volonteri2020} and \citetalias{Vasiliev2015} \citep[e.g.][]{Chen2022_orb} timescales. 

To further quantify the coalescence time distributions for the galaxy-informed timescale and the two comparison timescales, we show the median and 25/75th percentiles in the coalescence times vs. host galaxy stellar mass and MBHB chirp mass ($M_{c} \equiv (M_1 M_2)^{3/5} / (M_1 + M_2)^{1/5}$) in Fig. \ref{fig:tcoal_mstar_mchirp}. The galaxy-informed timescale is the quickest of the three across almost the whole $M_{\star}$ and $M_c$ ranges, only surpassed by \citetalias{Sesana2015} in the range where extrapolated densities are highest ($M_{\star}\sim10^{10} M_{\odot}, M_c\sim10^7 M_{\odot}$). The galaxy-informed timescale is shortest in the highest mass bins ($M_{\star} \gtrsim10^{11} M_{\odot}$) as these correspond to systems with the highest $R_{\text{eff}}$ and $v/\sigma$, and $M_{\text{tot}}$ values, even in the non-NSC cases. The \citetalias{Vasiliev2015} is the most constrained across $M_{\star}$ and $M_c$, likely because the timescale relies on the power law index $\gamma$ and subsequently-calculated $r_{\text{inf}}$ (see Eq. \ref{eq:sesana2010_rinf}), which span a smaller range than the $\rho_{\text{inf}}$ values they produce. The \citetalias{Sesana2015} is the most dependent on $\rho_{\text{inf}}$ and thus most greatly feels the effect of poor stellar resolution. This can be seen in low $M_{\star}$ galaxies where the central stellar densities are lowest and likely only represented by tens to hundreds of star particles. Fig. \ref{fig:tcoal_mstar_mchirp} shows how the \citetalias{Sesana2015} predicts that mergers in these low $M_{\star}$ galaxies would be exceedingly rare, with the median $T_{\text{coal}}$ taking much longer than the age of the Universe. Since we do expect MBH mergers to occur in low-mass galaxies \citep[e.g.][]{Volonteri2020, DeCun2023}, especially at high redshift where low-mass mergers are assembling the earliest massive galaxies \citep{Bhowmick2025}, it is possible that density-dependent models that work well in isolated galaxy environments with appropriate stellar resolution may unfairly penalize low stellar mass galaxies when applied to large volume cosmological simulations. This, in turn, may penalize mergers of low $M_c$ (here, $ \lesssim 10^7 M_{\odot}$), as these are hosted in low $M_{\star}$ galaxies, which sit at the very center of the expected mass range detectable by LISA \citep{AmaroSeoane2023, Colpi2024}. 

The median $T_{\text{coal}}$ produced here by the galaxy-informed timescale is approximately an order of magnitude faster than those produced by the comparison timescales, which fall closer to $\sim1$ Gyr when considering physically realistic systems. As it is the general understanding that stellar scattering depends nonlinearly on many different galaxy properties, some common approaches in studies of MBH evolution (of which detailed stellar hardening models fall outside of the scope) is to apply a $\sim 1$ Gyr time delay across the board \citep[e.g.][]{Bellovary2025}, or to predict that stellar scattering times are always negligible compared to dynamical friction times \citep[e.g.][]{IzquierdoVillalba2026_MBHCat}. Our work here supports the latter more than the former, though even the galaxy-informed timescale produces instances where scattering takes $\sim1$ Gyr or longer. We emphasize the relative ease of implementation of the timescale may be a selling point for these cases where a deep dynamical understanding of any one given system can be foregone for scattering times that globally reflect the influence of the galaxy hosts on the MBHB population.

\begin{figure}
    \centering
    \includegraphics[width=1.0\linewidth]{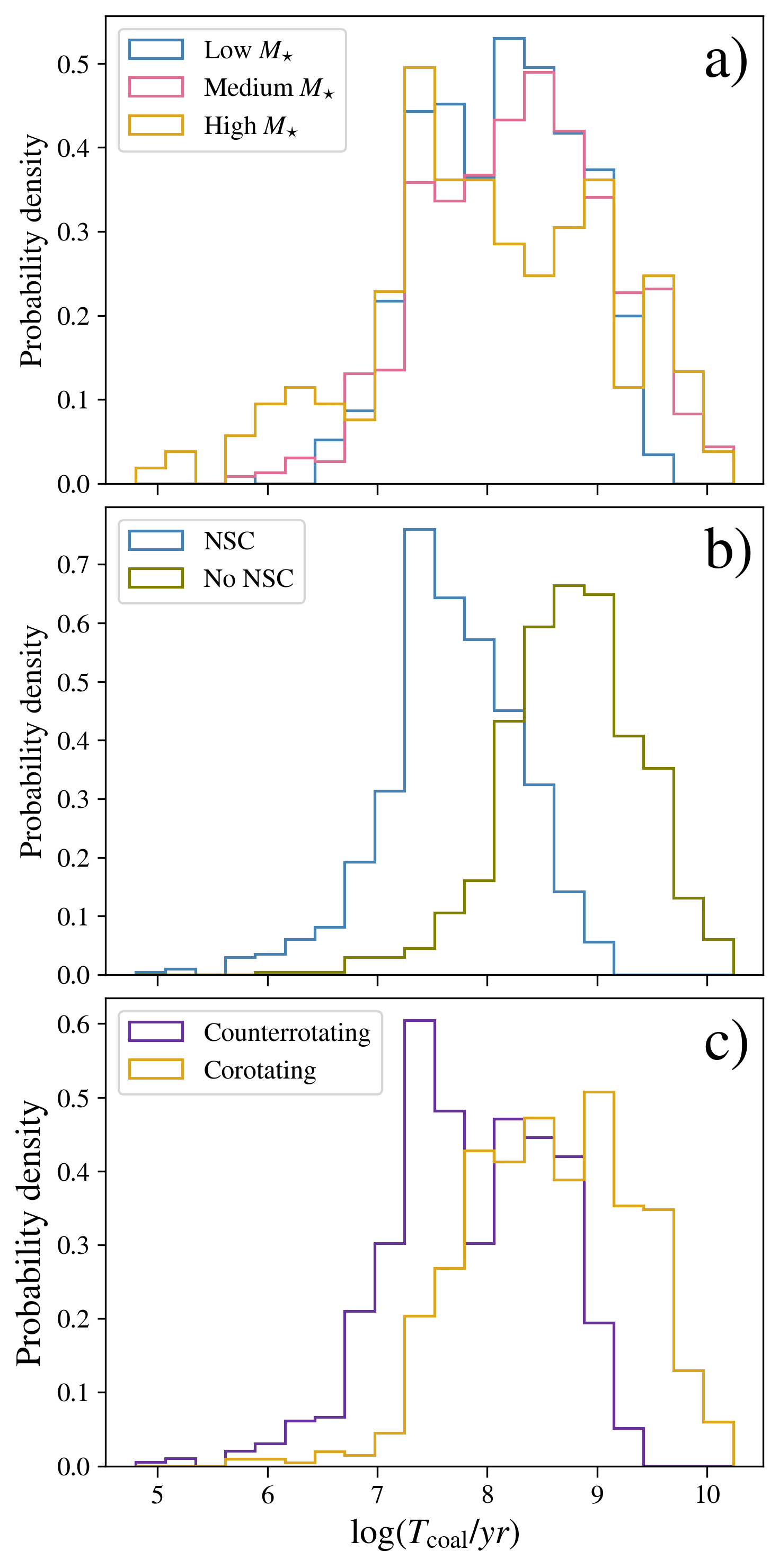}{\phantomsubcaption\label{fig:tcoal_binned_by3_Mstar}%
     \phantomsubcaption\label{fig:tcoal_binned_by3_NSC}%
     \phantomsubcaption\label{fig:tcoal_binned_by3_rot}%
     }
    \caption{Distribution of the galaxy-informed coalescence times binned by a) stellar mass of the remnant galaxy, divided into low mass ($M_{\star} \in 10^{[7.5-9]}M_{\odot}$), medium mass ($M_{\star} \in 10^{(9-10.5]} M_{\odot}$), and high mass ($M_{\star} > 10^{10.5} M_{\odot}$) galaxies, b) NSC assignment and c) corotation or counterrotation of the MBHB with respect to the host.}
    \label{fig:tcoal_binned_by3}
\end{figure}

\subsection{Coalescence times vs. galactic and orbital properties}\label{subsec:tcoal_dists_gal_orb}

We are also interested in the individual effect of certain galaxy properties on $T_{\text{coal}}$ for the galaxy-informed timescale. In Fig. \ref{fig:tcoal_binned_by3}, we show the galaxy-informed coalescence times binned by a) host galaxy stellar mass, b) whether or not the galaxy was assigned a NSC, and c) the nature of the MBHB rotation with respect to the stellar angular momentum of the host.
Fig. \ref{fig:tcoal_binned_by3_Mstar} divides the hosts into three stellar mass bins: low mass ($M_{\star} \in 10^{[7.5-9]}M_{\odot}$), medium mass ($M_{\star} \in 10^{(9-10.5]} M_{\odot}$), and high mass ($M_{\star} > 10^{10.5} M_{\odot}$). We see that the high $M_{\star}$ hosts tend toward having shorter coalescence times, with a median $T_{\text{coal}}\sim10^{7.9}$yr compared to the medium $M_{\star}$ ($\sim10^{8.3}$yr) and low $M_{\star}$ ($\sim10^{8.2}$yr) cases.

Fig. \ref{fig:tcoal_binned_by3_NSC} displays the effect of the NSC term on the coalescence times. The MBHBs in hosts assigned a NSC $\sim13$ times more quickly than those without; the NSC case has a median $T_{\text{coal}}$ of $\sim 10^{7.6}$yr, and the no NSC case has a median of $\sim 10^{8.8}$yr. We know from isolated merger models \citep{Ogiya2020, Khan2021} and zoom-in simulations of galactic centers \citep{Mukherjee2025} that NSCs are extremely efficient in accelerating MBHBs through the stellar scattering phase due to high stellar densities and tidal torques. Our result here serves to demonstrate that even a simple NSC implementation to account for this effect makes an appreciable difference to global stellar scattering times due to the high likelihood of NSCs in the galaxy stellar mass range that MBHs tend to occupy. For the hosts not assigned a NSC, the median $T_{\text{coal}}\sim 10^{8.8}$yr is similar to that produced by the comparison timescales; even when excluding unphysically dense systems ($\rho_{\text{inf}}>10^6 M_{\odot}\text{pc}^{-3}$), the medians come to within a factor of 2, indicating the NSC term to be one of the biggest differentiators of the galaxy-informed timescale. We remind the reader that these NSCs are not physically seeded and evolved, as NSCs are not resolved in \texttt{Romulus25}. We discuss the caveats of our NSC prescription in Sec. \ref{subsec:caveats_NSCs}.

Finally, we display the coalescence times divided into the corotating ($0^{\circ} \leq \Theta \leq 90^{\circ}$) and counterrotating ($90^{\circ} < \Theta \leq 180^{\circ}$) cases in Fig. \ref{fig:tcoal_binned_by3_rot}. The counterrotating cases have a median $T_{\text{coal}}$ of $\sim10^{7.8}$yr, $\sim6$ times faster than the corotating cases, with a median $T_{\text{coal}}$ of $\sim10^{8.6}$yr. The physical interpretation is that MBHBs that are counterrotating at the point of numerical merger in \texttt{Romulus25} would go on to be pumped to high eccentricities by interaction with the retrograde stellar orbits, which extracts angular momentum efficiently \citep{Sesana2011, KHB2015, Mirza2017}. The counterrotating sample shown in Fig. \ref{fig:tcoal_binned_by3_rot} also displays a double peak caused by the subsample that are assigned NSCs moving towards shorter $T_{\text{coal}}$. Naturally, then, the fastest cases in the entire sample correspond to MBHBs whose hosts are assigned NSCs and which are counterrotating. 

\begin{figure}
    \centering
    \includegraphics[width=1.0\linewidth]{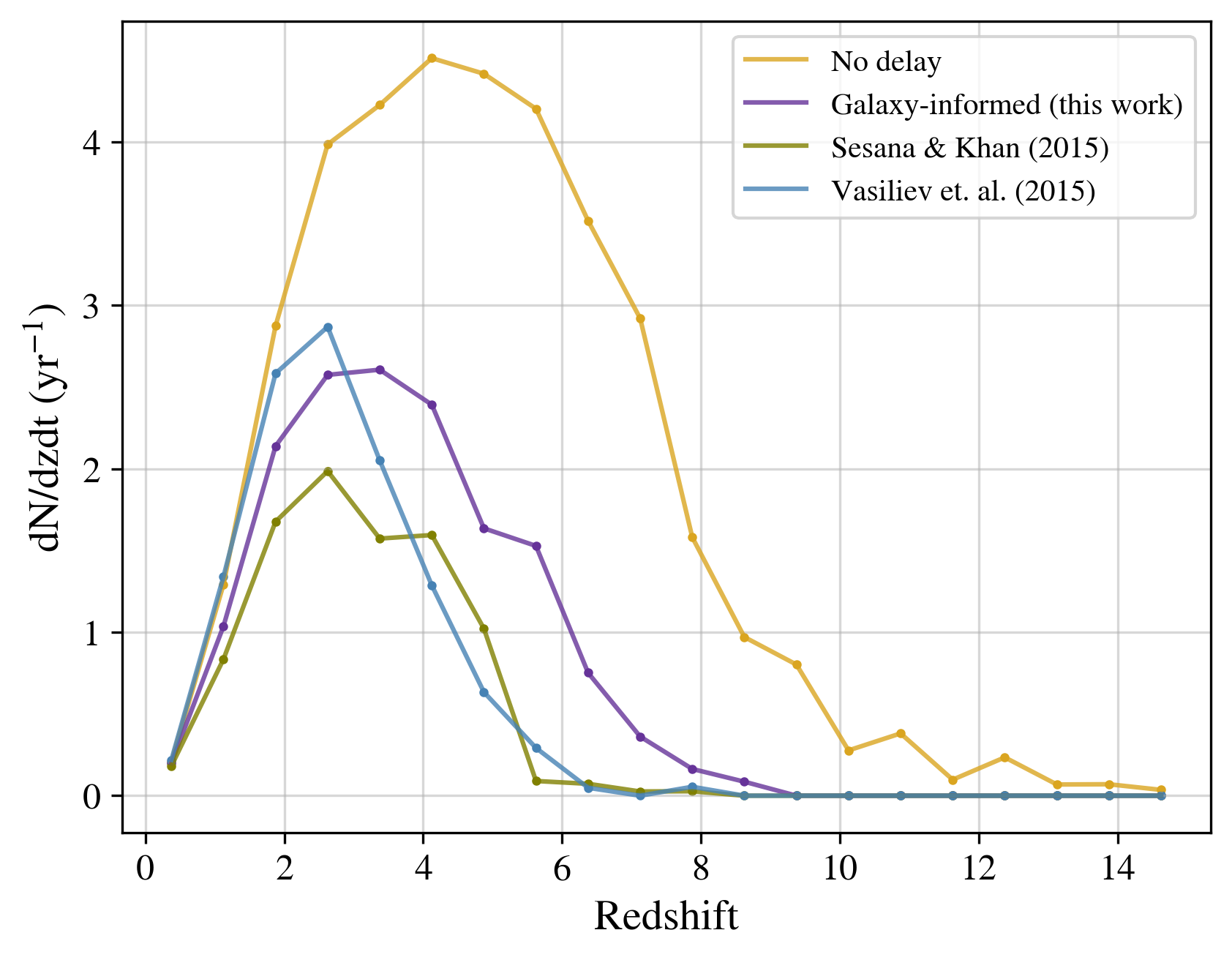}
    \caption{Merger rate of MBHs in \texttt{Romulus25} with no time delay (orange), the galaxy-informed delay (purple), the delay by \citetalias{Sesana2015} (green), and the delay by \citetalias{Vasiliev2015} (blue), binned into 20 redshift bins between $z=0$ and $z=15$.}
    \label{fig:merger_rate}
\end{figure}

\subsection{Merger rate and expected LISA detections}\label{subsec:merger_rate}

Since we attribute $T_{\text{coal}}$ as the time spent in the late dynamical friction phase and the stellar scattering phase, which occur at approximately the MBHB separations where \texttt{Romulus25} binaries undergo numerical merger (see Fig. \ref{fig:sep_and_offsets}), we added the respective coalescence times to the numerical merging times in the simulation. We can calculate the astrophysical MBH merger rate as in \citet{Chen2022_orb}: 

\begin{equation}\label{eq:merger_rate}
    \frac{\text{d}N}{\text{d}z\text{d}t} = \frac{\text{d}^2n(z)}{\text{d}z\text{d}V_{c}} \frac{\text{d}z}{\text{d}t} \frac{\text{d}V_{c}}{\text{d}z} \frac{1}{1+z}
\end{equation}
where $V_{c}$ is the comoving volume of the \texttt{Romulus25} box at the time of merger, $n(z)$ is number of MBH mergers occuring in a given redshift bin, and $1/(1+z)$ transforms the time element into the observer frame. We take the finite-interval approximation that

\begin{equation}\label{eq:merger_rate_finiteapprox}
    \frac{\text{d}^2n(z)}{\text{d}z\text{d}V_{c}} = \frac{N(z)}{\Delta z V_{\text{sim}}}
\end{equation}
where $N(z)$ is the number of mergers occuring in a given redshift bin with width $\Delta z$, and $V_{\text{sim}} = (25 \text{cMpc})^3$ is the full \texttt{Romulus25} volume in comoving units. 

In Fig. \ref{fig:merger_rate}, we show the merger rates for undelayed mergers and the mergers delayed by the the galaxy-informed timescale and the timescales by \citetalias{Sesana2015} and \citetalias{Vasiliev2015}. We determine the redshift at which the merger rate peaks for each timescale ($z_{\text{peak}}$), finding $z_{\text{peak}}\sim 4.1$ for the no delay case, $z_{\text{peak}} \sim 3.3$ for the galaxy-informed timescale, and $z_{\text{peak}} \sim 2.6$ for both the \citetalias{Sesana2015} and \citetalias{Vasiliev2015} timescales. Generally, longer coalescence times should correspond to a $z_{\text{peak}}$ closer to zero; however, the vast range of $T_{\text{coal}}$ resulting from the \citetalias{Sesana2015} timescale, with many taking longer than the age of the Universe, reduces the number of mergers drastically. This limits the effective distribution of $T_{\text{coal}}$ to those $<10^{10}$ yr, under which the median coalescence time by \citetalias{Sesana2015} becomes quicker ($\sim 10^{7.6}$ yr) than that of \citetalias{Vasiliev2015} ($\sim 10^{8.9}$ yr).

To calculate the number of expected mergers ($N_{\text{exp}}$) detectable by LISA, we integrate over all redshifts and assume a 4-year LISA mission lifetime ($T_{\text{obs}} = 4$yr), which is the low-end estimate of LISA's capabilities \citep{AmaroSeoane2023, Colpi2024}. We find $N_{\text{exp}}\sim 109$ for the undelayed mergers, $N_{\text{exp}}\sim 46$ for the galaxy-informed timescale, $N_{\text{exp}}\sim 22$ for the \citetalias{Sesana2015} timescale, and $N_{\text{exp}}\sim 36$ for the \citetalias{Vasiliev2015} timescale. The galaxy-informed timescale provides the most optimistic estimate for such LISA detections of the three timescales, though even this should still be considered an underestimate; MBH mergers with total masses as low as $\sim 10^5 M_{\odot}$ will be observable by LISA \citep{Colpi2024}. These low-mass MBHs are expected to increase the merger rate, but fall below the resolution that is seeded in the \texttt{Romulus25} simulation. Still, our event rate predictions fit squarely within the larger landscape of LISA event rate predictions, which span from $\lesssim 1$ up to thousands \citep{IzquierdoVillalba2026_MBHCat}. 

\begin{figure*}
    \centering
    \includegraphics[width=1.0\linewidth]{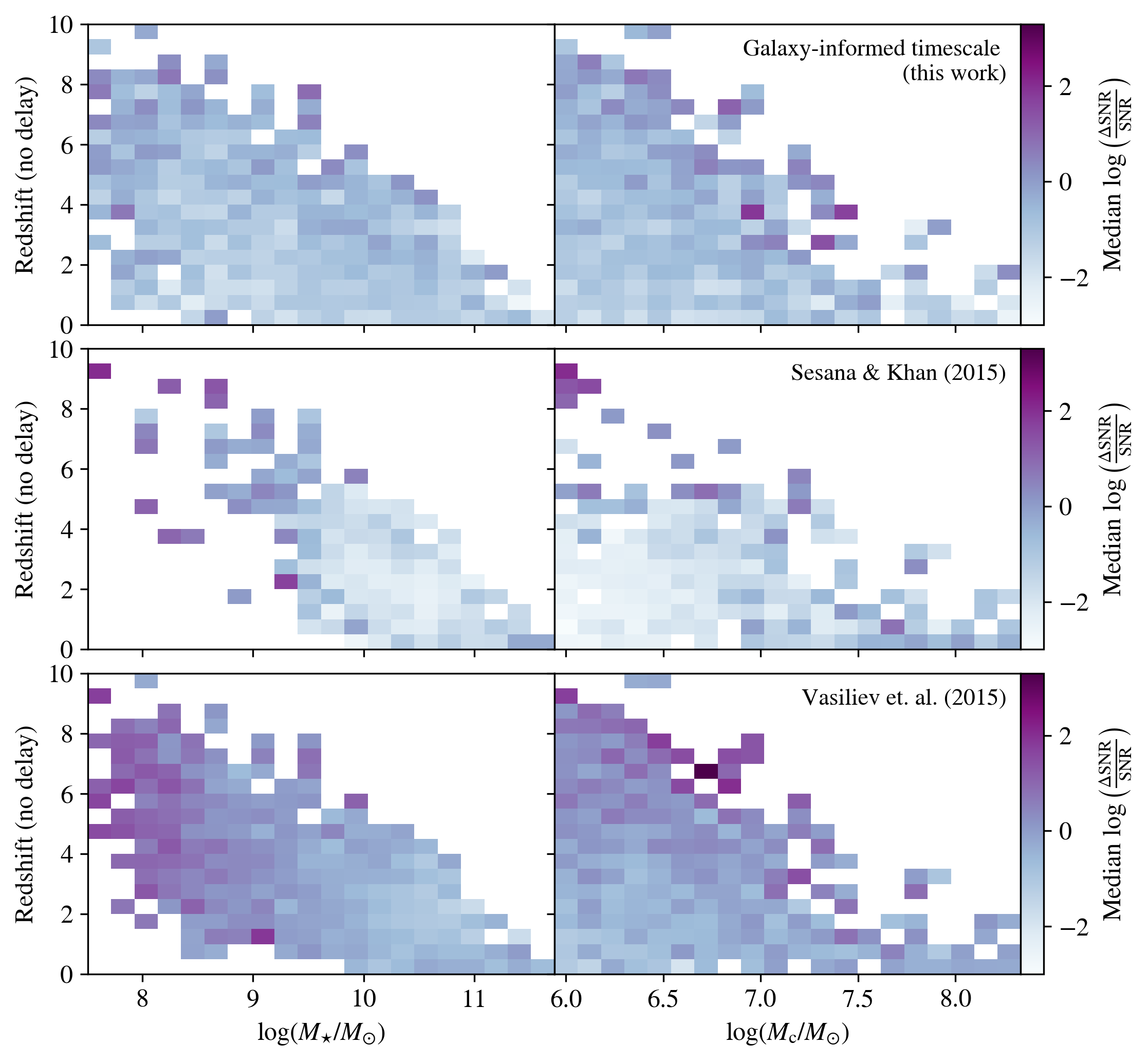}
    \caption{Median fractional change in signal-to-noise ratio (SNR) across redshift of numerical merger, host stellar mass ($M_{\star}$, left column), and binary chirp mass ($M_c$, right column) for the three timescales. Top: our galaxy-informed coalescence timescale (Eq. \ref{eq:timescale_4term}, \citetalias{KHB2025}). Middle: timescale set by the~\citet{Sesana2015} formalism. Bottom: timescale prescribed by~\citet{Vasiliev2015}. The color bar indicates the log of the fractional change in SNR (assuming a 4-year LISA mission) compared to the no-delay case. The \citetalias{Sesana2015} and \citetalias{Vasiliev2015} timescales induce high fractional SNR changes in the high redshift and low stellar mass hosts, with the typical merger being $\sim10-100$ times louder than the no-delay SNR in those regimes. The galaxy-informed timescale induces SNRs that are $\sim1.1-2$ times louder than the no-delay SNR, implying the use of the numerical merger population may be sufficient for basic SNR calculations (providing dynamical friction delays have been accounted for). The steeper color gradient in the middle panel is reflective of the much longer delay times experienced in lower-mass galaxies, and hints that some MBH mergers have been pushed to the future, as shown in Figure~\ref{fig:merger_rate}.}
    \label{fig:frac_snr}
\end{figure*}

For each merger, we calculate the sky-averaged signal-to-noise ratio (SNR) for a four-year mission using \texttt{lisacode} \citep{Petiteau2008}. We display the median fractional change in SNR from the no-delay SNR ($\Delta \text{SNR}/\text{SNR}$) in the $z_{\text{merge}}-M_{\star}$ plane (left column) and $z_{\text{merge}}-M_{c}$ plane (right column) for each timescale in Fig. \ref{fig:frac_snr}, where $z_{\text{merge}}$ is the redshift at the time of numerical merger in the simulations. Longer coalescence times push mergers closer (or even beyond) $z=0$, resulting in a high fractional change in SNR from the no-delay case. This is visible in the middle panel, where the \citetalias{Sesana2015} timescale displays high fractional SNR increases at low $M_{\star}$ and low $M_c$, and cases delayed beyond $z=0$ are not shown. The bottom panel displays the \citetalias{Vasiliev2015} timescale, where fewer mergers are delayed beyond $z=0$, but long $T_{\text{coal}}$ at high redshift and in the low stellar mass range (see Fig. \ref{fig:tcoal_mstar_mchirp}) produce high fractional SNR changes. The fractional SNR increases for the galaxy-informed timescale, shown in the top panel, are the least dependent on $M_{\star}$ and $M_c$ of the three timescales, and generally small, with most mergers only $\sim1.1-2$ times louder than the no-delay SNR. All three timescales result in distributions with median SNRs near $\sim 10^3$, extending as far as SNR $\sim 10^5$; the small fractional SNR changes found by the galaxy-informed timescale then imply the use of the numerical merger population in a given simulation, provided that delays due to dynamical friction have been accounted for, may be sufficient to achieve SNR distributions for basic uses.

\subsection{Implications for host galaxy identification}\label{subsec:merger_signatures}

The relatively short stellar scattering times we find, if combined with conditions that foster fast MBH pairing and sinking, may be beneficial in the attempt to identify the host galaxy of merging MBHs. It is the current understanding that a substantial time delay between the galaxy merger and subsequent MBH merger implies the loss of any merger-induced morphological signatures due to relaxation \citep{Volonteri2020, DeGraf2021, Bardati2024_morphology}. In a study using the IllustrisTNG simulations, \citet{DeGraf2021} finds that morphological disturbances in galaxies hosting high mass MBH mergers can remain apparent on the order of $\sim 500$ Myr. Estimates of the dynamical friction timescale for MBH hosts vary greatly, from extremely efficient mergers (often due to NSCs) on the order $\sim100$ Myr \citep[e.g.][]{Ogiya2020, Mukherjee2025} to the main bottleneck at $\sim$ hundreds of Gyr \citep[e.g.][]{Volonteri2020, Ma2021, Bortolas2022}.

In \texttt{Romulus25}, most MBHs are able to pair on the order of $\sim$ Gyr \citep{Tremmel2018_pairs}. This implies that even the very quick stellar scattering timescales found here would not necessarily justify searching for tidal signatures from a galaxy merger. In a study of MBH merger host morphology within \texttt{Romulus25}, \citet{Bardati2024_morphology} found that classical bulges may be more useful signposts for MBH merger hosts than tidal features like rings, tails, and streams \citep[see however][]{IzquierdoVillalba2023_properties}. In any case, when using cosmological simulations to compare MBH merger hosts to the general galaxy population, it is important to not overestimate the delay time between galaxy and MBH merger, as bulk galaxy properties such as color, star formation rate, and morphology may be artificially erased.  

Rather than looking for morphologies that tend to host MBH mergers in our simulations, we may wish to prioritize our searches to include galaxy features we know to assist MBH mergers. For example, in the local universe, we may wish to rank a nucleated galaxy higher in the ranking scheme. The electromagnetic signatures of counterrotating MBHBs are less clear, though orbital plane orientation is likely related to the orientations of the accretion disk and radio jets \citep{Gualandris2012}. Our findings that most merger hosts in \texttt{Romulus25} exhibit some degree of ordered rotation suggests that kinematic information may be relevant to the search for LISA host galaxies \citep[see also][]{Bardati2024_kinematics, Horlaville2026}. Finally, efficient stellar scattering may provide reason to search for stellar cores in galaxy centers; this is an active area of research \citep[e.g.][]{Dosopoulou2021, Harris2024} with intriguing observational prospects \citep[e.g.][]{McDonald2026}. Our understanding of MBH host demographics is still evolving, as are our strategies for electromagnetic followup and archival searches for merger hosts of GW sources across the frequency spectrum \citep{Petrov2024, Drake2025}. 

\section{Caveats \& Limitations}\label{sec:caveats}
In Sec. \ref{sec:results_and_discussion}, we found that the galaxy-informed merger timescale allows for relatively quick stellar scattering times for the \texttt{Romulus25} galaxies compared to other estimates. In this section, we discuss the limitations of our method. 

\subsection{Nuclear star cluster treatment}\label{subsec:caveats_NSCs}

Our current model for the NSC assignment is crude in that we take the stellar mass at the timestep most closely following the MBH merger and compare that to the local universe occupation fraction by \citet{Neumayer2020}, which almost surely does not hold at higher redshift where most of the our MBH mergers occur before applying delays (see the no delay case of Fig. \ref{fig:merger_rate}). The fact that we predict short delay times makes it that much more difficult to constrain the influence of NSCs at the peak in the MBH merger rate where NSC observations become sparse. Our method of obtaining masses for these NSC assigments is also flawed in that the $M_{\star}-M_{\text{NSC}}$ scaling relations differ among galaxy type and almost certainly change with redshift as well \citep{Neumayer2020, Hannah2024}.

Our model also assigns a NSC for every merger instance, regardless of any previous MBH merger history in that galaxy. It would be preferable to evolve the mass of a NSC with its host throughout cosmic time semi-analytically \citep[e.g.][]{Antonini2015_imprint, Antonini2015_coevo, Kritos2025}.  It may be the case that NSCs are entirely disrupted by extremely efficient stellar scattering; \citet{Khan2021} find stellar mass deficits ($M_{\text{def}}$) up to $4-5$ times the total mass of the MBHB, which is much larger than the more canonical picture which have $M_{\text{def}} \approx 0.5 M_{\text{tot}}$ \citep{Merritt2006}. In future studies, we will track the evolution of NSCs under the influence of MBH mergers and investigate the impact of high $M_{\text{def}}$ on global MBH merger timescales.

It is also important to acknowledge that the terms of Eq. \ref{eq:timescale_4term} are by no means independent, and that omitting the NSC term without also modifying the other features is a slight statistical faux-pas. The presence of a NSC may also provide physical reason to modify certain unresolved processes in the simulation; for example, the nature of NSC rotation not only may signal something about its formation channel \citep{Tsatsi2017, Pinna2021}, but is also likely affected by the presence of the MBHB \citep{Mastrobuono2023}. 

\subsection{Kinematics and orbital setup}\label{sec:caveats_rotation}
In Fig. \ref{fig:ingredients_cosTheta}, we presented a measured distribution of MBHB orbital plane inclinations with respect to the stellar angular momentum of each MBH merger host at the moment of numerical merger, and found the distribution to be roughly uniform. We found that this was also the case when measuring $\vec{l}_{\text{orb}}$ when the MBHB center-of-mass came within $2\epsilon_{\text{soft}}$ of the galaxy center, implying that the result is not sensitive to the radius of measurement within the central few kpc. It is not necessarily true that the inclinations would be isotropic on larger scales; satellite infalls towards larger galaxies are expected to be mainly radial, as they may be infalling along larger dark matter filaments \citep{Benson2005, Wetzel2011, Li2020} and would then have preferential infall angles. For an appropriate comparison in \texttt{Romulus25}, we would need to ensure that each secondary MBH is associated with a subhalo; this isn't necessarily true for every MBH merger in our sample (see the example galaxy in Fig. \ref{fig:orbits}) as BHs are seeded according to properties of local gas particles rather than placed at the centers of halos/subhalos.

The sense of rotation as described by $\cos \Theta$ found itself in the galaxy-informed timescale because it helped to maximize the $R^2$ of the coalescence time fit, even surpassing the initial ($e_i$) and final eccentricities ($e_f$) as potential terms. In the simulations by \citetalias{KHB2025}, $\cos \Theta$ and $e_f$ were highly correlated, as counterrotating cases quickly had their eccentricities pumped to high values, a phenomenon observed in other numerical setups thought to caused by many cumulative retrograde stellar encounters which bleed angular momentum from the binary \citep{Iwasawa2011, Mirza2017}. Because $\cos \Theta$ described this phenomenon well for the galaxy-informed timescale, we chose not to perform a detailed study of eccentricities in the \texttt{Romulus25} situation, saving this for a future investigation. We instead assume a high $e$ distribution for use in the \citetalias{Sesana2015} and \citetalias{Vasiliev2015} timescales motivated by existing studies of MBHB eccentricities in cosmological simulations \citep{Chen2022_orb, Francois2026}.

In a recent work aiming to describe $T_{\text{coal}}$ using scaling relations of host galaxy properties, \citet{Francois2026} found that incorporating the eccentricity when the MBHB becomes bound ($e_b$) was necessary to achieve high predictive power for $T_{\text{coal}}$. They argue that the choice to fold orbital information into the kinematic term of the timescale ($v/\sigma |_{\text{proj}} = v / \sigma \cos \Theta$) is what enables the high predictive power of Eq. \ref{eq:timescale_4term}, and that the coupling of $v/ \sigma$ with $\cos \Theta$ may not hold in systems with less rotational support. The $v / \sigma$ values that we display in Fig. \ref{fig:ingredients_vsigma} are even higher (more rotationally supported) than the those measured in \citetalias{KHB2025}. While it's possible that a more detailed $v/\sigma$ treatment may result in less rotational support \citep[see $v/\sigma$ values found in][]{Bardati2024_kinematics}, the kinematic makeup of our hosts we find in this work implies that we remain in the regime where rotational support and orbital inclination are coupled.

Lastly, we are not tracking the evolution of $\cos \Theta$ throughout through the dynamical friction phase within \texttt{Romulus25}, nor are we prescribing an evolution for $\cos \Theta$ throughout the stellar scattering phase, despite evidence that the orbital plane may flip from counterrotating to corotating after a few hardening times \citep{Gualandris2012, Mirza2017}. Instead, we report $\cos \Theta$ for the MBH merger population only at the time of numerical merger (when the secondary MBHs stop existing in the simulation), which should approximately correspond to the initial infall angle of the secondary MBH in the simulations by \citetalias{KHB2025}.

\subsection{Subgrid model limitations}

It is important to note that the subgrid model deployed here was based on a suite of direct $N$-body simulations. Direct $N$-body simulations are built to model galactic centers, and therefore certainly do not span the range of stellar structures that may influence the overall sinking timescale. Specifically, spiral arms and bars are lacking, and these have been shown to potentially stall the secondary MBH by capturing it in resonance orbits on the kiloparsec scale for $\sim$ Gyrs \citep{Bortolas2022,Noahabstract}. The simulations were also constructed in equilibrium, which is not realistic for a post-merger galaxy remnant, especially if a NSC is deposited into a galaxy from a satellite merger. Even for the galactic center, the direct $N$-body simulation suite did not model the most diffuse non-nucleated dwarf galaxies, which were shown to have exceptionally long merger timescale, if they were to host MBHs in the first place~\citep[e.g.][]{Khan2024}. However, perhaps the most critical missing component is gas, which has long been thought accelerate MBH mergers in most configurations (\citealp{Mayer2007, Cuadra2009,Haiman2009}; see however, \citealp{Dorazio2021, Siwek2023, Dittmann2024, Clyburn2026}). The interplay between stellar and gas hardening is still an open question, though semi-analytic models \citep{Bortolas2021}, hydrodynamical simulations \citep{Shihong2024}, and comparison studies \citep{KelleyBlechaHernquist2017,IzquierdoVillalba2026_MBHCat}
are making strides in this area.

\subsection{Cosmological limitations}

As with any cosmological simulation, there are inherent limitations from resolution, which manifests in missing low mass objects and physics on small scales, and from box size, which manifests in the limited ability to generalize results due to cosmic variance. We chose \texttt{Romulus25} as our testbed because it is a uniform volume simulation that specializes in calculating accurate MBH orbits, necessary for the coalescence timescale prescription used here. However, it cannot resolve black holes with $M_{\text{BH}}  < 10^6 M_{\odot}$ and galaxies with $M_{\star} \lesssim 10^{7.5}M_{\odot}$, the inclusion of which would increase our MBH merger rate and tip the typical MBHB mass ratio toward more unequal masses. The volume, $V_{\text{sim}} = (25 \text{cMpc})^3$ also restricts the number of high mass BHs and galaxies, limiting the ability to extrapolate our results to the supermassive black hole range detectable by pulsar timing. This volume and mass range does, however, make \texttt{Romulus25} an excellent test bed for MBHBs detectable by LISA, though it may overrepresent the true number of MBHs with exactly $M_{\text{BH}} = 10^6 M_{\odot}$ due to the seeding prescription. In recent years, cosmological simulations have made headway in the implementation of low-mass seeds and multiple seeding channels \citep{Bhowmick2025, Koudmani2026, Prole2026}, as well as more physical seeding sites \citep{Dunn2018, Bhowmick2022, Prole2024}, which may raise the MBH merger rate as well \citep[see][and references therein]{IzquierdoVillalba2026_MBHCat}.

Aside from the seeding prescription, other choices in the treatment of black hole/galaxy co-evolution may trickle down to bias the MBHB populations we see here. Feedback may be among the largest source of astrophysical uncertainty, manifesting in MBH mass, certainly, but also in galaxy stellar mass and central density \citep{Wright2026}, morphology, gas content, and color \citep{Barre2026}. In addition, these simulations, like most cosmological simulations, neglect black hole spin and gravitational wave recoil; black hole spin is known to change the feedback efficiency \citep{Nemmen2007, Tchekhovskoy2012, Unal2020}, and recoil is expected to alter the MBH occupation fraction, growth history and bulge co-evolution \citep{Blecha2011, IzquierdoVillalba2026_recoil}. Finally, as we discussed previously, our seeding mechanism for NSC was done in post-processing, which bypasses a self-consistent treatment of MBH-bulge co-evolution \citep{Kormendy2013}. 

\section{Conclusions}\label{sec:conclusions}

In this work, we applied a galaxy-informed timescale (Eq. \ref{eq:timescale_4term}, \citetalias{KHB2025}) for the stellar scattering phase of MBH coalescence that depends on host galaxy properties to a population of MBH merger hosts in the \texttt{Romulus25} simulation \citep{Tremmel2017}. We present our major takeaways below.

\begin{enumerate}
    \item The galaxy-informed merger timescale predicts a median $T_{\text{coal}}$ for the stellar scattering phase of $\sim10^{8.2}\,$yr, roughly an order of magnitude faster than timescales which rely  on extrapolations of the stellar density profile. The range of $T_{\text{coal}}$ spans $\sim 10^{4.8}$-$10^{10.2}$yr for a population of 1457 merger hosts.
    \item Approximately $50\%$ of our MBH merger hosts were assigned a NSC; we predict these MBHBs would progress through stellar scattering $\sim13$ times faster than the non-NSC cases due to increased effective masses, tidal torques, and stellar densities that aid sinking and hardening \citep[see][]{Ogiya2020, Khan2021, Mukherjee2025}.
    \item At the time of numerical merger (where the BHs are separated by $\Delta r \lesssim0.7$kpc, see Fig. \ref{fig:sep_and_offsets}), MBHB orbital planes appear to be isotropically oriented with respect to their host galaxy disks. This means that 50\% of our sample is counterrotating at the start of the stellar scattering phase; we predict these MBHBs would progress through stellar scattering $\sim6$ times faster than the corotating MBHBs due to the early onset of strong GW emission aided by high eccentricities \citep[see][]{Iwasawa2011, KHB2015, Mirza2017}.\
    \item All MBH merger host galaxies in our sample had some degree of ordered rotation near the center (4$\epsilon_{\rm soft}$), with $\sim40\%$ being rotationally-dominated. This may be counterintuitive to a common expectation that dispersion increases in post-merger galaxies~\citep[e.g.][]{Bournaud04, Cox06}, and underlines this importance of taking galaxy rotation into account when modeling the MBHB stellar scattering phase. The kinematic distribution we see here is reflective of the prevalence of gas accretion and more unequal mass mergers in the buildup of these MBHB hosts~\citep[e.g.][and references therein]{Lagos18}.
    \item The galaxy-informed timescale predicts a MBH merger rate of $\sim11.5$ $\text{yr}^{-1}$ with the peak MBH merger rate occuring at $z\sim3.3$. We predict $\sim 46$ MBH merger events over a 4-year LISA mission, with the typical SNR in the thousands. These are optimistic predictions when compared to other timescales applied to the same MBHB population, though more work will be required to address the large variance seen across different simulations \citep[see][]{IzquierdoVillalba2026_MBHCat}. 
\end{enumerate}

The method we describe here is based on higher resolution simulation suites and is easy to implement, but it is certainly not the last word in modeling MBHB coalescence times. Future work for modeling galactic centers includes modeling realistic shapes, kinematics, and stellar populations, including stellar mass binaries and post-Newtonian gravity; astrophysically-motivated hydrodynamics, including AGN and stellar feedback; and a live potential to capture the interplay between gas, stars, and MBHs at the subparsec scale. As our understanding of MBHB evolution in high resolution studies continues to improve, so then should our post-processing and sub-grid modeling in cosmological environments. Practically, any models should strike the balance of capturing the relevant physics while maintaining relative ease of implementation. Future work in this area includes more robust semi-analytic and sub-grid models for NSCs to capture their global effect on MBH sinking and hardening timescales across cosmic time.

\section*{Acknowledgements}
JSR would like to acknowledge M. Rizzo-Lopez, K. Hayes, Z. Hoelscher, R. Nowicki, and H. Blake-Goszyk for helpful comments on this manuscript. No AI was used in the making of this manuscript; AI was used sparingly in the debugging process. JSR acknowledges funding support from the National Science Foundation (NSF) EMIT Program (NSF2125764). This material is based upon work supported by Tamkeen under the NYUAD Research Institute grant CG014.

\section*{Data Availability}

A file containing the properties of the MBHBs and hosts collected here is available upon request, as well as the scripts used to generate figures for this paper. The \texttt{Romulus25} database can be made available upon reasonable request. \texttt{Romulus25} was run using \texttt{ChaNGa} which has a public version: \url{https://github.com/N-BodyShop/changa}.



\bibliographystyle{mnras}
\bibliography{refs} 

@ARTICLE{Bournaud04,
       author = {{Bournaud}, F. and {Combes}, F. and {Jog}, C.~J.},
        title = "{Unequal-mass galaxy merger remnants: Spiral-like morphology but elliptical-like kinematics}",
      journal = {\aap},
         year = 2004,
        month = apr,
       volume = {418},
        pages = {L27-L30},
          doi = {10.1051/0004-6361:20040114},
archivePrefix = {arXiv},
       eprint = {astro-ph/0403307},
 primaryClass = {astro-ph},
       adsurl = {https://ui.adsabs.harvard.edu/abs/2004A&A...418L..27B}
}

@ARTICLE{Lagos18,
       author = {{Lagos}, Claudia del P. and {Schaye}, Joop and {Bah{\'e}}, Yannick and {van de Sande}, Jesse and {Kay}, Scott T. and {Barnes}, David and {Davis}, Timothy A. and {Dalla Vecchia}, Claudio},
        title = "{The connection between mass, environment, and slow rotation in simulated galaxies}",
      journal = {\mnras},
         year = 2018,
        month = jun,
       volume = {476},
       number = {4},
        pages = {4327-4345},
          doi = {10.1093/mnras/sty489},
archivePrefix = {arXiv},
       eprint = {1712.01398},
 primaryClass = {astro-ph.GA},
       adsurl = {https://ui.adsabs.harvard.edu/abs/2018MNRAS.476.4327L}
}

@ARTICLE{Cox06,
       author = {{Cox}, T.~J. and {Dutta}, Suvendra N. and {Di Matteo}, Tiziana and {Hernquist}, Lars and {Hopkins}, Philip F. and {Robertson}, Brant and {Springel}, Volker},
        title = "{The Kinematic Structure of Merger Remnants}",
      journal = {\apj},
         year = 2006,
        month = oct,
       volume = {650},
       number = {2},
        pages = {791-811},
          doi = {10.1086/507474},
archivePrefix = {arXiv},
       eprint = {astro-ph/0607446},
 primaryClass = {astro-ph},
       adsurl = {https://ui.adsabs.harvard.edu/abs/2006ApJ...650..791C}
}

@INPROCEEDINGS{Noahabstract,
       author = {{Herrero}, Noah and {Holley-Bockelmann}, Kelly and {Khan}, Fazeel Mahmood and {Valluri}, Monica and {Tahmasebzadeh}, Behzad},
        title = "{A Black Hole Walks into a Bar: Massive Black Hole Dynamics in Weakly-Barred Galaxies}",
    booktitle = {American Astronomical Society Meeting Abstracts \#245},
         year = 2025,
       series = {American Astronomical Society Meeting Abstracts},
       volume = {245},
        month = jan,
          eid = {402.20},
        pages = {402.20},
       adsurl = {https://ui.adsabs.harvard.edu/abs/2025AAS...24540220H}
}

@ARTICLE{KHB2025,
       author = {{Holley-Bockelmann}, Kelly and {Khan}, Fazeel Mahmood and {Williams}, Isaiah and {Roth}, Jaelyn and {Rizzo Smith}, Michael and {Porter}, Kaitlin and {Bellovary}, Jillian and {Derdzinski}, Andrea and {Macci{\`o}}, Andrea V.},
        title = "{Handy Relation between Binary Black Hole Merger Times and Host Galaxy Properties}",
      journal = {\apjl},
         year = 2025,
        month = dec,
       volume = {995},
       number = {1},
          eid = {L32},
        pages = {L32},
          doi = {10.3847/2041-8213/ae1ccd},
archivePrefix = {arXiv},
       eprint = {2508.14253},
 primaryClass = {astro-ph.GA},
       adsurl = {https://ui.adsabs.harvard.edu/abs/2025ApJ...995L..32H}
}

@ARTICLE{Dittmann2024,
       author = {{Dittmann}, Alexander J. and {Ryan}, Geoffrey},
        title = "{The Evolution of Accreting Binaries: From Brown Dwarfs to Supermassive Black Holes}",
      journal = {\apj},
         year = 2024,
        month = may,
       volume = {967},
       number = {1},
          eid = {12},
        pages = {12},
          doi = {10.3847/1538-4357/ad2f1e},
archivePrefix = {arXiv},
       eprint = {2310.07758},
 primaryClass = {astro-ph.GA},
       adsurl = {https://ui.adsabs.harvard.edu/abs/2024ApJ...967...12D}
}

@ARTICLE{Duffell2024,
       author = {{Duffell}, Paul C. and {Dittmann}, Alexander J. and {D'Orazio}, Daniel J. and {Franchini}, Alessia and {Kratter}, Kaitlin M. and {Penzlin}, Anna B.~T. and {Ragusa}, Enrico and {Siwek}, Magdalena and {Tiede}, Christopher and {Wang}, Haiyang and {Zrake}, Jonathan and {Dempsey}, Adam M. and {Haiman}, Zoltan and {Lupi}, Alessandro and {Pirog}, Michal and {Ryan}, Geoffrey},
        title = "{The Santa Barbara Binary{\ensuremath{-}}disk Code Comparison}",
      journal = {\apj},
         year = 2024,
        month = aug,
       volume = {970},
       number = {2},
          eid = {156},
        pages = {156},
          doi = {10.3847/1538-4357/ad5a7e},
archivePrefix = {arXiv},
       eprint = {2402.13039},
 primaryClass = {astro-ph.SR},
       adsurl = {https://ui.adsabs.harvard.edu/abs/2024ApJ...970..156D}
}

@ARTICLE{Begelman1980,
       author = {{Begelman}, M.~C. and {Blandford}, R.~D. and {Rees}, M.~J.},
        title = "{Massive black hole binaries in active galactic nuclei}",
      journal = {\nat},
         year = 1980,
        month = sep,
       volume = {287},
       number = {5780},
        pages = {307-309},
          doi = {10.1038/287307a0},
       adsurl = {https://ui.adsabs.harvard.edu/abs/1980Natur.287..307B}
}

@ARTICLE{KHB2015,
       author = {{Holley-Bockelmann}, Kelly and {Khan}, Fazeel Mahmood},
        title = "{Galaxy Rotation and Rapid Supermassive Binary Coalescence}",
      journal = {\apj},
         year = 2015,
        month = sep,
       volume = {810},
       number = {2},
          eid = {139},
        pages = {139},
          doi = {10.1088/0004-637X/810/2/139},
archivePrefix = {arXiv},
       eprint = {1505.06203},
 primaryClass = {astro-ph.GA},
       adsurl = {https://ui.adsabs.harvard.edu/abs/2015ApJ...810..139H}
}

@ARTICLE{Khan2015,
       author = {{Khan}, Fazeel Mahmood and {Holley-Bockelmann}, Kelly and {Berczik}, Peter},
        title = "{Ultramassive Black Hole Coalescence}",
      journal = {\apj},
         year = 2015,
        month = jan,
       volume = {798},
       number = {2},
          eid = {103},
        pages = {103},
          doi = {10.1088/0004-637X/798/2/103},
archivePrefix = {arXiv},
       eprint = {1405.6425},
 primaryClass = {astro-ph.GA},
       adsurl = {https://ui.adsabs.harvard.edu/abs/2015ApJ...798..103K}
}

@ARTICLE{Khan2021,
       author = {{Khan}, Fazeel Mahmood and {Holley-Bockelmann}, Kelly},
        title = "{Extremely efficient mergers of intermediate-mass black hole binaries in nucleated dwarf galaxies}",
      journal = {\mnras},
         year = 2021,
        month = nov,
       volume = {508},
       number = {1},
        pages = {1174-1188},
          doi = {10.1093/mnras/stab2646},
archivePrefix = {arXiv},
       eprint = {2109.12129},
 primaryClass = {astro-ph.GA},
       adsurl = {https://ui.adsabs.harvard.edu/abs/2021MNRAS.508.1174K}
}

@ARTICLE{Khan2024,
       author = {{Khan}, Fazeel Mahmood and {Javed}, Fiza and {Holley-Bockelmann}, Kelly and {Mayer}, Lucio and {Berczik}, Peter and {Macci{\`o}}, Andrea V.},
        title = "{The Potential for Long-lived Intermediate-mass Black Hole Binaries in the Lowest Density Dwarf Galaxies}",
      journal = {\apj},
         year = 2024,
        month = nov,
       volume = {976},
       number = {1},
          eid = {22},
        pages = {22},
          doi = {10.3847/1538-4357/ad8082},
archivePrefix = {arXiv},
       eprint = {2408.14541},
 primaryClass = {astro-ph.GA},
       adsurl = {https://ui.adsabs.harvard.edu/abs/2024ApJ...976...22K}
}

@ARTICLE{Tremmel2017,
       author = {{Tremmel}, M. and {Karcher}, M. and {Governato}, F. and {Volonteri}, M. and {Quinn}, T.~R. and {Pontzen}, A. and {Anderson}, L. and {Bellovary}, J.},
        title = "{The Romulus cosmological simulations: a physical approach to the formation, dynamics and accretion models of SMBHs}",
      journal = {\mnras},
         year = 2017,
        month = sep,
       volume = {470},
       number = {1},
        pages = {1121-1139},
          doi = {10.1093/mnras/stx1160},
archivePrefix = {arXiv},
       eprint = {1607.02151},
 primaryClass = {astro-ph.GA},
       adsurl = {https://ui.adsabs.harvard.edu/abs/2017MNRAS.470.1121T}
}

@ARTICLE{Tremmel2015,
       author = {{Tremmel}, M. and {Governato}, F. and {Volonteri}, M. and {Quinn}, T.~R.},
        title = "{Off the beaten path: a new approach to realistically model the orbital decay of supermassive black holes in galaxy formation simulations}",
      journal = {\mnras},
         year = 2015,
        month = aug,
       volume = {451},
       number = {2},
        pages = {1868-1874},
          doi = {10.1093/mnras/stv1060},
archivePrefix = {arXiv},
       eprint = {1501.07609},
 primaryClass = {astro-ph.GA},
       adsurl = {https://ui.adsabs.harvard.edu/abs/2015MNRAS.451.1868T}
}

@ARTICLE{Menon2015,
       author = {{Menon}, Harshitha and {Wesolowski}, Lukasz and {Zheng}, Gengbin and {Jetley}, Pritish and {Kale}, Laxmikant and {Quinn}, Thomas and {Governato}, Fabio},
        title = "{Adaptive techniques for clustered N-body cosmological simulations}",
      journal = {Computational Astrophysics and Cosmology},
         year = 2015,
        month = mar,
       volume = {2},
          eid = {1},
        pages = {1},
          doi = {10.1186/s40668-015-0007-9},
archivePrefix = {arXiv},
       eprint = {1409.1929},
 primaryClass = {astro-ph.IM},
       adsurl = {https://ui.adsabs.harvard.edu/abs/2015ComAC...2....1M}
}

@ARTICLE{Bellovary2010,
       author = {{Bellovary}, Jillian M. and {Governato}, Fabio and {Quinn}, Thomas R. and {Wadsley}, James and {Shen}, Sijing and {Volonteri}, Marta},
        title = "{Wandering Black Holes in Bright Disk Galaxy Halos}",
      journal = {\apjl},
         year = 2010,
        month = oct,
       volume = {721},
       number = {2},
        pages = {L148-L152},
          doi = {10.1088/2041-8205/721/2/L148},
archivePrefix = {arXiv},
       eprint = {1008.5147},
 primaryClass = {astro-ph.CO},
       adsurl = {https://ui.adsabs.harvard.edu/abs/2010ApJ...721L.148B}
}

@ARTICLE{Ricarte2021,
       author = {{Ricarte}, Angelo and {Tremmel}, Michael and {Natarajan}, Priyamvada and {Zimmer}, Charlotte and {Quinn}, Thomas},
        title = "{Origins and demographics of wandering black holes}",
      journal = {\mnras},
         year = 2021,
        month = jun,
       volume = {503},
       number = {4},
        pages = {6098-6111},
          doi = {10.1093/mnras/stab866},
archivePrefix = {arXiv},
       eprint = {2103.12124},
 primaryClass = {astro-ph.GA},
       adsurl = {https://ui.adsabs.harvard.edu/abs/2021MNRAS.503.6098R}
}

@ARTICLE{Neumayer2020,
       author = {{Neumayer}, Nadine and {Seth}, Anil and {B{\"o}ker}, Torsten},
        title = "{Nuclear star clusters}",
      journal = {\aapr},
         year = 2020,
        month = jul,
       volume = {28},
       number = {1},
          eid = {4},
        pages = {4},
          doi = {10.1007/s00159-020-00125-0},
archivePrefix = {arXiv},
       eprint = {2001.03626},
 primaryClass = {astro-ph.GA},
       adsurl = {https://ui.adsabs.harvard.edu/abs/2020A&ARv..28....4N}
}

@ARTICLE{Mukherjee2025,
       author = {{Mukherjee}, Diptajyoti and {Zhou}, Yihao and {Chen}, Nianyi and {Di Carlo}, Ugo Niccol{\`o} and {Di Matteo}, Tiziana},
        title = "{MAGICS. III. Seeds Sink Swiftly: Nuclear Star Clusters Dramatically Accelerate Seed Black Hole Mergers}",
      journal = {\apj},
         year = 2025,
        month = mar,
       volume = {981},
       number = {2},
          eid = {203},
        pages = {203},
          doi = {10.3847/1538-4357/adb1b0},
archivePrefix = {arXiv},
       eprint = {2409.19095},
 primaryClass = {astro-ph.GA},
       adsurl = {https://ui.adsabs.harvard.edu/abs/2025ApJ...981..203M}
}

@ARTICLE{Bortolas2022,
       author = {{Bortolas}, Elisa and {Bonetti}, Matteo and {Dotti}, Massimo and {Lupi}, Alessandro and {Capelo}, Pedro R. and {Mayer}, Lucio and {Sesana}, Alberto},
        title = "{The role of bars on the dynamical-friction-driven inspiral of massive objects}",
      journal = {\mnras},
         year = 2022,
        month = may,
       volume = {512},
       number = {3},
        pages = {3365-3382},
          doi = {10.1093/mnras/stac645},
archivePrefix = {arXiv},
       eprint = {2103.07486},
 primaryClass = {astro-ph.GA},
       adsurl = {https://ui.adsabs.harvard.edu/abs/2022MNRAS.512.3365B}
}

@ARTICLE{Berczik2006,
       author = {{Berczik}, Peter and {Merritt}, David and {Spurzem}, Rainer and {Bischof}, Hans-Peter},
        title = "{Efficient Merger of Binary Supermassive Black Holes in Nonaxisymmetric Galaxies}",
      journal = {\apjl},
         year = 2006,
        month = may,
       volume = {642},
       number = {1},
        pages = {L21-L24},
          doi = {10.1086/504426},
archivePrefix = {arXiv},
       eprint = {astro-ph/0601698},
 primaryClass = {astro-ph},
       adsurl = {https://ui.adsabs.harvard.edu/abs/2006ApJ...642L..21B}
}

@ARTICLE{Khan2013,
       author = {{Khan}, Fazeel Mahmood and {Holley-Bockelmann}, Kelly and {Berczik}, Peter and {Just}, Andreas},
        title = "{Supermassive Black Hole Binary Evolution in Axisymmetric Galaxies: The Final Parsec Problem is Not a Problem}",
      journal = {\apj},
         year = 2013,
        month = aug,
       volume = {773},
       number = {2},
          eid = {100},
        pages = {100},
          doi = {10.1088/0004-637X/773/2/100},
archivePrefix = {arXiv},
       eprint = {1302.1871},
 primaryClass = {astro-ph.GA},
       adsurl = {https://ui.adsabs.harvard.edu/abs/2013ApJ...773..100K}
}

@ARTICLE{Yu2002,
       author = {{Yu}, Qingjuan},
        title = "{Evolution of massive binary black holes}",
      journal = {\mnras},
         year = 2002,
        month = apr,
       volume = {331},
       number = {4},
        pages = {935-958},
          doi = {10.1046/j.1365-8711.2002.05242.x},
archivePrefix = {arXiv},
       eprint = {astro-ph/0109530},
 primaryClass = {astro-ph},
       adsurl = {https://ui.adsabs.harvard.edu/abs/2002MNRAS.331..935Y}
}

@ARTICLE{Derdzinski2025,
       author = {{Derdzinski}, Andrea and {Dittmann}, Alexander J. and {Franchini}, Alessia and {Lupi}, Alessandro and {Brucy}, No{\'e} and {Capelo}, Pedro R. and {Masset}, Fr{\'e}d{\'e}ric S. and {Mignon-Risse}, Rapha{\"e}l and {Rizzo Smith}, Michael and {Santiago-Leandro}, Edwin and {Toscani}, Martina and {Velasco-Romero}, David A. and {Wissing}, Robert and {Garg}, Mudit and {Mayer}, Lucio and {Serafinelli}, Roberto and {Souvaitzis}, Lazaros and {D'Orazio}, Daniel J. and {Menu}, Jonathan},
        title = "{The LISA Astrophysics ``Disc-IMRI'' Code Comparison Project: Intermediate-Mass-Ratio Binaries in AGN-Like Discs}",
      journal = {arXiv e-prints},
         year = 2025,
        month = dec,
          eid = {arXiv:2512.10893},
        pages = {arXiv:2512.10893},
          doi = {10.48550/arXiv.2512.10893},
archivePrefix = {arXiv},
       eprint = {2512.10893},
 primaryClass = {astro-ph.HE},
       adsurl = {https://ui.adsabs.harvard.edu/abs/2025arXiv251210893D}
}

@ARTICLE{Mirza2017,
       author = {{Mirza}, M.~A. and {Tahir}, A. and {Khan}, F.~M. and {Holley-Bockelmann}, H. and {Baig}, A.~M. and {Berczik}, P. and {Chishtie}, F.},
        title = "{Galaxy rotation and supermassive black hole binary evolution}",
      journal = {\mnras},
         year = 2017,
        month = sep,
       volume = {470},
       number = {1},
        pages = {940-947},
          doi = {10.1093/mnras/stx1248},
archivePrefix = {arXiv},
       eprint = {1704.03490},
 primaryClass = {astro-ph.GA},
       adsurl = {https://ui.adsabs.harvard.edu/abs/2017MNRAS.470..940M}
}

@ARTICLE{Sesana2015,
       author = {{Sesana}, Alberto and {Khan}, Fazeel Mahmood},
        title = "{Scattering experiments meet N-body - I. A practical recipe for the evolution of massive black hole binaries in stellar environments}",
      journal = {\mnras},
         year = 2015,
        month = nov,
       volume = {454},
       number = {1},
        pages = {L66-L70},
          doi = {10.1093/mnrasl/slv131},
archivePrefix = {arXiv},
       eprint = {1505.02062},
 primaryClass = {astro-ph.GA},
       adsurl = {https://ui.adsabs.harvard.edu/abs/2015MNRAS.454L..66S}
}

@ARTICLE{Vasiliev2015,
       author = {{Vasiliev}, Eugene and {Antonini}, Fabio and {Merritt}, David},
        title = "{The Final-parsec Problem in the Collisionless Limit}",
      journal = {\apj},
         year = 2015,
        month = sep,
       volume = {810},
       number = {1},
          eid = {49},
        pages = {49},
          doi = {10.1088/0004-637X/810/1/49},
archivePrefix = {arXiv},
       eprint = {1505.05480},
 primaryClass = {astro-ph.GA},
       adsurl = {https://ui.adsabs.harvard.edu/abs/2015ApJ...810...49V}
}

@ARTICLE{Chen2022_orb,
       author = {{Chen}, Nianyi and {Ni}, Yueying and {Holgado}, A. Miguel and {Di Matteo}, Tiziana and {Tremmel}, Michael and {DeGraf}, Colin and {Bird}, Simeon and {Croft}, Rupert and {Feng}, Yu},
        title = "{Massive black hole mergers with orbital information: predictions from the ASTRID simulation}",
      journal = {\mnras},
         year = 2022,
        month = aug,
       volume = {514},
       number = {2},
        pages = {2220-2238},
          doi = {10.1093/mnras/stac1432},
archivePrefix = {arXiv},
       eprint = {2112.08555},
 primaryClass = {astro-ph.GA},
       adsurl = {https://ui.adsabs.harvard.edu/abs/2022MNRAS.514.2220C}
}

@ARTICLE{Varisco2021,
       author = {{Varisco}, Ludovica and {Bortolas}, Elisa and {Dotti}, Massimo and {Sesana}, Alberto},
        title = "{Stellar hardening of massive black hole binaries: the impact of the host rotation}",
      journal = {\mnras},
         year = 2021,
        month = nov,
       volume = {508},
       number = {1},
        pages = {1533-1542},
          doi = {10.1093/mnras/stab2649},
archivePrefix = {arXiv},
       eprint = {2104.14570},
 primaryClass = {astro-ph.GA},
       adsurl = {https://ui.adsabs.harvard.edu/abs/2021MNRAS.508.1533V}
}

@ARTICLE{Sijacki2015,
       author = {{Sijacki}, Debora and {Vogelsberger}, Mark and {Genel}, Shy and {Springel}, Volker and {Torrey}, Paul and {Snyder}, Gregory F. and {Nelson}, Dylan and {Hernquist}, Lars},
        title = "{The Illustris simulation: the evolving population of black holes across cosmic time}",
      journal = {\mnras},
         year = 2015,
        month = sep,
       volume = {452},
       number = {1},
        pages = {575-596},
          doi = {10.1093/mnras/stv1340},
archivePrefix = {arXiv},
       eprint = {1408.6842},
 primaryClass = {astro-ph.GA},
       adsurl = {https://ui.adsabs.harvard.edu/abs/2015MNRAS.452..575S}
}

@ARTICLE{Schaye2015,
       author = {{Schaye}, Joop and {Crain}, Robert A. and {Bower}, Richard G. and {Furlong}, Michelle and {Schaller}, Matthieu and {Theuns}, Tom and {Dalla Vecchia}, Claudio and {Frenk}, Carlos S. and {McCarthy}, I.~G. and {Helly}, John C. and {Jenkins}, Adrian and {Rosas-Guevara}, Y.~M. and {White}, Simon D.~M. and {Baes}, Maarten and {Booth}, C.~M. and {Camps}, Peter and {Navarro}, Julio F. and {Qu}, Yan and {Rahmati}, Alireza and {Sawala}, Till and {Thomas}, Peter A. and {Trayford}, James},
        title = "{The EAGLE project: simulating the evolution and assembly of galaxies and their environments}",
      journal = {\mnras},
         year = 2015,
        month = jan,
       volume = {446},
       number = {1},
        pages = {521-554},
          doi = {10.1093/mnras/stu2058},
archivePrefix = {arXiv},
       eprint = {1407.7040},
 primaryClass = {astro-ph.GA},
       adsurl = {https://ui.adsabs.harvard.edu/abs/2015MNRAS.446..521S}
}

@ARTICLE{Gualandris2017,
       author = {{Gualandris}, Alessia and {Read}, Justin I. and {Dehnen}, Walter and {Bortolas}, Elisa},
        title = "{Collisionless loss-cone refilling: there is no final parsec problem}",
      journal = {\mnras},
         year = 2017,
        month = jan,
       volume = {464},
       number = {2},
        pages = {2301-2310},
          doi = {10.1093/mnras/stw2528},
archivePrefix = {arXiv},
       eprint = {1609.09383},
 primaryClass = {astro-ph.GA},
       adsurl = {https://ui.adsabs.harvard.edu/abs/2017MNRAS.464.2301G}
}

@ARTICLE{Peters1963,
       author = {{Peters}, P.~C. and {Mathews}, J.},
        title = "{Gravitational Radiation from Point Masses in a Keplerian Orbit}",
      journal = {Physical Review},
         year = 1963,
        month = jul,
       volume = {131},
       number = {1},
        pages = {435-440},
          doi = {10.1103/PhysRev.131.435},
       adsurl = {https://ui.adsabs.harvard.edu/abs/1963PhRv..131..435P}
}

@ARTICLE{IzquierdoVillalba2026_MBHCat,
       author = {{Izquierdo-Villalba}, David and {Habouzit}, Melanie and {Bonetti}, Matteo and {Bonoli}, Silvia and {Gualandris}, Alessia and {Volonteri}, Marta and {Angeloni}, Federico and {Barausse}, Enrico and {Bhowmick}, Aklant and {Blecha}, Laura and {Bonilla Rivera}, Alexander and {Bortolas}, Elisa and {Caliskan}, Mesut and {Capelo}, Pedro R. and {Caramete}, Ana and {Caramete}, Laurentiu and {Chen}, Nianyi and {Colpi}, Monica and {Contini}, Thierry and {Dav{\'e}}, Romeel and {Dayal}, Pratika and {DeGraf}, Colin and {Deane}, Roger and {Decarli}, Roberto and {Delpech}, R{\'e}mi and {Di Matteo}, Tiziana and {Dong-P{\'a}ez}, Chi An and {Graham}, Alister W. and {Haggard}, Daryl and {Irodotou}, Dimitrios and {Johansson}, Peter H. and {Keitaanranta}, Atte and {Kelley}, Luke Zoltan and {Khan}, Fazeel Mahmood and {Langen}, Vivienne and {Li}, Kunyang and {Liao}, Shihong and {Mangiagli}, Alberto and {Marsat}, Sylvain and {McCaffrey}, Joe and {Ni}, Yueying and {Pillay}, Coral and {Pislan}, Florentina-Crenguta and {Rawlings}, Alex and {Regan}, John and {Reinoso}, Basti{\'a}n and {Roth}, Jaelyn and {Ruiz}, Milton and {Sergijenko}, Olga and {Sesana}, Alberto and {Shaifullah}, Golam and {Singh}, Jasbir and {Spinoso}, Daniele and {Toubiana}, Alexandre and {Tremmel}, Michael and {Trinca}, Alessandro and {Valiante}, Rosa and {Zhou}, Yihao and {Dubois}, Yohan and {Graziani}, Luca and {Lovell}, Christopher C. and {Peirani}, Sebastien and {Roper}, William J. and {Schaye}, Joop and {Schneider}, Raffaella and {Trebitsch}, Maxime and {Vijayan}, Aswin and {Vogelsberger}, Mark and {Wilkins}, Stephen and {Wise}, John},
        title = "{The LISA Astrophysics MBHcatalogues Project: A comparison of predictions of simulated massive black hole binaries}",
      journal = {arXiv e-prints},
         year = 2026,
        month = apr,
          eid = {arXiv:2605.00092},
        pages = {arXiv:2605.00092},
          doi = {10.48550/arXiv.2605.00092},
archivePrefix = {arXiv},
       eprint = {2605.00092},
 primaryClass = {astro-ph.GA},
       adsurl = {https://ui.adsabs.harvard.edu/abs/2026arXiv260500092I}
}

@ARTICLE{AmaroSeoane2023,
       author = {{Amaro-Seoane}, Pau and {Andrews}, Jeff and {Arca Sedda}, Manuel and {Askar}, Abbas and {Baghi}, Quentin and {Balasov}, Razvan and {Bartos}, Imre and {Bavera}, Simone S. and {Bellovary}, Jillian and {Berry}, Christopher P.~L. and {Berti}, Emanuele and {Bianchi}, Stefano and {Blecha}, Laura and {Blondin}, St{\'e}phane and {Bogdanovi{\'c}}, Tamara and {Boissier}, Samuel and {Bonetti}, Matteo and {Bonoli}, Silvia and {Bortolas}, Elisa and {Breivik}, Katelyn and {Capelo}, Pedro R. and {Caramete}, Laurentiu and {Cattorini}, Federico and {Charisi}, Maria and {Chaty}, Sylvain and {Chen}, Xian and {Chru{\'s}li{\'n}ska}, Martyna and {Chua}, Alvin J.~K. and {Church}, Ross and {Colpi}, Monica and {D'Orazio}, Daniel and {Danielski}, Camilla and {Davies}, Melvyn B. and {Dayal}, Pratika and {De Rosa}, Alessandra and {Derdzinski}, Andrea and {Destounis}, Kyriakos and {Dotti}, Massimo and {Du{\c{t}}an}, Ioana and {Dvorkin}, Irina and {Fabj}, Gaia and {Foglizzo}, Thierry and {Ford}, Saavik and {Fouvry}, Jean-Baptiste and {Franchini}, Alessia and {Fragos}, Tassos and {Fryer}, Chris and {Gaspari}, Massimo and {Gerosa}, Davide and {Graziani}, Luca and {Groot}, Paul and {Habouzit}, Melanie and {Haggard}, Daryl and {Haiman}, Zoltan and {Han}, Wen-Biao and {Istrate}, Alina and {Johansson}, Peter H. and {Khan}, Fazeel Mahmood and {Kimpson}, Tomas and {Kokkotas}, Kostas and {Kong}, Albert and {Korol}, Valeriya and {Kremer}, Kyle and {Kupfer}, Thomas and {Lamberts}, Astrid and {Larson}, Shane and {Lau}, Mike and {Liu}, Dongliang and {Lloyd-Ronning}, Nicole and {Lodato}, Giuseppe and {Lupi}, Alessandro and {Ma}, Chung-Pei and {Maccarone}, Tomas and {Mandel}, Ilya and {Mangiagli}, Alberto and {Mapelli}, Michela and {Mathis}, St{\'e}phane and {Mayer}, Lucio and {McGee}, Sean and {McKernan}, Berry and {Miller}, M. Coleman and {Mota}, David F. and {Mumpower}, Matthew and {Nasim}, Syeda S. and {Nelemans}, Gijs and {Noble}, Scott and {Pacucci}, Fabio and {Panessa}, Francesca and {Paschalidis}, Vasileios and {Pfister}, Hugo and {Porquet}, Delphine and {Quenby}, John and {Ricarte}, Angelo and {R{\"o}pke}, Friedrich K. and {Regan}, John and {Rosswog}, Stephan and {Ruiter}, Ashley and {Ruiz}, Milton and {Runnoe}, Jessie and {Schneider}, Raffaella and {Schnittman}, Jeremy and {Secunda}, Amy and {Sesana}, Alberto and {Seto}, Naoki and {Shao}, Lijing and {Shapiro}, Stuart and {Sopuerta}, Carlos and {Stone}, Nicholas C. and {Suvorov}, Arthur and {Tamanini}, Nicola and {Tamfal}, Tomas and {Tauris}, Thomas and {Temmink}, Karel and {Tomsick}, John and {Toonen}, Silvia and {Torres-Orjuela}, Alejandro and {Toscani}, Martina and {Tsokaros}, Antonios and {Unal}, Caner and {V{\'a}zquez-Aceves}, Ver{\'o}nica and {Valiante}, Rosa and {van Putten}, Maurice and {van Roestel}, Jan and {Vignali}, Christian and {Volonteri}, Marta and {Wu}, Kinwah and {Younsi}, Ziri and {Yu}, Shenghua and {Zane}, Silvia and {Zwick}, Lorenz and {Antonini}, Fabio and {Baibhav}, Vishal and {Barausse}, Enrico and {Bonilla Rivera}, Alexander and {Branchesi}, Marica and {Branduardi-Raymont}, Graziella and {Burdge}, Kevin and {Chakraborty}, Srija and {Cuadra}, Jorge and {Dage}, Kristen and {Davis}, Benjamin and {de Mink}, Selma E. and {Decarli}, Roberto and {Doneva}, Daniela and {Escoffier}, Stephanie and {Gandhi}, Poshak and {Haardt}, Francesco and {Lousto}, Carlos O. and {Nissanke}, Samaya and {Nordhaus}, Jason and {O'Shaughnessy}, Richard and {Portegies Zwart}, Simon and {Pound}, Adam and {Schussler}, Fabian and {Sergijenko}, Olga and {Spallicci}, Alessandro and {Vernieri}, Daniele and {Vigna-G{\'o}mez}, Alejandro},
        title = "{Astrophysics with the Laser Interferometer Space Antenna}",
      journal = {Living Reviews in Relativity},
         year = 2023,
        month = dec,
       volume = {26},
       number = {1},
          eid = {2},
        pages = {2},
          doi = {10.1007/s41114-022-00041-y},
archivePrefix = {arXiv},
       eprint = {2203.06016},
 primaryClass = {gr-qc},
       adsurl = {https://ui.adsabs.harvard.edu/abs/2023LRR....26....2A}
}

@ARTICLE{Katz2020,
       author = {{Katz}, Michael L. and {Kelley}, Luke Zoltan and {Dosopoulou}, Fani and {Berry}, Samantha and {Blecha}, Laura and {Larson}, Shane L.},
        title = "{Probing massive black hole binary populations with LISA}",
      journal = {\mnras},
         year = 2020,
        month = jan,
       volume = {491},
       number = {2},
        pages = {2301-2317},
          doi = {10.1093/mnras/stz3102},
archivePrefix = {arXiv},
       eprint = {1908.05779},
 primaryClass = {astro-ph.HE},
       adsurl = {https://ui.adsabs.harvard.edu/abs/2020MNRAS.491.2301K}
}

@ARTICLE{Bortolas2021,
       author = {{Bortolas}, Elisa and {Franchini}, Alessia and {Bonetti}, Matteo and {Sesana}, Alberto},
        title = "{The Competing Effect of Gas and Stars in the Evolution of Massive Black Hole Binaries}",
      journal = {\apjl},
         year = 2021,
        month = sep,
       volume = {918},
       number = {1},
          eid = {L15},
        pages = {L15},
          doi = {10.3847/2041-8213/ac1c0c},
archivePrefix = {arXiv},
       eprint = {2108.13436},
 primaryClass = {astro-ph.HE},
       adsurl = {https://ui.adsabs.harvard.edu/abs/2021ApJ...918L..15B}
}

@ARTICLE{Sesana2010,
       author = {{Sesana}, Alberto},
        title = "{Self Consistent Model for the Evolution of Eccentric Massive Black Hole Binaries in Stellar Environments: Implications for Gravitational Wave Observations}",
      journal = {\apj},
         year = 2010,
        month = aug,
       volume = {719},
       number = {1},
        pages = {851-864},
          doi = {10.1088/0004-637X/719/1/851},
archivePrefix = {arXiv},
       eprint = {1006.0730},
 primaryClass = {astro-ph.CO},
       adsurl = {https://ui.adsabs.harvard.edu/abs/2010ApJ...719..851S}
}

@ARTICLE{Dittmann2026,
       author = {{Dittmann}, Alexander J. and {Ryan}, Geoffrey and {Combi}, Luciano},
        title = "{Eccentric Binaries Accreting from Thin Disks: Orbital Evolution}",
      journal = {\apj},
         year = 2026,
        month = apr,
       volume = {1001},
       number = {2},
          eid = {132},
        pages = {132},
          doi = {10.3847/1538-4357/ae4a9c},
archivePrefix = {arXiv},
       eprint = {2512.11954},
 primaryClass = {astro-ph.GA},
       adsurl = {https://ui.adsabs.harvard.edu/abs/2026ApJ..1001..132D}
}

@ARTICLE{DeCun2023,
       author = {{De Cun}, Victoria I. and {Bellovary}, Jillian M. and {Katz}, Michael L.},
        title = "{Merging black holes in dwarf galaxies: calculating binary black hole coalescence time-scales from simulations for LISA detection}",
      journal = {\mnras},
         year = 2023,
        month = apr,
       volume = {520},
       number = {3},
        pages = {3916-3922},
          doi = {10.1093/mnras/stad311},
archivePrefix = {arXiv},
       eprint = {2111.15035},
 primaryClass = {astro-ph.GA},
       adsurl = {https://ui.adsabs.harvard.edu/abs/2023MNRAS.520.3916D}
}

@ARTICLE{Tiede2020,
       author = {{Tiede}, Christopher and {Zrake}, Jonathan and {MacFadyen}, Andrew and {Haiman}, Zoltan},
        title = "{Gas-driven Inspiral of Binaries in Thin Accretion Disks}",
      journal = {\apj},
         year = 2020,
        month = sep,
       volume = {900},
       number = {1},
          eid = {43},
        pages = {43},
          doi = {10.3847/1538-4357/aba432},
archivePrefix = {arXiv},
       eprint = {2005.09555},
 primaryClass = {astro-ph.GA},
       adsurl = {https://ui.adsabs.harvard.edu/abs/2020ApJ...900...43T}
}

@ARTICLE{Dorazio2021,
       author = {{D'Orazio}, Daniel J. and {Duffell}, Paul C.},
        title = "{Orbital Evolution of Equal-mass Eccentric Binaries due to a Gas Disk: Eccentric Inspirals and Circular Outspirals}",
      journal = {\apjl},
         year = 2021,
        month = jun,
       volume = {914},
       number = {1},
          eid = {L21},
        pages = {L21},
          doi = {10.3847/2041-8213/ac0621},
archivePrefix = {arXiv},
       eprint = {2103.09251},
 primaryClass = {astro-ph.HE},
       adsurl = {https://ui.adsabs.harvard.edu/abs/2021ApJ...914L..21D}
}

@ARTICLE{Colpi2024,
       author = {{Colpi}, Monica and {Danzmann}, Karsten and {Hewitson}, Martin and {Holley-Bockelmann}, Kelly and {Jetzer}, Philippe and {Nelemans}, Gijs and {Petiteau}, Antoine and {Shoemaker}, David and {Sopuerta}, Carlos and {Stebbins}, Robin and {Tanvir}, Nial and {Ward}, Henry and {Weber}, William Joseph and {Thorpe}, Ira and {Daurskikh}, Anna and {Deep}, Atul and {Fern{\'a}ndez N{\'u}{\~n}ez}, Ignacio and {Garc{\'\i}a Marirrodriga}, C{\'e}sar and {Gehler}, Martin and {Halain}, Jean-Philippe and {Jennrich}, Oliver and {Lammers}, Uwe and {Larra{\~n}aga}, Jonan and {Lieser}, Maike and {L{\"u}tzgendorf}, Nora and {Martens}, Waldemar and {Mondin}, Linda and {Piris Ni{\~n}o}, Ana and {Amaro-Seoane}, Pau and {Arca Sedda}, Manuel and {Auclair}, Pierre and {Babak}, Stanislav and {Baghi}, Quentin and {Baibhav}, Vishal and {Baker}, Tessa and {Bayle}, Jean-Baptiste and {Berry}, Christopher and {Berti}, Emanuele and {Boileau}, Guillaume and {Bonetti}, Matteo and {Brito}, Richard and {Buscicchio}, Riccardo and {Calcagni}, Gianluca and {Capelo}, Pedro R. and {Caprini}, Chiara and {Caputo}, Andrea and {Castelli}, Eleonora and {Chen}, Hsin-Yu and {Chen}, Xian and {Chua}, Alvin and {Davies}, Gareth and {Derdzinski}, Andrea and {Domcke}, Valerie Fiona and {Doneva}, Daniela and {Dvorkin}, Irna and {Mar{\'\i}a Ezquiaga}, Jose and {Gair}, Jonathan and {Haiman}, Zoltan and {Harry}, Ian and {Hartwig}, Olaf and {Hees}, Aurelien and {Heffernan}, Anna and {Husa}, Sascha and {Izquierdo-Villalba}, David and {Karnesis}, Nikolaos and {Klein}, Antoine and {Korol}, Valeriya and {Korsakova}, Natalia and {Kupfer}, Thomas and {Laghi}, Danny and {Lamberts}, Astrid and {Larson}, Shane and {Le Jeune}, Maude and {Lewicki}, Marek and {Littenberg}, Tyson and {Madge}, Eric and {Mangiagli}, Alberto and {Marsat}, Sylvain and {Vilchez}, Ivan Martin and {Maselli}, Andrea and {Mathews}, Josh and {van de Meent}, Maarten and {Muratore}, Martina and {Nardini}, Germano and {Pani}, Paolo and {Peloso}, Marco and {Pieroni}, Mauro and {Pound}, Adam and {Quelquejay-Leclere}, Hippolyte and {Ricciardone}, Angelo and {Rossi}, Elena Maria and {Sartirana}, Andrea and {Savalle}, Etienne and {Sberna}, Laura and {Sesana}, Alberto and {Shoemaker}, Deirdre and {Slutsky}, Jacob and {Sotiriou}, Thomas and {Speri}, Lorenzo and {Staab}, Martin and {Steer}, Dani{\`e}le and {Tamanini}, Nicola and {Tasinato}, Gianmassimo and {Torrado}, Jesus and {Torres-Orjuela}, Alejandro and {Toubiana}, Alexandre and {Vallisneri}, Michele and {Vecchio}, Alberto and {Volonteri}, Marta and {Yagi}, Kent and {Zwick}, Lorenz},
        title = "{LISA Definition Study Report}",
      journal = {arXiv e-prints},
         year = 2024,
        month = feb,
          eid = {arXiv:2402.07571},
        pages = {arXiv:2402.07571},
          doi = {10.48550/arXiv.2402.07571},
archivePrefix = {arXiv},
       eprint = {2402.07571},
 primaryClass = {astro-ph.CO},
       adsurl = {https://ui.adsabs.harvard.edu/abs/2024arXiv240207571C}
}

@ARTICLE{Volonteri2020,
       author = {{Volonteri}, Marta and {Pfister}, Hugo and {Beckmann}, Ricarda S. and {Dubois}, Yohan and {Colpi}, Monica and {Conselice}, Christopher J. and {Dotti}, Massimo and {Martin}, Garreth and {Jackson}, Ryan and {Kraljic}, Katarina and {Pichon}, Christophe and {Trebitsch}, Maxime and {Yi}, Sukyoung K. and {Devriendt}, Julien and {Peirani}, S{\'e}bastien},
        title = "{Black hole mergers from dwarf to massive galaxies with the NewHorizon and Horizon-AGN simulations}",
      journal = {\mnras},
         year = 2020,
        month = oct,
       volume = {498},
       number = {2},
        pages = {2219-2238},
          doi = {10.1093/mnras/staa2384},
archivePrefix = {arXiv},
       eprint = {2005.04902},
 primaryClass = {astro-ph.GA},
       adsurl = {https://ui.adsabs.harvard.edu/abs/2020MNRAS.498.2219V}
}

@ARTICLE{DeGraf2021,
       author = {{DeGraf}, Colin and {Sijacki}, Debora and {Di Matteo}, Tiziana and {Holley-Bockelmann}, Kelly and {Snyder}, Greg and {Springel}, Volker},
        title = "{Morphological evolution of supermassive black hole merger hosts and multimessenger signatures}",
      journal = {\mnras},
         year = 2021,
        month = may,
       volume = {503},
       number = {3},
        pages = {3629-3642},
          doi = {10.1093/mnras/stab721},
archivePrefix = {arXiv},
       eprint = {2012.00775},
 primaryClass = {astro-ph.GA},
       adsurl = {https://ui.adsabs.harvard.edu/abs/2021MNRAS.503.3629D}
}

@ARTICLE{Ogiya2020,
       author = {{Ogiya}, Go and {Hahn}, Oliver and {Mingarelli}, Chiara M.~F. and {Volonteri}, Marta},
        title = "{Accelerated orbital decay of supermassive black hole binaries in merging nuclear star clusters}",
      journal = {\mnras},
         year = 2020,
        month = apr,
       volume = {493},
       number = {3},
        pages = {3676-3689},
          doi = {10.1093/mnras/staa444},
archivePrefix = {arXiv},
       eprint = {1911.11526},
 primaryClass = {astro-ph.GA},
       adsurl = {https://ui.adsabs.harvard.edu/abs/2020MNRAS.493.3676O}
}

@ARTICLE{Bardati2024_morphology,
       author = {{Bardati}, Jaeden and {Ruan}, John J. and {Haggard}, Daryl and {Tremmel}, Michael},
        title = "{Signatures of Massive Black Hole Merger Host Galaxies from Cosmological Simulations. I. Unique Galaxy Morphologies in Imaging}",
      journal = {\apj},
         year = 2024,
        month = jan,
       volume = {961},
       number = {1},
          eid = {34},
        pages = {34},
          doi = {10.3847/1538-4357/ad055a},
archivePrefix = {arXiv},
       eprint = {2308.03828},
 primaryClass = {astro-ph.GA},
       adsurl = {https://ui.adsabs.harvard.edu/abs/2024ApJ...961...34B}
}

@ARTICLE{Gualandris2022,
       author = {{Gualandris}, Alessia and {Khan}, Fazeel Mahmood and {Bortolas}, Elisa and {Bonetti}, Matteo and {Sesana}, Alberto and {Berczik}, Peter and {Holley-Bockelmann}, Kelly},
        title = "{Eccentricity evolution of massive black hole binaries from formation to coalescence}",
      journal = {\mnras},
         year = 2022,
        month = apr,
       volume = {511},
       number = {4},
        pages = {4753-4765},
          doi = {10.1093/mnras/stac241},
archivePrefix = {arXiv},
       eprint = {2201.08646},
 primaryClass = {astro-ph.GA},
       adsurl = {https://ui.adsabs.harvard.edu/abs/2022MNRAS.511.4753G}
}

@ARTICLE{Ma2023,
       author = {{Ma}, Linhao and {Hopkins}, Philip F. and {Kelley}, Luke Zoltan and {Faucher-Gigu{\`e}re}, Claude-Andr{\'e}},
        title = "{A new discrete dynamical friction estimator based on N-body simulations}",
      journal = {\mnras},
         year = 2023,
        month = mar,
       volume = {519},
       number = {4},
        pages = {5543-5553},
          doi = {10.1093/mnras/stad036},
archivePrefix = {arXiv},
       eprint = {2208.12275},
 primaryClass = {astro-ph.GA},
       adsurl = {https://ui.adsabs.harvard.edu/abs/2023MNRAS.519.5543M}
}

@ARTICLE{Sesana2011,
       author = {{Sesana}, Alberto and {Gualandris}, Alessia and {Dotti}, Massimo},
        title = "{Massive black hole binary eccentricity in rotating stellar systems}",
      journal = {\mnras},
         year = 2011,
        month = jul,
       volume = {415},
       number = {1},
        pages = {L35-L39},
          doi = {10.1111/j.1745-3933.2011.01073.x},
archivePrefix = {arXiv},
       eprint = {1105.0670},
 primaryClass = {astro-ph.GA},
       adsurl = {https://ui.adsabs.harvard.edu/abs/2011MNRAS.415L..35S}
}

@ARTICLE{Bellovary2025,
       author = {{Bellovary}, Jillian and {Luo}, Yuantong and {Quinn}, Thomas R. and {Munshi}, Ferah and {Tremmel}, Michael and {Wadsley}, James},
        title = "{Intermediate Mass Ratio Inspirals in Milky Way Galaxies}",
      journal = {\apj},
         year = 2025,
        month = jun,
       volume = {986},
       number = {2},
          eid = {208},
        pages = {208},
          doi = {10.3847/1538-4357/add9a0},
archivePrefix = {arXiv},
       eprint = {2411.12117},
 primaryClass = {astro-ph.GA},
       adsurl = {https://ui.adsabs.harvard.edu/abs/2025ApJ...986..208B}
}

@ARTICLE{Tremmel2018_pairs,
       author = {{Tremmel}, M. and {Governato}, F. and {Volonteri}, M. and {Quinn}, T.~R. and {Pontzen}, A.},
        title = "{Dancing to CHANGA: a self-consistent prediction for close SMBH pair formation time-scales following galaxy mergers}",
      journal = {\mnras},
         year = 2018,
        month = apr,
       volume = {475},
       number = {4},
        pages = {4967-4977},
          doi = {10.1093/mnras/sty139},
archivePrefix = {arXiv},
       eprint = {1708.07126},
 primaryClass = {astro-ph.GA},
       adsurl = {https://ui.adsabs.harvard.edu/abs/2018MNRAS.475.4967T}
}

@ARTICLE{Blecha2011,
       author = {{Blecha}, Laura and {Cox}, Thomas J. and {Loeb}, Abraham and {Hernquist}, Lars},
        title = "{Recoiling black holes in merging galaxies: relationship to active galactic nucleus lifetimes, starbursts and the M$_{BH}$-{\ensuremath{\sigma}}$_{*}$ relation}",
      journal = {\mnras},
         year = 2011,
        month = apr,
       volume = {412},
       number = {4},
        pages = {2154-2182},
          doi = {10.1111/j.1365-2966.2010.18042.x},
archivePrefix = {arXiv},
       eprint = {1009.4940},
 primaryClass = {astro-ph.CO},
       adsurl = {https://ui.adsabs.harvard.edu/abs/2011MNRAS.412.2154B}
}

@ARTICLE{Petrov2024,
       author = {{Petrov}, Polina and {Taylor}, Stephen R. and {Charisi}, Maria and {Ma}, Chung-Pei},
        title = "{Identifying the Host Galaxies of Supermassive Black Hole Binaries Found by Pulsar Timing Arrays}",
      journal = {\apj},
         year = 2024,
        month = nov,
       volume = {976},
       number = {1},
          eid = {129},
        pages = {129},
          doi = {10.3847/1538-4357/ad7b14},
archivePrefix = {arXiv},
       eprint = {2406.04409},
 primaryClass = {astro-ph.GA},
       adsurl = {https://ui.adsabs.harvard.edu/abs/2024ApJ...976..129P}
}

@ARTICLE{Drake2025,
       author = {{Drake}, Carolyn L. and {Runnoe}, Jessie C. and {Stemo}, Aaron and {Bogdanovi{\'c}}, Tamara and {Eracleous}, Michael and {Ruan}, John and {Szekerczes}, Kaitlyn and {Futrowsky}, Kate and {Yu}, Weixiang},
        title = "{Preliminary census of galaxies in the LISA localization volume ─ I. Searching for LISA candidate massive black hole binary merger hosts using Sloan Digital Sky Survey photometry}",
      journal = {\mnras},
         year = 2025,
        month = oct,
       volume = {543},
       number = {1},
        pages = {375-389},
          doi = {10.1093/mnras/staf1375},
archivePrefix = {arXiv},
       eprint = {2508.01069},
 primaryClass = {astro-ph.GA},
       adsurl = {https://ui.adsabs.harvard.edu/abs/2025MNRAS.543..375D}
}

@ARTICLE{McDonald2026,
       author = {{McDonald}, Michael and {Khullar}, Gourav and {Lagattuta}, David and {Mahler}, Guillaume and {Dattathri}, Shashank and {Diego}, Jose M. and {Edge}, Alastair C. and {Floyd}, Benjamin and {Gladders}, Michael D. and {Hughes}, Scott A. and {Jauzac}, Mathilde and {Khonji}, Nader and {Leroy}, Gavin and {Massey}, Richard and {Montes}, Mireia and {Natarajan}, Priyamvada and {Reefe}, Michael and {Sharon}, Keren and {van den Bosch}, Frank and {Werner}, Stephane V. and {Zitrin}, Adi},
        title = "{A Kiloparsec-scale Stellar Cavity in the Center of A402-BCG May Be Caused by Dynamic Interactions with an Ultramassive Black Hole}",
      journal = {\apjl},
         year = 2026,
        month = may,
       volume = {1002},
       number = {1},
          eid = {L19},
        pages = {L19},
          doi = {10.3847/2041-8213/ae5bbe},
       adsurl = {https://ui.adsabs.harvard.edu/abs/2026ApJ..1002L..19M}
}

@ARTICLE{Harris2024,
       author = {{Harris}, C.~J. and {G{\"u}ltekin}, Kayhan},
        title = "{Connecting core galaxy properties to the massive black hole binary population}",
      journal = {\mnras},
         year = 2024,
        month = feb,
       volume = {528},
       number = {1},
        pages = {1-10},
          doi = {10.1093/mnras/stad3589},
archivePrefix = {arXiv},
       eprint = {2311.04877},
 primaryClass = {astro-ph.GA},
       adsurl = {https://ui.adsabs.harvard.edu/abs/2024MNRAS.528....1H}
}

@ARTICLE{Dosopoulou2021,
       author = {{Dosopoulou}, Fani and {Greene}, Jenny E. and {Ma}, Chung-Pei},
        title = "{Galaxy Core Formation by Supermassive Black Hole Binaries: The Importance of Realistic Initial Conditions and Galaxy Morphology}",
      journal = {\apj},
         year = 2021,
        month = nov,
       volume = {922},
       number = {1},
          eid = {40},
        pages = {40},
          doi = {10.3847/1538-4357/ac1fe4},
archivePrefix = {arXiv},
       eprint = {2108.08317},
 primaryClass = {astro-ph.GA},
       adsurl = {https://ui.adsabs.harvard.edu/abs/2021ApJ...922...40D}
}

@ARTICLE{Hannah2024,
       author = {{Hannah}, Christian H. and {Seth}, Anil C. and {Stone}, Nicholas C. and {van Velzen}, Sjoert},
        title = "{Counting the Unseen. I. Nuclear Density Scaling Relations for Nucleated Galaxies}",
      journal = {\aj},
         year = 2024,
        month = sep,
       volume = {168},
       number = {3},
          eid = {137},
        pages = {137},
          doi = {10.3847/1538-3881/ad630a},
archivePrefix = {arXiv},
       eprint = {2407.10911},
 primaryClass = {astro-ph.GA},
       adsurl = {https://ui.adsabs.harvard.edu/abs/2024AJ....168..137H}
}

@ARTICLE{Pinna2021,
       author = {{Pinna}, Francesca and {Neumayer}, Nadine and {Seth}, Anil and {Emsellem}, Eric and {Nguyen}, Dieu D. and {B{\"o}ker}, Torsten and {Cappellari}, Michele and {McDermid}, Richard M. and {Voggel}, Karina and {Walcher}, C. Jakob},
        title = "{Resolved Nuclear Kinematics Link the Formation and Growth of Nuclear Star Clusters with the Evolution of Their Early- and Late-type Hosts}",
      journal = {\apj},
         year = 2021,
        month = nov,
       volume = {921},
       number = {1},
          eid = {8},
        pages = {8},
          doi = {10.3847/1538-4357/ac158f},
archivePrefix = {arXiv},
       eprint = {2107.08903},
 primaryClass = {astro-ph.GA},
       adsurl = {https://ui.adsabs.harvard.edu/abs/2021ApJ...921....8P}
}

@ARTICLE{Tsatsi2017,
       author = {{Tsatsi}, Athanasia and {Mastrobuono-Battisti}, Alessandra and {van de Ven}, Glenn and {Perets}, Hagai B. and {Bianchini}, Paolo and {Neumayer}, Nadine},
        title = "{On the rotation of nuclear star clusters formed by cluster inspirals}",
      journal = {\mnras},
         year = 2017,
        month = jan,
       volume = {464},
       number = {3},
        pages = {3720-3727},
          doi = {10.1093/mnras/stw2593},
archivePrefix = {arXiv},
       eprint = {1610.01162},
 primaryClass = {astro-ph.GA},
       adsurl = {https://ui.adsabs.harvard.edu/abs/2017MNRAS.464.3720T}
}

@ARTICLE{Benson2005,
       author = {{Benson}, A.~J.},
        title = "{Orbital parameters of infalling dark matter substructures}",
      journal = {\mnras},
         year = 2005,
        month = apr,
       volume = {358},
       number = {2},
        pages = {551-562},
          doi = {10.1111/j.1365-2966.2005.08788.x},
archivePrefix = {arXiv},
       eprint = {astro-ph/0407428},
 primaryClass = {astro-ph},
       adsurl = {https://ui.adsabs.harvard.edu/abs/2005MNRAS.358..551B}
}

@ARTICLE{Wetzel2011,
       author = {{Wetzel}, Andrew R.},
        title = "{On the orbits of infalling satellite haloes}",
      journal = {\mnras},
         year = 2011,
        month = mar,
       volume = {412},
       number = {1},
        pages = {49-58},
          doi = {10.1111/j.1365-2966.2010.17877.x},
archivePrefix = {arXiv},
       eprint = {1001.4792},
 primaryClass = {astro-ph.CO},
       adsurl = {https://ui.adsabs.harvard.edu/abs/2011MNRAS.412...49W}
}

@ARTICLE{Francois2026,
       author = {{Francois}, Thibaut L. and {Gualandris}, Alessia and {Dehnen}, Walter},
        title = "{Scaling Relations for Binary Black Hole Merger Times from Cosmological Initial Conditions}",
      journal = {arXiv e-prints},
         year = 2026,
        month = jul,
          eid = {arXiv:2607.07813},
        pages = {arXiv:2607.07813},
          doi = {10.48550/arXiv.2607.07813},
archivePrefix = {arXiv},
       eprint = {2607.07813},
 primaryClass = {astro-ph.GA},
       adsurl = {https://ui.adsabs.harvard.edu/abs/2026arXiv260707813F}
}

@ARTICLE{Schramm2013,
       author = {{Schramm}, Malte and {Silverman}, John D.},
        title = "{The Black Hole-Bulge Mass Relation of Active Galactic Nuclei in the Extended Chandra Deep Field-South Survey}",
      journal = {\apj},
         year = 2013,
        month = apr,
       volume = {767},
       number = {1},
          eid = {13},
        pages = {13},
          doi = {10.1088/0004-637X/767/1/13},
archivePrefix = {arXiv},
       eprint = {1212.2999},
 primaryClass = {astro-ph.CO},
       adsurl = {https://ui.adsabs.harvard.edu/abs/2013ApJ...767...13S}
}

@BOOK{Binney1987,
       author = {{Binney}, James and {Tremaine}, Scott},
        title = "{Galactic dynamics}",
         year = 1987,
       adsurl = {https://ui.adsabs.harvard.edu/abs/1987gady.book.....B}
}

@ARTICLE{Chandrasekhar1943,
       author = {{Chandrasekhar}, S.},
        title = "{Dynamical Friction. I. General Considerations: the Coefficient of Dynamical Friction.}",
      journal = {\apj},
         year = 1943,
        month = mar,
       volume = {97},
        pages = {255},
          doi = {10.1086/144517},
       adsurl = {https://ui.adsabs.harvard.edu/abs/1943ApJ....97..255C}
}

@INPROCEEDINGS{Milosavljevic2003,
       author = {{Milosavljevi{\'c}}, Milo{\v{s}} and {Merritt}, David},
        title = "{The Final Parsec Problem}",
    booktitle = {The Astrophysics of Gravitational Wave Sources},
         year = 2003,
       editor = {{Centrella}, Joan M.},
       series = {American Institute of Physics Conference Series},
       volume = {686},
        month = oct,
    publisher = {AIP},
        pages = {201-210},
          doi = {10.1063/1.1629432},
archivePrefix = {arXiv},
       eprint = {astro-ph/0212270},
 primaryClass = {astro-ph},
       adsurl = {https://ui.adsabs.harvard.edu/abs/2003AIPC..686..201M}
}

@ARTICLE{Mayer2007,
       author = {{Mayer}, L. and {Kazantzidis}, S. and {Madau}, P. and {Colpi}, M. and {Quinn}, T. and {Wadsley}, J.},
        title = "{Rapid Formation of Supermassive Black Hole Binaries in Galaxy Mergers with Gas}",
      journal = {Science},
         year = 2007,
        month = jun,
       volume = {316},
       number = {5833},
        pages = {1874},
          doi = {10.1126/science.1141858},
archivePrefix = {arXiv},
       eprint = {0706.1562},
 primaryClass = {astro-ph},
       adsurl = {https://ui.adsabs.harvard.edu/abs/2007Sci...316.1874M}
}

@ARTICLE{Haiman2009,
       author = {{Haiman}, Zolt{\'a}n and {Kocsis}, Bence and {Menou}, Kristen},
        title = "{The Population of Viscosity- and Gravitational Wave-driven Supermassive Black Hole Binaries Among Luminous Active Galactic Nuclei}",
      journal = {\apj},
         year = 2009,
        month = aug,
       volume = {700},
       number = {2},
        pages = {1952-1969},
          doi = {10.1088/0004-637X/700/2/1952},
archivePrefix = {arXiv},
       eprint = {0904.1383},
 primaryClass = {astro-ph.CO},
       adsurl = {https://ui.adsabs.harvard.edu/abs/2009ApJ...700.1952H}
}

@ARTICLE{Alonso2024,
       author = {{Alonso-{\'A}lvarez}, Gonzalo and {Cline}, James M. and {Dewar}, Caitlyn},
        title = "{Self-Interacting Dark Matter Solves the Final Parsec Problem of Supermassive Black Hole Mergers}",
      journal = {\prl},
         year = 2024,
        month = jul,
       volume = {133},
       number = {2},
          eid = {021401},
        pages = {021401},
          doi = {10.1103/PhysRevLett.133.021401},
archivePrefix = {arXiv},
       eprint = {2401.14450},
 primaryClass = {astro-ph.CO},
       adsurl = {https://ui.adsabs.harvard.edu/abs/2024PhRvL.133b1401A}
}

@ARTICLE{Ma2021,
       author = {{Ma}, Linhao and {Hopkins}, Philip F. and {Ma}, Xiangcheng and {Angl{\'e}s-Alc{\'a}zar}, Daniel and {Faucher-Gigu{\`e}re}, Claude-Andr{\'e} and {Kelley}, Luke Zoltan},
        title = "{Seeds don't sink: even massive black hole 'seeds' cannot migrate to galaxy centres efficiently}",
      journal = {\mnras},
         year = 2021,
        month = dec,
       volume = {508},
       number = {2},
        pages = {1973-1985},
          doi = {10.1093/mnras/stab2713},
archivePrefix = {arXiv},
       eprint = {2101.02727},
 primaryClass = {astro-ph.GA},
       adsurl = {https://ui.adsabs.harvard.edu/abs/2021MNRAS.508.1973M}
}

@ARTICLE{Zhou2025,
       author = {{Zhou}, Yihao and {Mukherjee}, Diptajyoti and {Chen}, Nianyi and {Di Matteo}, Tiziana and {Johansson}, Peter H. and {Rantala}, Antti and {Partmann}, Christian and {Di Carlo}, Ugo Niccol{\`o} and {Bird}, Simeon and {Ni}, Yueying},
        title = "{MAGICS. II. Seed Black Holes Stripped of Their Surrounding Stars Do Not Sink}",
      journal = {\apj},
         year = 2025,
        month = feb,
       volume = {980},
       number = {1},
          eid = {79},
        pages = {79},
          doi = {10.3847/1538-4357/ada283},
archivePrefix = {arXiv},
       eprint = {2409.19914},
 primaryClass = {astro-ph.GA},
       adsurl = {https://ui.adsabs.harvard.edu/abs/2025ApJ...980...79Z}
}

@ARTICLE{Chen2022_DF,
       author = {{Chen}, Nianyi and {Ni}, Yueying and {Tremmel}, Michael and {Di Matteo}, Tiziana and {Bird}, Simeon and {DeGraf}, Colin and {Feng}, Yu},
        title = "{Dynamical friction modelling of massive black holes in cosmological simulations and effects on merger rate predictions}",
      journal = {\mnras},
         year = 2022,
        month = feb,
       volume = {510},
       number = {1},
        pages = {531-550},
          doi = {10.1093/mnras/stab3411},
archivePrefix = {arXiv},
       eprint = {2104.00021},
 primaryClass = {astro-ph.GA},
       adsurl = {https://ui.adsabs.harvard.edu/abs/2022MNRAS.510..531C}
}

@ARTICLE{Petiteau2008,
       author = {{Petiteau}, Antoine and {Auger}, G{\'e}rard and {Halloin}, Hubert and {Jeannin}, Olivier and {Plagnol}, Eric and {Pireaux}, Sophie and {Regimbau}, Tania and {Vinet}, Jean-Yves},
        title = "{LISACode: A scientific simulator of LISA}",
      journal = {\prd},
         year = 2008,
        month = jan,
       volume = {77},
       number = {2},
          eid = {023002},
        pages = {023002},
          doi = {10.1103/PhysRevD.77.023002},
archivePrefix = {arXiv},
       eprint = {0802.2023},
 primaryClass = {gr-qc},
       adsurl = {https://ui.adsabs.harvard.edu/abs/2008PhRvD..77b3002P}
}

@ARTICLE{Cuadra2009,
       author = {{Cuadra}, J. and {Armitage}, P.~J. and {Alexander}, R.~D. and {Begelman}, M.~C.},
        title = "{Massive black hole binary mergers within subparsec scale gas discs}",
      journal = {\mnras},
         year = 2009,
        month = mar,
       volume = {393},
       number = {4},
        pages = {1423-1432},
          doi = {10.1111/j.1365-2966.2008.14147.x},
archivePrefix = {arXiv},
       eprint = {0809.0311},
 primaryClass = {astro-ph},
       adsurl = {https://ui.adsabs.harvard.edu/abs/2009MNRAS.393.1423C}
}

@ARTICLE{Siwek2023,
       author = {{Siwek}, Magdalena and {Weinberger}, Rainer and {Hernquist}, Lars},
        title = "{Orbital evolution of binaries in circumbinary discs}",
      journal = {\mnras},
         year = 2023,
        month = jun,
       volume = {522},
       number = {2},
        pages = {2707-2717},
          doi = {10.1093/mnras/stad1131},
archivePrefix = {arXiv},
       eprint = {2302.01785},
 primaryClass = {astro-ph.HE},
       adsurl = {https://ui.adsabs.harvard.edu/abs/2023MNRAS.522.2707S}
}

@ARTICLE{Clyburn2026,
       author = {{Clyburn}, Madeline and {Zrake}, Jonathan},
        title = "{Unequal mass binary evolution driven by high-Mach circumbinary discs}",
      journal = {\mnras},
         year = 2026,
        month = may,
       volume = {548},
       number = {1},
          eid = {stag567},
        pages = {stag567},
          doi = {10.1093/mnras/stag567},
archivePrefix = {arXiv},
       eprint = {2603.27022},
 primaryClass = {astro-ph.HE},
       adsurl = {https://ui.adsabs.harvard.edu/abs/2026MNRAS.548ag567C}
}

@ARTICLE{Shihong2024,
       author = {{Liao}, Shihong and {Irodotou}, Dimitrios and {Johansson}, Peter H. and {Naab}, Thorsten and {Rizzuto}, Francesco Paolo and {Hislop}, Jessica M. and {Rawlings}, Alexander and {Wright}, Ruby J.},
        title = "{RABBITS - I. The crucial role of nuclear star formation in driving the coalescence of supermassive black hole binaries}",
      journal = {\mnras},
         year = 2024,
        month = mar,
       volume = {528},
       number = {3},
        pages = {5080-5097},
          doi = {10.1093/mnras/stae360},
archivePrefix = {arXiv},
       eprint = {2311.01499},
 primaryClass = {astro-ph.GA},
       adsurl = {https://ui.adsabs.harvard.edu/abs/2024MNRAS.528.5080L}
}

@ARTICLE{Wright2026,
       author = {{Wright}, Ruby J. and {Heiskanen}, Roosa and {Liao}, Shihong and {Rawlings}, Alexander and {Johansson}, Peter H. and {Mattero}, Max and {Panther}, Fiona H. and {Keitaanranta}, Atte},
        title = "{RABBITS IV: Stellar feedback and SMBH merging time-scales in the sub-Milky Way mass regime}",
      journal = {arXiv e-prints},
         year = 2026,
        month = jul,
          eid = {arXiv:2607.13786},
        pages = {arXiv:2607.13786},
          doi = {10.48550/arXiv.2607.13786},
archivePrefix = {arXiv},
       eprint = {2607.13786},
 primaryClass = {astro-ph.GA},
       adsurl = {https://ui.adsabs.harvard.edu/abs/2026arXiv260713786W}
}

@ARTICLE{Barre2026,
       author = {{Barr{\'e}}, Ruxin and {Babul}, Arif and {Gozaliasl}, Ghassem and {Finoguenov}, Alexis and {Dav{\'e}}, Romeel and {Padawer-Blatt}, Aviv and {Rennehan}, Douglas and {Saeedzadeh}, Vida and {Hough}, Renier T. and {Quinn}, Thomas R.},
        title = "{Forged by Feedback: Stellar Properties of Brightest Group Galaxies in Cosmological Simulations}",
      journal = {\apj},
         year = 2026,
        month = apr,
       volume = {1001},
       number = {1},
          eid = {122},
        pages = {122},
          doi = {10.3847/1538-4357/ae4e2a},
archivePrefix = {arXiv},
       eprint = {2602.10192},
 primaryClass = {astro-ph.GA},
       adsurl = {https://ui.adsabs.harvard.edu/abs/2026ApJ..1001..122B}
}

@ARTICLE{IzquierdoVillalba2026_recoil,
       author = {{Izquierdo-Villalba}, David},
        title = "{Overmassive and Undermassive Massive Black Holes: The Role of Environment and Gravitational-Wave Recoils}",
      journal = {\aap},
         year = 2026,
        month = apr,
       volume = {708},
          eid = {A303},
        pages = {A303},
          doi = {10.1051/0004-6361/202659681},
       adsurl = {https://ui.adsabs.harvard.edu/abs/2026A&A...708A.303I}
}

@ARTICLE{Quinlan1996,
       author = {{Quinlan}, Gerald D.},
        title = "{The dynamical evolution of massive black hole binaries I. Hardening in a fixed stellar background}",
      journal = {\na},
         year = 1996,
        month = jul,
       volume = {1},
       number = {1},
        pages = {35-56},
          doi = {10.1016/S1384-1076(96)00003-6},
archivePrefix = {arXiv},
       eprint = {astro-ph/9601092},
 primaryClass = {astro-ph},
       adsurl = {https://ui.adsabs.harvard.edu/abs/1996NewA....1...35Q}
}

@ARTICLE{Biava2019,
       author = {{Biava}, Nadia and {Colpi}, Monica and {Capelo}, Pedro R. and {Bonetti}, Matteo and {Volonteri}, Marta and {Tamfal}, Tomas and {Mayer}, Lucio and {Sesana}, Alberto},
        title = "{The lifetime of binary black holes in S{\'e}rsic galaxy models}",
      journal = {\mnras},
         year = 2019,
        month = aug,
       volume = {487},
       number = {4},
        pages = {4985-4994},
          doi = {10.1093/mnras/stz1614},
archivePrefix = {arXiv},
       eprint = {1903.05682},
 primaryClass = {astro-ph.GA},
       adsurl = {https://ui.adsabs.harvard.edu/abs/2019MNRAS.487.4985B}
}

@ARTICLE{KelleyBlechaHernquist2017,
       author = {{Kelley}, Luke Zoltan and {Blecha}, Laura and {Hernquist}, Lars},
        title = "{Massive black hole binary mergers in dynamical galactic environments}",
      journal = {\mnras},
         year = 2017,
        month = jan,
       volume = {464},
       number = {3},
        pages = {3131-3157},
          doi = {10.1093/mnras/stw2452},
archivePrefix = {arXiv},
       eprint = {1606.01900},
 primaryClass = {astro-ph.HE},
       adsurl = {https://ui.adsabs.harvard.edu/abs/2017MNRAS.464.3131K}
}

@ARTICLE{Kelley2017,
       author = {{Kelley}, Luke Zoltan and {Blecha}, Laura and {Hernquist}, Lars and {Sesana}, Alberto and {Taylor}, Stephen R.},
        title = "{The gravitational wave background from massive black hole binaries in Illustris: spectral features and time to detection with pulsar timing arrays}",
      journal = {\mnras},
         year = 2017,
        month = nov,
       volume = {471},
       number = {4},
        pages = {4508-4526},
          doi = {10.1093/mnras/stx1638},
archivePrefix = {arXiv},
       eprint = {1702.02180},
 primaryClass = {astro-ph.HE},
       adsurl = {https://ui.adsabs.harvard.edu/abs/2017MNRAS.471.4508K}
}

@ARTICLE{Mastrobuono2023,
       author = {{Mastrobuono-Battisti}, Alessandra and {Ogiya}, Go and {Hahn}, Oliver and {Schultheis}, Mathias},
        title = "{Searching for clues of past binary supermassive black hole mergers in nuclear star clusters}",
      journal = {\mnras},
         year = 2023,
        month = jun,
       volume = {521},
       number = {4},
        pages = {6089-6104},
          doi = {10.1093/mnras/stad898},
archivePrefix = {arXiv},
       eprint = {2303.12826},
 primaryClass = {astro-ph.GA},
       adsurl = {https://ui.adsabs.harvard.edu/abs/2023MNRAS.521.6089M}
}

@ARTICLE{Wadsley2004,
       author = {{Wadsley}, J.~W. and {Stadel}, J. and {Quinn}, T.},
        title = "{Gasoline: a flexible, parallel implementation of TreeSPH}",
      journal = {\na},
         year = 2004,
        month = feb,
       volume = {9},
       number = {2},
        pages = {137-158},
          doi = {10.1016/j.newast.2003.08.004},
archivePrefix = {arXiv},
       eprint = {astro-ph/0303521},
 primaryClass = {astro-ph},
       adsurl = {https://ui.adsabs.harvard.edu/abs/2004NewA....9..137W}
}

@ARTICLE{Haardt2012,
       author = {{Haardt}, Francesco and {Madau}, Piero},
        title = "{Radiative Transfer in a Clumpy Universe. IV. New Synthesis Models of the Cosmic UV/X-Ray Background}",
      journal = {\apj},
         year = 2012,
        month = feb,
       volume = {746},
       number = {2},
          eid = {125},
        pages = {125},
          doi = {10.1088/0004-637X/746/2/125},
       adsurl = {https://ui.adsabs.harvard.edu/abs/2012ApJ...746..125H}
}

@ARTICLE{Guedes2011,
       author = {{Guedes}, Javiera and {Callegari}, Simone and {Madau}, Piero and {Mayer}, Lucio},
        title = "{Forming Realistic Late-type Spirals in a {\ensuremath{\Lambda}}CDM Universe: The Eris Simulation}",
      journal = {\apj},
         year = 2011,
        month = dec,
       volume = {742},
       number = {2},
          eid = {76},
        pages = {76},
          doi = {10.1088/0004-637X/742/2/76},
archivePrefix = {arXiv},
       eprint = {1103.6030},
 primaryClass = {astro-ph.CO},
       adsurl = {https://ui.adsabs.harvard.edu/abs/2011ApJ...742...76G}
}

@ARTICLE{Stinson2006,
       author = {{Stinson}, Greg and {Seth}, Anil and {Katz}, Neal and {Wadsley}, James and {Governato}, Fabio and {Quinn}, Tom},
        title = "{Star formation and feedback in smoothed particle hydrodynamic simulations - I. Isolated galaxies}",
      journal = {\mnras},
         year = 2006,
        month = dec,
       volume = {373},
       number = {3},
        pages = {1074-1090},
          doi = {10.1111/j.1365-2966.2006.11097.x},
archivePrefix = {arXiv},
       eprint = {astro-ph/0602350},
 primaryClass = {astro-ph},
       adsurl = {https://ui.adsabs.harvard.edu/abs/2006MNRAS.373.1074S}
}

@ARTICLE{Genina2024,
       author = {{Genina}, Anna and {Springel}, Volker and {Rantala}, Antti},
        title = "{A calibrated model for N-body dynamical friction acting on supermassive black holes}",
      journal = {\mnras},
         year = 2024,
        month = oct,
       volume = {534},
       number = {1},
        pages = {957-977},
          doi = {10.1093/mnras/stae2144},
archivePrefix = {arXiv},
       eprint = {2405.08870},
 primaryClass = {astro-ph.GA},
       adsurl = {https://ui.adsabs.harvard.edu/abs/2024MNRAS.534..957G}
}

@ARTICLE{Bellovary2011,
       author = {{Bellovary}, Jillian and {Volonteri}, Marta and {Governato}, Fabio and {Shen}, Sijing and {Quinn}, Thomas and {Wadsley}, James},
        title = "{The First Massive Black Hole Seeds and Their Hosts}",
      journal = {\apj},
         year = 2011,
        month = nov,
       volume = {742},
       number = {1},
          eid = {13},
        pages = {13},
          doi = {10.1088/0004-637X/742/1/13},
archivePrefix = {arXiv},
       eprint = {1104.3858},
 primaryClass = {astro-ph.CO},
       adsurl = {https://ui.adsabs.harvard.edu/abs/2011ApJ...742...13B}
}

@ARTICLE{Gualandris2012,
       author = {{Gualandris}, Alessia and {Dotti}, Massimo and {Sesana}, Alberto},
        title = "{Massive black hole binary plane reorientation in rotating stellar systems}",
      journal = {\mnras},
         year = 2012,
        month = feb,
       volume = {420},
       number = {1},
        pages = {L38-L42},
          doi = {10.1111/j.1745-3933.2011.01188.x},
archivePrefix = {arXiv},
       eprint = {1109.3707},
 primaryClass = {astro-ph.GA},
       adsurl = {https://ui.adsabs.harvard.edu/abs/2012MNRAS.420L..38G}
}

@ARTICLE{Antonini2015_coevo,
       author = {{Antonini}, Fabio and {Barausse}, Enrico and {Silk}, Joseph},
        title = "{The Coevolution of Nuclear Star Clusters, Massive Black Holes, and Their Host Galaxies}",
      journal = {\apj},
         year = 2015,
        month = oct,
       volume = {812},
       number = {1},
          eid = {72},
        pages = {72},
          doi = {10.1088/0004-637X/812/1/72},
archivePrefix = {arXiv},
       eprint = {1506.02050},
 primaryClass = {astro-ph.GA},
       adsurl = {https://ui.adsabs.harvard.edu/abs/2015ApJ...812...72A}
}

@ARTICLE{Kritos2025,
       author = {{Kritos}, Konstantinos and {Beckmann}, Ricarda S. and {Silk}, Joseph and {Berti}, Emanuele and {Yi}, Sophia and {Volonteri}, Marta and {Dubois}, Yohan and {Devriendt}, Julien},
        title = "{Supermassive Black Hole Growth in Hierarchically Merging Nuclear Star Clusters}",
      journal = {\apj},
         year = 2025,
        month = sep,
       volume = {991},
       number = {1},
          eid = {58},
        pages = {58},
          doi = {10.3847/1538-4357/adeb44},
archivePrefix = {arXiv},
       eprint = {2412.15334},
 primaryClass = {astro-ph.GA},
       adsurl = {https://ui.adsabs.harvard.edu/abs/2025ApJ...991...58K}
}

@ARTICLE{Merritt2006,
       author = {{Merritt}, David},
        title = "{Mass Deficits, Stalling Radii, and the Merger Histories of Elliptical Galaxies}",
      journal = {\apj},
         year = 2006,
        month = sep,
       volume = {648},
       number = {2},
        pages = {976-986},
          doi = {10.1086/506139},
archivePrefix = {arXiv},
       eprint = {astro-ph/0603439},
 primaryClass = {astro-ph},
       adsurl = {https://ui.adsabs.harvard.edu/abs/2006ApJ...648..976M}
}

@ARTICLE{Li2020,
       author = {{Li}, Zhao-Zhou and {Zhao}, Dong-Hai and {Jing}, Y.~P. and {Han}, Jiaxin and {Dong}, Fu-Yu},
        title = "{Orbital Distribution of Infalling Satellite Halos across Cosmic Time}",
      journal = {\apj},
         year = 2020,
        month = dec,
       volume = {905},
       number = {2},
          eid = {177},
        pages = {177},
          doi = {10.3847/1538-4357/abc481},
archivePrefix = {arXiv},
       eprint = {2008.05710},
 primaryClass = {astro-ph.CO},
       adsurl = {https://ui.adsabs.harvard.edu/abs/2020ApJ...905..177L}
}

@ARTICLE{Antonini2015_imprint,
       author = {{Antonini}, Fabio and {Barausse}, Enrico and {Silk}, Joseph},
        title = "{The Imprint of Massive Black Hole Mergers on the Correlation between Nuclear Star Clusters and Their Host Galaxies}",
      journal = {\apjl},
         year = 2015,
        month = jun,
       volume = {806},
       number = {1},
          eid = {L8},
        pages = {L8},
          doi = {10.1088/2041-8205/806/1/L8},
archivePrefix = {arXiv},
       eprint = {1504.04033},
 primaryClass = {astro-ph.GA},
       adsurl = {https://ui.adsabs.harvard.edu/abs/2015ApJ...806L...8A}
}

@ARTICLE{Munshi2013,
       author = {{Munshi}, Ferah and {Governato}, F. and {Brooks}, A.~M. and {Christensen}, C. and {Shen}, S. and {Loebman}, S. and {Moster}, B. and {Quinn}, T. and {Wadsley}, J.},
        title = "{Reproducing the Stellar Mass/Halo Mass Relation in Simulated {\ensuremath{\Lambda}}CDM Galaxies: Theory versus Observational Estimates}",
      journal = {\apj},
         year = 2013,
        month = mar,
       volume = {766},
       number = {1},
          eid = {56},
        pages = {56},
          doi = {10.1088/0004-637X/766/1/56},
archivePrefix = {arXiv},
       eprint = {1209.1389},
 primaryClass = {astro-ph.CO},
       adsurl = {https://ui.adsabs.harvard.edu/abs/2013ApJ...766...56M}
}

@ARTICLE{Bardati2024_kinematics,
       author = {{Bardati}, Jaeden and {Ruan}, John J. and {Haggard}, Daryl and {Tremmel}, Michael and {Horlaville}, Patrick},
        title = "{Signatures of Massive Black Hole Merger Host Galaxies from Cosmological Simulations. II. Unique Stellar Kinematics in Integral Field Unit Spectroscopy}",
      journal = {\apj},
         year = 2024,
        month = dec,
       volume = {977},
       number = {2},
          eid = {265},
        pages = {265},
          doi = {10.3847/1538-4357/ad9471},
archivePrefix = {arXiv},
       eprint = {2407.14061},
 primaryClass = {astro-ph.GA},
       adsurl = {https://ui.adsabs.harvard.edu/abs/2024ApJ...977..265B}
}

@ARTICLE{Georgiev2016,
       author = {{Georgiev}, Iskren Y. and {B{\"o}ker}, Torsten and {Leigh}, Nathan and {L{\"u}tzgendorf}, Nora and {Neumayer}, Nadine},
        title = "{Masses and scaling relations for nuclear star clusters, and their co-existence with central black holes}",
      journal = {\mnras},
         year = 2016,
        month = apr,
       volume = {457},
       number = {2},
        pages = {2122-2138},
          doi = {10.1093/mnras/stw093},
archivePrefix = {arXiv},
       eprint = {1601.02613},
 primaryClass = {astro-ph.GA},
       adsurl = {https://ui.adsabs.harvard.edu/abs/2016MNRAS.457.2122G}
}

@ARTICLE{Tremmel2024,
       author = {{Tremmel}, Michael and {Ricarte}, Angelo and {Natarajan}, Priyamvada and {Bellovary}, Jillian and {Sharma}, Ray and {Quinn}, Thomas R.},
        title = "{An Enhanced Massive Black Hole Occupation Fraction Predicted in Cluster Dwarf Galaxies}",
      journal = {The Open Journal of Astrophysics},
         year = 2024,
        month = apr,
       volume = {7},
          eid = {26},
        pages = {26},
          doi = {10.33232/001c.116617},
archivePrefix = {arXiv},
       eprint = {2306.12813},
 primaryClass = {astro-ph.GA},
       adsurl = {https://ui.adsabs.harvard.edu/abs/2024OJAp....7E..26T}
}

@ARTICLE{Hoyer2021,
       author = {{Hoyer}, Nils and {Neumayer}, Nadine and {Georgiev}, Iskren Y. and {Seth}, Anil C. and {Greene}, Jenny E.},
        title = "{The nucleation fraction of local volume galaxies}",
      journal = {\mnras},
         year = 2021,
        month = nov,
       volume = {507},
       number = {3},
        pages = {3246-3266},
          doi = {10.1093/mnras/stab2277},
archivePrefix = {arXiv},
       eprint = {2107.05313},
 primaryClass = {astro-ph.GA},
       adsurl = {https://ui.adsabs.harvard.edu/abs/2021MNRAS.507.3246H}
}

@ARTICLE{Carlsten2022,
       author = {{Carlsten}, Scott G. and {Greene}, Jenny E. and {Beaton}, Rachael L. and {Greco}, Johnny P.},
        title = "{ELVES II: Globular Clusters and Nuclear Star Clusters of Dwarf Galaxies: the Importance of Environment}",
      journal = {\apj},
         year = 2022,
        month = mar,
       volume = {927},
       number = {1},
          eid = {44},
        pages = {44},
          doi = {10.3847/1538-4357/ac457e},
archivePrefix = {arXiv},
       eprint = {2105.03440},
 primaryClass = {astro-ph.GA},
       adsurl = {https://ui.adsabs.harvard.edu/abs/2022ApJ...927...44C}
}

@ARTICLE{Pechetti2020,
       author = {{Pechetti}, Renuka and {Seth}, Anil and {Neumayer}, Nadine and {Georgiev}, Iskren and {Kacharov}, Nikolay and {den Brok}, Mark},
        title = "{Luminosity Models and Density Profiles for Nuclear Star Clusters for a Nearby Volume-limited Sample of 29 Galaxies}",
      journal = {\apj},
         year = 2020,
        month = sep,
       volume = {900},
       number = {1},
          eid = {32},
        pages = {32},
          doi = {10.3847/1538-4357/abaaa7},
archivePrefix = {arXiv},
       eprint = {1911.09686},
 primaryClass = {astro-ph.GA},
       adsurl = {https://ui.adsabs.harvard.edu/abs/2020ApJ...900...32P}
}

@ARTICLE{Bhowmick2025,
       author = {{Bhowmick}, Aklant K. and {Blecha}, Laura and {Kelley}, Luke Z. and {Sivasankaran}, Aneesh and {Torrey}, Paul and {Weinberger}, Rainer and {Chen}, Nianyi and {Vogelsberger}, Mark and {Hernquist}, Lars and {Natarajan}, Priyamvada and {Di Matteo}, Tiziana},
        title = "{Dynamics of Low-mass Black Hole Seeds in the BRAHMA Simulations Using Subgrid Dynamical Friction: Impact on Merger-driven Black Hole Growth in the High-redshift Universe}",
      journal = {\apj},
         year = 2025,
        month = sep,
       volume = {991},
       number = {1},
          eid = {81},
        pages = {81},
          doi = {10.3847/1538-4357/adf96b},
archivePrefix = {arXiv},
       eprint = {2506.09184},
 primaryClass = {astro-ph.GA},
       adsurl = {https://ui.adsabs.harvard.edu/abs/2025ApJ...991...81B}
}

@ARTICLE{Horlaville2026,
       author = {{Horlaville}, Patrick and {Ruan}, John J. and {Eracleous}, Michael and {Bardati}, Jaeden and {Runnoe}, Jessie C. and {Haggard}, Daryl},
        title = "{Predicting Potential Host Galaxies of Supermassive Black Hole Binaries Based on Stellar Kinematics in Archival IFU Surveys}",
      journal = {\apj},
         year = 2026,
        month = jun,
       volume = {1003},
       number = {2},
          eid = {229},
        pages = {229},
          doi = {10.3847/1538-4357/ae6662},
archivePrefix = {arXiv},
       eprint = {2504.21145},
 primaryClass = {astro-ph.GA},
       adsurl = {https://ui.adsabs.harvard.edu/abs/2026ApJ..1003..229H}
}

@ARTICLE{Iwasawa2011,
       author = {{Iwasawa}, Masaki and {An}, Sangyong and {Matsubayashi}, Tatsushi and {Funato}, Yoko and {Makino}, Junichiro},
        title = "{Eccentric Evolution of Supermassive Black Hole Binaries}",
      journal = {\apjl},
         year = 2011,
        month = apr,
       volume = {731},
       number = {1},
          eid = {L9},
        pages = {L9},
          doi = {10.1088/2041-8205/731/1/L9},
archivePrefix = {arXiv},
       eprint = {1011.4017},
 primaryClass = {astro-ph.GA},
       adsurl = {https://ui.adsabs.harvard.edu/abs/2011ApJ...731L...9I}
}

@ARTICLE{Baile15,
   author = {{Li}, B. and {Holley-Bockelmann}, K. and {Khan}, F.~M.},
    title = "{Classification of Stellar Orbits in Axisymmetric Galaxies}",
  journal = {\apj},
archivePrefix = "arXiv",
   eprint = {1412.2134},
     year = 2015,
    month = sep,
   volume = 811,
      eid = {25},
    pages = {25},
      doi = {10.1088/0004-637X/811/1/25},
   adsurl = {http://adsabs.harvard.edu/abs/2015ApJ...811...25L}
}

@ARTICLE{Khan2026,
       author = {{Khan}, Fazeel Mahmood and {Berczik}, Peter and {Sobolenko}, Margaryta and {Just}, Andreas and {Spurzem}, Rainer and {Holley-Bockelmann}, Kelly and {Macci{\`o}}, Andrea V.},
        title = "{Intermediate-mass black hole binary evolution in nuclear star clusters: The effect of the stellar-mass black hole population}",
      journal = {\aap},
         year = 2026,
        month = feb,
       volume = {706},
          eid = {A354},
        pages = {A354},
          doi = {10.1051/0004-6361/202557283},
archivePrefix = {arXiv},
       eprint = {2512.11038},
 primaryClass = {astro-ph.GA},
       adsurl = {https://ui.adsabs.harvard.edu/abs/2026A&A...706A.354K}
}

@ARTICLE{Koudmani2026,
       author = {{Koudmani}, Sophie and {Scholtz}, Jan and {Taylor}, Anthony J. and {Juod{\v{z}}balis}, Ignas and {Sijacki}, Debora and {Somerville}, Rachel S. and {Maiolino}, Roberto and {Curtis-Lake}, Emma and {Finkelstein}, Steven L. and {Bourne}, Martin A. and {D'Eugenio}, Francesco and {Geris}, Sophia and {Ivey}, Lucy R. and {{\"U}bler}, Hannah},
        title = "{How to raise a supermassive black hole: interpreting early JWST AGN with the AESOPICA simulations}",
      journal = {arXiv e-prints},
         year = 2026,
        month = jul,
          eid = {arXiv:2607.26177},
        pages = {arXiv:2607.26177},
          doi = {10.48550/arXiv.2607.26177},
archivePrefix = {arXiv},
       eprint = {2607.26177},
 primaryClass = {astro-ph.GA},
       adsurl = {https://ui.adsabs.harvard.edu/abs/2026arXiv260726177K}
}

@ARTICLE{Dunn2018,
       author = {{Dunn}, Glenna and {Bellovary}, Jillian and {Holley-Bockelmann}, Kelly and {Christensen}, Charlotte and {Quinn}, Thomas},
        title = "{Sowing Black Hole Seeds: Direct Collapse Black Hole Formation with Realistic Lyman-Werner Radiation in Cosmological Simulations}",
      journal = {\apj},
         year = 2018,
        month = jul,
       volume = {861},
       number = {1},
          eid = {39},
        pages = {39},
          doi = {10.3847/1538-4357/aac7c2},
archivePrefix = {arXiv},
       eprint = {1803.01007},
 primaryClass = {astro-ph.GA},
       adsurl = {https://ui.adsabs.harvard.edu/abs/2018ApJ...861...39D}
}

@ARTICLE{Bhowmick2022,
       author = {{Bhowmick}, Aklant K. and {Blecha}, Laura and {Torrey}, Paul and {Kelley}, Luke Zoltan and {Vogelsberger}, Mark and {Nelson}, Dylan and {Weinberger}, Rainer and {Hernquist}, Lars},
        title = "{Impact of gas spin and Lyman-Werner flux on black hole seed formation in cosmological simulations: implications for direct collapse}",
      journal = {\mnras},
         year = 2022,
        month = feb,
       volume = {510},
       number = {1},
        pages = {177-196},
          doi = {10.1093/mnras/stab3439},
archivePrefix = {arXiv},
       eprint = {2107.06899},
 primaryClass = {astro-ph.GA},
       adsurl = {https://ui.adsabs.harvard.edu/abs/2022MNRAS.510..177B}
}

@ARTICLE{Prole2024,
       author = {{Prole}, Lewis R. and {Regan}, John A. and {Glover}, Simon C.~O. and {Klessen}, Ralf S. and {Priestley}, Felix D. and {Clark}, Paul C.},
        title = "{Heavy black hole seed formation in high-z atomic cooling halos}",
      journal = {\aap},
         year = 2024,
        month = may,
       volume = {685},
          eid = {A31},
        pages = {A31},
          doi = {10.1051/0004-6361/202348903},
archivePrefix = {arXiv},
       eprint = {2312.06769},
 primaryClass = {astro-ph.GA},
       adsurl = {https://ui.adsabs.harvard.edu/abs/2024A&A...685A..31P}
}

@ARTICLE{Prole2026,
       author = {{Prole}, Lewis Robert and {Regan}, John A. and {Mehta}, Daxal and {Pakmor}, Rudiger and {Koudmani}, Sophie and {Bourne}, Martin A. and {Glover}, Simon C.~O. and {Wise}, John H. and {Klessen}, Ralf S. and {Tremmel}, Michael and {Sijacki}, Debora and {Beckmann}, Ricarda S. and {Haehnelt}, Martin G. and {Brennan}, John and {van de Bor}, Pelle and {Clark}, Paul C.},
        title = "{The SEEDZ Simulations: Methodology and First Results on Massive Black Hole Seeding and Early Galaxy Growth}",
      journal = {The Open Journal of Astrophysics},
         year = 2026,
        month = mar,
       volume = {9},
        pages = {59236},
          doi = {10.33232/001c.159236},
archivePrefix = {arXiv},
       eprint = {2511.09640},
 primaryClass = {astro-ph.GA},
       adsurl = {https://ui.adsabs.harvard.edu/abs/2026OJAp....959236P}
}

@ARTICLE{Unal2020,
       author = {{{\"U}nal}, Caner and {Loeb}, Abraham},
        title = "{On Spin dependence of the Fundamental Plane of black hole activity}",
      journal = {\mnras},
         year = 2020,
        month = jun,
       volume = {495},
       number = {1},
        pages = {278-284},
          doi = {10.1093/mnras/staa1119},
archivePrefix = {arXiv},
       eprint = {2002.11778},
 primaryClass = {astro-ph.HE},
       adsurl = {https://ui.adsabs.harvard.edu/abs/2020MNRAS.495..278U}
}

@ARTICLE{Nemmen2007,
       author = {{Nemmen}, R.~S. and {Bower}, R.~G. and {Babul}, A. and {Storchi-Bergmann}, T.},
        title = "{Models for jet power in elliptical galaxies: a case for rapidly spinning black holes}",
      journal = {\mnras},
         year = 2007,
        month = jun,
       volume = {377},
       number = {4},
        pages = {1652-1662},
          doi = {10.1111/j.1365-2966.2007.11726.x},
archivePrefix = {arXiv},
       eprint = {astro-ph/0612354},
 primaryClass = {astro-ph},
       adsurl = {https://ui.adsabs.harvard.edu/abs/2007MNRAS.377.1652N}
}

@INPROCEEDINGS{Tchekhovskoy2012,
       author = {{Tchekhovskoy}, Alexander and {McKinney}, Jonathan C. and {Narayan}, Ramesh},
        title = "{General Relativistic Modeling of Magnetized Jets from Accreting Black Holes}",
    booktitle = {Journal of Physics Conference Series},
         year = 2012,
       series = {Journal of Physics Conference Series},
       volume = {372},
        month = jul,
    publisher = {IOP},
          eid = {012040},
        pages = {012040},
          doi = {10.1088/1742-6596/372/1/012040},
archivePrefix = {arXiv},
       eprint = {1202.2864},
 primaryClass = {astro-ph.HE},
       adsurl = {https://ui.adsabs.harvard.edu/abs/2012JPhCS.372a2040T}
}

@ARTICLE{Kormendy2013,
       author = {{Kormendy}, John and {Ho}, Luis C.},
        title = "{Coevolution (Or Not) of Supermassive Black Holes and Host Galaxies}",
      journal = {\araa},
         year = 2013,
        month = aug,
       volume = {51},
       number = {1},
        pages = {511-653},
          doi = {10.1146/annurev-astro-082708-101811},
archivePrefix = {arXiv},
       eprint = {1304.7762},
 primaryClass = {astro-ph.CO},
       adsurl = {https://ui.adsabs.harvard.edu/abs/2013ARA&A..51..511K}
}

@ARTICLE{Begelman2006,
       author = {{Begelman}, Mitchell C. and {Volonteri}, Marta and {Rees}, Martin J.},
        title = "{Formation of supermassive black holes by direct collapse in pre-galactic haloes}",
      journal = {\mnras},
         year = 2006,
        month = jul,
       volume = {370},
       number = {1},
        pages = {289-298},
          doi = {10.1111/j.1365-2966.2006.10467.x},
archivePrefix = {arXiv},
       eprint = {astro-ph/0602363},
 primaryClass = {astro-ph},
       adsurl = {https://ui.adsabs.harvard.edu/abs/2006MNRAS.370..289B}
}






it can be placed in an Appendix which appears after the list of references.


\bsp	
\label{lastpage}
\end{document}